\documentclass[aps,prx,a4paper,superscriptaddress,floatfix,notitlepage,longbibliography]{revtex4-1}

\usepackage{graphicx,color}
\usepackage{graphics}
\usepackage{bm} % bold math
\usepackage{amsfonts,amsmath,amssymb,amsbsy,amscd,latexsym}

\definecolor{burntorange}{rgb}{0.8, 0.33, 0.0}

\newcommand{\bc}{\begin{center}}
\newcommand{\ec}{\end{center}}
\newcommand{\be}{\begin{equation}}
\newcommand{\ee}{\end{equation}}
\newcommand{\ba}{\begin{eqnarray*}}
\newcommand{\ea}{\end{eqnarray*}}
\newcommand{\bna}{\begin{eqnarray}}
\newcommand{\ena}{\end{eqnarray}}

\newcommand{\mpaa}{\begin{minipage}[t]{7.5cm}}
\newcommand{\mpea}{\end{minipage}}
\newcommand{\ud}{\mathop{}\!\mathrm{d}}

\begin{document}

\title{Extended Poisson-Kac theory:\\ A unifying framework for
  stochastic processes with finite propagation velocity}

\author{Massimiliano Giona}
\email{massimiliano.giona@uniroma1.it}
\affiliation{Facolt\`{a} di Ingegneria, La Sapienza Universit\`{a} di Roma,
via Eudossiana 18, 00184, Roma, Italy}
  
\author{Andrea Cairoli}
\email{andrea.cairoli@crick.ac.uk}
\affiliation{The Francis Crick Institute, 1 Midland Road, London NW1 1AT, United Kingdom}

\author{Rainer Klages}
\email{r.klages@qmul.ac.uk}
\affiliation{Queen Mary University of London, School of Mathematical Sciences, Mile End Road, London E1 4NS, United Kingdom}
\affiliation{Institut f\"ur Theoretische Physik, Technische Universit\"at Berlin, Hardenbergstra{\ss}e 36, 10623 Berlin, Germany}

\date{\today}

\begin{abstract}

%BASIC INTRODUCTION (1-2 SENTENCES)
Stochastic processes play a key role for modeling a huge variety
of transport problems out of equilibrium, with manifold applications
throughout the natural and social sciences.
%MORE DETAILED BACKGROUND (2-3 SENTENCES)
  To formulate models of stochastic dynamics the conventional approach
  consists in superimposing random fluctuations on a suitable
  deterministic evolution. These fluctuations are sampled from
  probability distributions that are prescribed a priori, most
  commonly as Gaussian or L\'evy.
%GENERAL PROBLEM (1 SENTENCE)
While these distributions are motivated by (generalised) central limit
theorems they are nevertheless \textit{unbounded}, meaning that
arbitrarily large fluctuations can be obtained with finite
probability. This property implies the violation of fundamental
physical principles such as special relativity and may yield
divergencies for basic physical quantities like energy. 
%MAIN RESULTS (1 SENTENCE)
Here we solve the fundamental problem of unbounded random
fluctuations by constructing a comprehensive theoretical framework of
stochastic processes possessing physically realistic finite
propagation velocity.
%COMPARISON WITH PREVIOUS RESULTS (2-3 SENTENCES)
Our approach is motivated by the theory of L\'evy walks, which we
embed into an extension of conventional Poisson-Kac processes.
The resulting extended theory employs generalised transition
  rates to model subtle microscopic dynamics, which reproduces
  non-trivial spatio-temporal correlations on macroscopic scales. It
thus enables the modelling of many different kinds of dynamical
features, as we demonstrate by three physically and biologically
  motivated examples. The corresponding stochastic models capture the
whole spectrum of diffusive dynamics from normal to anomalous
diffusion, including the striking ``Brownian yet non Gaussian"
diffusion, and more sophisticated phenomena such as senescence.
%BROADER PERSPECTIVE (2-3 SENTENCES)
Extended Poisson-Kac theory can therefore be used to model a wide
range of finite velocity dynamical phenomena that are observed
experimentally.

\end{abstract}

%\keywords{random dynamical systems $|$ anomalous diffusion $|$ chaos $|$ intermittency $|$ ergodicity breaking} 

\maketitle

\section{Introduction}
\label{sec:intro}

\subsection{From infinite to finite velocity stochastic processes}

Stochastic processes are used extensively as theoretical models in the
natural and social sciences \cite{BCKV16}. They enable powerful
coarse-grained mathematical descriptions of generic dynamical
phenomena over a wide range of time and length scales
\cite{vK,Risk,Gard09}, where all the underlying microscopic physical
processes are effectively integrated out. To illustrate this concept,
we consider the famous example of a tracer particle immersed in a
fluid.  The motion of this tracer can be determined, in principle, by
specifying its own deterministic Newtonian equation of motion and
those of all fluid particles, as well as a suitable potential
describing their mutual interactions.  Solving these equations
equipped with initial conditions for all particle velocities and
positions yields the exact temporal evolution of the tracer kinematic
variables \cite{Zwan01}. Nevertheless, this approach is often
analytically intractable and numerically extremely
demanding. Alternatively, the tracer motion can be modelled by a much
simpler equation where a stochastic noise term with prescribed
statistical properties is introduced, which describes effectively the
force on the tracer resulting from its microscopic interactions with
the fluid particles.  This approach has the advantage that we do not
need to resolve the motion of the fluid particles. In particular, by
assuming their velocities to be Gaussian distributed, the noise term
can be shown to yield a Wiener process \cite{Lang08,CKW04}.  In the
absence of additional external forces the tracer position distribution
is also Gaussian, which is expected as a result of the averaging of
independent and identically distributed random displacements with
finite variances that are induced by the microscopic interactions
between the tracer and the fluid particles.  This is a manifestation
of the celebrated \textit{central limit theorem} \cite{Gard09}.  When
these displacements follow instead distributions with infinite
variances, the \textit{generalised central limit theorem} prescribes
that the statistics of the position process is modelled by a L\'evy
stable distribution \cite{Gnedenko1954}.  A wide spectrum of
stochastic processes is thus modelled by drawing random variables from
one of these special probability density functions (PDFs), depending
on the underlying physical properties of the system under
investigation. All these distributions share the property of being
\textit{unbounded}, i.e., of non compact support, which means that
arbitrarily large random variables can potentially be sampled with
finite probability.

However, this property is never satisfied in physical reality.  For
instance, in the previous example it is clear that the velocities of
the fluid particles cannot be arbitrarily large.  {Indeed, by
  sampling the propagation velocities from the unbounded tails of a
  Gaussian PDF one may generate rare random realisations of the
  particle velocity that exceed the speed of light, thus violating
  fundamental principles of physics, most notably the theory of
  special relativity \cite{Catt48}, even though the probability of
  such events is small enough that they are never realised in
  practice.  While the relativistic constraint of a finite
propagation velocity is most prominent on large astrophysical scales
\cite{DuHa09,ReZa13} there are also major consequences on small
scales.  Examples are the ballistic motion of a tracer in a rarified
gas observed at very small time scales \cite{Blum2006}, the deviations
from the diffusion approximation for photon scattering in random media
\cite{YLA90}, the breakdown of Fourier's law in nanosystems
\cite{COGMZ08,SCJ16} and the propagation of heat waves in superfluidic
helium \cite{JoPr89}.

These violations become particularly prominent for all stochastic
processes generating anomalous diffusion \cite{MeKl00,KRS08}, where
deviations from the normal diffusive behaviour characteristic of
Brownian motion are often modelled by sampling random variables from
power law tailed PDFs \cite{SZK93,KSZ96,KRS08,MJCB14,ZDK15}.  When
these distributions particularly describe fluctuations in the position
of a random walker, the second moment of the resulting position
distribution may grow faster in the long-time limit than for
conventional Brownian motion, $\langle|{\bf r}|^2\rangle\sim t^a$ with $a>1$ instead of %compared to
$a=1$. This ``superdiffusive" spreading has been observed
experimentally for a huge variety of natural phenomena in physical,
chemical and biological systems (see Refs.~\cite{ZDK15,MeKl04,VLRS11}
for reviews).  Historically, such striking anomalous dynamics was
first modelled by L\'evy flights \cite{Man82,KSZ96}.  These are
Markovian random walks with instantaneous jumps, whose lengths are
sampled from a stable L\'evy distribution
\cite{SZK93,KSZ96,Kla16}.  However, because of the power law tails of
this distribution, the second and all higher order moments of the
walker position statistics are mathematically not well defined
\cite{ZDK15}.  Consequently, all corresponding physical quantities
like energy would diverge. 

To cure this deficiency L\'evy walks (LWs) were introduced. In this
model, the random walker is required to spend an amount of time for
each jump that is proportional to the sampled jump length
\cite{SKW82,GNZ85,ShlKl85,KBS87,Shles87,GZR88}. From a different
perspective, this is equivalent to require the random walker to move
with constant velocity (the proportionality constant above) and change
its direction after a random time sampled from a prescribed power law
tailed distribution (this is then the counterpart of the jump length
distribution in the L\'evy flight model).  These processes are most
conveniently modelled as a special case of the broad class of
continuous time random walks (CTRWs) with the additional constraint
that the velocity is constant \cite{ShlKl85,ZuKl93a,MeKl00,ZDK15}.
An extension of CTRWs to include persistent (or anti-persistent)
  motion as a memory effect was also developed \cite{shlesinger79}.
LWs thus provide a paradigmatic example of a stochastic process
exhibiting finite propagation velocities, a crucial requirement to
give this mathematical formalism physical meaning. Owing to the
intrinsic spatio-temporal coupling, these processes exhibit intricate
mathematical properties in terms of the shape of the corresponding
position PDFs as well as the generalised (fractional) diffusion
equations governing them \cite{BKMS04,RDHB14,Fed16,MaZo16,PBL19}. Over
the past two decades LWs have been used widely to understand a wealth
of phenomena particularly in the physical and biological sciences,
many of them being observed experimentally (see
Refs.~\cite{VLRS11,ZDK15,Reyn18} and further references therein).

A second fundamental class of stochastic dynamics possessing finite
propagation velocities, which has been developed in parallel to LWs,
is represented by Poisson-Kac (PK) processes.  These models were
originally formulated by Taylor in the context of turbulent diffusion
\cite{Tay21}. But their first mathematical characterisation was given
by Goldstein, who referred to them as persistent random walks.  He
showed that their statistics satisfy the telegrapher's equation
\cite{Gold51}.  These processes became later established in the
formulation proposed by Kac in a famous lecture from 1956 (reprinted
in 1974 \cite{Kac74}). A PK process is defined therein as a
one-dimensional random walk, where the direction of the walker's
velocity is flipped at random instances of time.  The switching of the
velocity direction is assumed to be governed by a Poisson counting
process, which thus induces an exponential PDF of the times between
successive direction changes, or transitions.  Kac then showed that in
one dimension the Cattaneo equation (here identical to the
  telegrapher's equation) can be derived for the walker position
distribution, thus providing a stochastic interpretation of this
equation \footnote{We note that any attempt to extend this equation to
higher spatial dimensions fails to ensure the positivity of the
corresponding PDF \cite{KoeBe98,GBC17}.}.  In contrast to the
classical parabolic diffusion equation, the Cattaneo equation is a
hyperbolic generalised diffusion equation stipulating a finite
propagation velocity by satisfying special relativity
\cite{Catt48,Catt58}.

Starting from this basic analysis PK processes were exploited in
different ways. Perhaps their most prominent application is as a model
to generate dichotomous noise, which is bounded and coloured (in
contrast to the classical Wiener-induced white noise) , as represented
by its exponentially decaying two-point correlation function (see
Refs.~\cite{Bena06,Weiss07} for comprehensive reviews).
Two-dimensional generalisations of PK processes have been studied in
mathematical works by Kolesnik and
collaborators~\cite{Kol08,Kol11}. On more physical grounds these
processes were used to derive the one-dimensional Dirac equation for a
free electron \cite{GJKS84} and for generalising conventional
hydrodynamic theories \cite{Ros93}. Furthermore, by means of the
Cattaneo equation an interesting relation between PK processes and the
theory of extended thermodynamics has been proposed
\cite{MuRu93}. This connection motivated, among others, the
formulation of generalised PK processes.  These processes extend the
conventional theory formalised by Kac to general $n$ spatial
dimensions, while accounting for a $d$-dimensional set ($d \leq n$;
either discrete or continuous) of different velocity states
parametrised by a stochastic parameter whose dynamical evolution is
modelled by a Poisson field, i.e., a continuous Markov chain process.
As such, each state transition of this Poisson field corresponds to a
transition of the generalised PK process to a different velocity
state.  These processes have been defined with the long-term goal to
provide a micro/mesoscopic stochastic dynamical basis for extended
thermodynamics by clarifying the consequences of a finite propagation
velocity \cite{GBC17,Giona17a,GBC17b}. Along these lines the modelling
of atomic processes in the presence of quantum fluctuations related to
transitions amongst the energy levels and to the second quantisation
of the electromagnetic field could also be investigated
\cite{einstein1917,petruccione}.  For a more detailed review of
generalised PK processes and their applications we refer to
Ref.~\cite{GBC17}.

\subsection{Towards a unifying theory of finite-velocity stochastic processes}

So far these two basic classes of stochastic models, LWs and
generalised PK processes, have co-existed independently, without
exploring any cross-links between them.  However, both share the same
fundamental feature that the propagation velocity is finite, which
crucially distinguishes them from other, more common coarse-grained
models of stochastic dynamics that instead can exhibit potentially
infinite propagation speed.  We may furthermore remark that even the
classical simple lattice random walk (respectively all lattice models
\cite{Weiss94}) can be formulated in terms of finite propagation
velocity processes \cite{Giona18}. The main purpose of our article is
therefore to first formally establish the connection between LWs and
PK processes. On this basis we formulate a comprehensive theory of
stochastic processes with finite propagation velocity and finite
transition rates. We then explore the mathematical and physical
consequences of such a theoretical framework.

We address the first problem in two different ways: We start by
enquiring to which extent PK processes can be understood within the
framework of LWs. A full answer to this question is obtained through
the statistical description of LWs in terms of partial probability
density waves (PPDWs) developed by Fedotov and collaborators
\cite{FTZ15,Fed16}.  Within this formalism, it can be demonstrated 
that, by assuming an exponential distribution of
transition times, a one-dimensional LW is equal to the classical
one-dimensional PK model with two states and equal-in-modulo and
opposite-in-direction velocities \cite{Kac74}.  From this argument, it
follows that the one-dimensional PK process can be viewed as a special
case of a LW; and Cattaneo-like fractional differential equations
(i.e., Cattaneo in time and fractional regarding spatial operators)
can be derived for LWs possessing power law statistics of the
transition times \cite{BKMS04,Fed16}. Clarifying this relation between
LWs and PK processes yields our first main result.

However, this represents only an application of the PPDW formalism
already established in Refs.~\cite{FTZ15,Fed16}. Much more inspiring
is the other direction of embedding LWs into a suitably amended theory
of generalised PK processes. The formulation of such a theoretical
framework is our second main result. 
We show that LWs can be viewed as a `non-autonomous' extension of PK processes,
reflecting the explicit dependence of the transition rates on the time
elapsed after the latest velocity transition. Upon a lift of the
transition time coordinate LWs in ${\mathbb R}^n$ can then be obtained
from a new form of generalised PK processes in ${\mathbb R}^{n+1}$.
Here the additional variable both behaves as a state variable and
modulates the stochastic Poisson field governing the randomisation
dynamics of the generalised PK process.  In practice, this
transitional age variable can perform discontinuous transitions at
each transition instant of the prescribed Poisson field.  To emphasise
the complementary nature of the lifted variable, we will refer to this
class of models as {\em overlapping PK} processes. Within this
formalism, the age-theory of LWs follows as a particular case
\cite{Giona2019age}.

By using this generalised theory of overlapping PK processes, we are
able to explore entirely new classes of stochastic processes
possessing finite propagation velocity, which we will denote
collectively as \textit{extended} PK processes (EPK).  These are
determined by spatio-temporal inhomogeneities, transitional
asynchronies among the state variables, and correlations of the
microscopic transition rates. The latter are quantities that can be
measured experimentally \cite{CWG15,SMJP18,KWFA18} and thus can be
specified \textit{ad hoc} for the particular system under study.  Our
third main result is thus to illustrate the power of this new
theoretical framework by presenting three relevant examples of EPK
processes, each characterised by different settings in their
transitional structure. First, we employ our formalism to model random
walks where the age of the walker after any velocity transition (i.e.,
the time elapsed) increases as a function of the number of transition
events occurred (a feature that we call \textit{senescence}). Second,
we show that the transitional structure characterising an EPK model
naturally allows to account for a hierarchical (multi-level) structure
of the fluctuations that can capture Brownian yet not Gaussian
diffusion (at short time scales) \cite{Wang:2012aa,Chechkin:2017aa}.
Third, we discuss how an EPK process possessing a continuous
distribution of transition rates, which undergo uncorrelated Markov
chain dynamics, also reproduces the long-term diffusive properties of
a standard LW. Even more intriguingly, by introducing correlations in
the transition dynamics of the rates, we demonstrate that the EPK
model generates a subdiffusive LW dynamics.

\subsection{Outline of the article}

%Our article is organised as follows. 
The presentation of our results is organised as follows.
In Sec.~\ref{sec:fvpA}-\ref{subsec:ppdw} we review PK processes and LWs.
%by then showing that the former ones can
We then show that the former ones can be regarded as a special case of
the latter ones.  In Sec.~\ref{subsec:disspk} we formalise this
connection by explicitly defining the concepts of \textit{state
  variables} and \textit{transitional parameters}. We discuss how the
structure of PK processes (and consequently also of LWs) can be
understood by identifying in the model what are the state variables
and what are instead the transitional parameters.  In
Secs.~\ref{subsec:nov}-\ref{sec:oPK} we generalise conventional PK
processes by defining \textit{overlapping} state variables, a
necessary conceptual equipment in order to formally embed LWs into a
generalised PK formalism.  Section~\ref{subset:markov} further
clarifies the concept of overlap in comparison to models where the
dynamics of the relevant variables (i.e., those previously
overlapping) are fully transitional independent. By modulating the
transitional asynchrony between these and the other main state
variables of the model, we define the most general form of EPK
processes.  In Sec.~\ref{sec:ex} we discuss in detail the three case
studies of EPK processes mentioned previously and investigate their
novel statistical features.  These examples demonstrate the modelling
power of our theoretical framework, which is sufficiently flexible to
accommodate many unique features and thus encompasses a wide variety
of stochastic models.
%\textcolor{red}{Moreover, the application of EPK theory to
%classical and fundamental problems of statistical
%physics is also briefly outlined.} 
We conclude with Sec.~\ref{sec:concl}, where we summarise our results
and outline a spectrum of further applications of our theoretical
framework to transport and collective phenomena in biology, as well as
to classical and fundamental problems in statistical physics.

For any reader who wants to learn only about the physical essence of
the new theory that we propose, we recommend to read through
Secs.~\ref{sec:fvpA}-\ref{subsec:ppdw} explaining the connection
between LWs and PK processes. Section \ref{subsec:nov} then gives the
basic idea of how to generalise ordinary PK processes leading to our
extended theory. The main message of our work is summarised in
Fig.~\ref{fig:stokin}.

\section{Poisson-Kac processes as a special case of L\'evy walks}
\label{sec:fvp}

In this section, we review the PPDW approach first introduced by
Fedotov and collaborators as a model to describe the stochastic
dynamic of a conventional LW \cite{FTZ15,Fed16}. We then employ it to
establish a connection between LWs and PK processes.  We demonstrate
this relation by showing that the Cattaneo equation, which describes
the temporal evolution of the PDF of a PK process, can be obtained as
a special case in this framework.  Without loss of generality, we only
discuss the one-dimensional setting.  Higher-dimensional extensions of
both these processes have been considered elsewhere
\cite{Pins91,Kol08,Kol11,MaZo16,ZFDB16,AlRa18,GBC17}, and our
considerations extend straightforwardly to these settings.  As a
preparatory step for the derivation of EPK processes in
Sec.~\ref{sec:generalised PK}, we also discuss here how to identify
(and formalise) the mathematical structure of PK processes (and
likewise LWs).  This discussion, although trivial when applied to such
simple models, will be valuable for constructing more complex
stochastic dynamics with finite propagation speed.

\subsection{Poisson-Kac processes in a nutshell}
\label{sec:fvpA}

A classical one-dimensional PK process is defined by the stochastic differential equation
\cite{Kac74,Bena06,Weiss07,GBC17}
\be
\ud x(t) = b (-1)^{\chi(t,\lambda)} \, \ud t \: ,
\label{eq:pk}
\ee
where $x$ denotes the position of the random walker on the line at
time $t$, $b$ is a positive constant that represents its propagation
speed and $\chi(t,\lambda)$ is a Poisson process characterised by the
transition rate $\lambda$.  For Eq.~\eqref{eq:pk} to specify a
temporal dynamic, at the initial time $t=0$ we must equip the Poisson
process $\chi$ with a suitable initial condition, which is specified
by choosing the probabilities for which $\chi(0,\lambda)$ is equal to
either zero or one.  It is illustrative to compare the dynamic modeled
by Eq.~\eqref{eq:pk} with the one generated by a standard Wiener
process. In the physics literature this is described by the
overdamped Langevin equation
\be
\ud x(t) = \sqrt{2D} \,\zeta(t) \, \ud t \: ,
\label{eq:wiener}
\ee
where $\zeta(t)$ is a Gaussian white noise with null ensemble average,
$\langle \zeta(t) \rangle=0$, and two-point correlation function
$\langle \zeta(t)\zeta(t^{\prime}) \rangle= \delta(t-t^{\prime})$.
Hence a Wiener walker moves over constant time intervals $\ud t$ with
Gaussian distributed random velocities possessing zero mean and
variance equal to $2D \ud t$.  By construction, therefore, the
propagation speed of the Wiener walker is unbounded, but the
probabilities of sampling large velocities decay exponentially.  In
contrast, the PK walker moves with constant propagation speed and
switches the direction of its velocity after random time intervals,
whose duration is determined by the change of parity governed by the
Poisson counting process $\chi$. By taking
$b,\lambda\to\infty$ while keeping the ratio $D =b^2/(2\lambda)$
fixed, the so called Kac limit, one can show that the Wiener process
can be recovered as a limiting case of the PK process
\cite{Kac74,GBC17}.  In that sense the PK process Eq.~\eqref{eq:pk}
can be considered as a generalisation of the Wiener process
Eq.~\eqref{eq:wiener}.

The PK process is characterised by an exponential distribution of
interevent times and an exponential correlation function decay. 
Its position PDF, $P(x,t)\equiv\prec\delta(x-X(t))\succ$, where $\prec
\cdot \succ$ denotes averaging over independent realisations of the
Poisson process $\chi$, obeys the Cattaneo equation \cite{Kac74}
\be
\frac{1}{2 \, \lambda} \frac{\partial^2 P(x,t)}{\partial t^2}
+ \frac{\partial P(x,t)}{\partial t} = D \frac{\partial^2 P(x,t)}
{\partial x^2}\:.
\label{eq:catt}
\ee
The second moment $\langle x^2\rangle=\int x^2 P(x,t)\,\ud t$ of this PDF
grows linearly in the long time limit; 
thus, it describes normal diffusive dynamics.  
Note that in the limit of
$\lambda\to\infty$ the ordinary diffusion equation is recovered from
Eq.~(\ref{eq:catt}).

\subsection{L\'evy walks as specific continuous time random walks}
\label{sec:fvpB}

A one-dimensional LW is a continuous stochastic processes possessing a
bounded, constant propagation speed $b$; thus, the walker velocity
attains values $\pm b$. With speed $b$ a L\'evy walker moves in one
direction for a ``running" time $\tau$ after which it {either
  definitely, or randomly, changes its direction of motion, called
  velocity model, or two-state model, respectively
  \cite{ZuKl93a}. Accordingly, $\tau$ may also be called the
transition time. In full generality, we assume this variable to be
sampled from a prescribed probability distribution $T(\tau)$, where
$\tau \in [0,\infty)$.  Crucially, for a LW, $T(\tau)$ is chosen to
  possess power law tails
  \cite{GNZ85,ShlKl85,GZR88,ZuKl93a,MeKl00,ZDK15}. These fat tails
  enhance the probability of long directed jumps, in contrast to the
  Wiener process Eq.~\eqref{eq:wiener} where the probability of such
  jumps decays exponentially.

To characterise the statistics of this process, the main object to
calculate is the position PDF $P(x,t)$.  Here the underlying ensemble
averaging is made over all random realisations of velocity
transitions.  Historically, for LWs the former has been achieved first
within the framework of CTRWs.  As is shown in
Appendix~\ref{sec:ctrw}, the CTRW description of a two-state
LW \footnote{Without loss of generality, we restrict ourselves
  here and in the following to two-state LWs. We remark, however, that
  the velocity LW model could easily be embedded into our general
  theory that we develop here by appropriately adjusting the
  transition matrix Eq.~(\ref{eq4_14}) in
  Appendix~\ref{subsec:LWrecovery}.}  is specified by the equations
for the walker position, $x_n$, and the total elapsed time at each
transition, $t_n$,
\begin{align}
x_{n+1}&= x_n + b\,s_0 \, (-1)^n \, \tau_n \:, &
t_{n+1}=t_{n}+\tau_n 
\label{eq:ctlw}
\end{align}
where $s_0$ is a random variable attaining values $\pm 1$ with
  equal probability that specifies the initial direction of motion of
  the random walker, with a power law tailed transition time PDF such
as, e.g.,
\begin{equation}
T(\tau)=\frac{\xi}{(1+\tau)^{\xi+1}}\:,\:\xi>0\:.
\label{eq:pltt}
\end{equation}
Clearly, we can choose many different functions for $T(\tau)$, but we
remark that the qualitative statistical behaviour of the resulting
stochastic dynamic is solely determined by the power law scaling of
$T(\tau)$ for $\tau\to\infty$.

On the level of the position PDF, $P(x,t)$, 
%the constraint above leads
%to a non-trivial spatio-temporal coupling of the stochastic dynamics
%expressed by the transition probability
%$\phi(\Delta,\tau)=\delta(|\Delta|-b \tau)T(\tau)/2$.
%Based on this $\phi(\Delta,\tau)$, 
a master equation for $P(x,t)$ can
be derived and solved spatially in Fourier and temporally in Laplace
transform.  By using this analytical result, one can calculate
straightforwardly the second moment of the process, $\langle x^2(t)
\rangle = \int_0^\infty x^2 \, P(x,t) \, \ud x$, and derive its
characteristic scaling in the long time limit
\cite{ZuKl93a,MeKl00,ZDK15}.  For $T(\tau)$ specified as in
Eq.~(\ref{eq:pltt}) the second moment, $\langle \tau^2 \rangle =
\int_0^\infty \tau^2 \, T(\tau) \, d \tau$, diverges if $\xi \leq 2$.
In this case the fundamental condition underlying the central limit
theorem is violated; consequently, LWs are characterised by the
superdiffusive scaling $\langle x^2(t) \rangle\sim t^\gamma$, where
$\gamma=2$ for $0 < \xi <1$ and $\gamma=3-\xi$ for $1 < \xi <2$
\cite{GNZ85,ShlKl85,KBS87,Shles87,GZR88,ZuKl93a,ZDK15}.  

%Interestingly, only later
%a closed temporal evolution equation for the position statistics of a
%LW, analogous to the Cattaneo equation~\eqref{eq:catt} for PK
%processes and to the diffusion equation for Wiener processes, has been
%derived (see details below) by Fedotov \cite{Fed16},
%\begin{multline}
%\dersecpar{}{t} P(x,t) + 
%\frac{1}{2} \left[ \derpar{}{t} - b \derpar{}{x} \right] \int_0^t K(t^{\prime}) P(x- b\, t^{\prime},t-t^{\prime}) \diff{t^{\prime}} + 
%\frac{1}{2} \left[ \derpar{}{t} + b \derpar{}{x} \right]  \int_0^t K(t^{\prime}) P(x+b\, t^{\prime},t-t^{\prime}) \diff{t^{\prime}} = 
%b^2 \dersecpar{}{x} P(x,t) \: ,  
%\label{LWFPE}
%\end{multline}
%where the memory kernel is defined in temporal Laplace transform as 
%$K(s)=s\, T(s)/(1-T(s))$. 
%The integral operators appearing in this equation, also known as ``fractional substantial derivatives" \cite{sokolov2003,Friedrich2006,cairoli2015}, are another evident manifestation of the spatio-temporal coupling characteristic of the LW dynamic. 

Although the traditional formalisms defining PK processes and LWs are
thus quite different, compare Eqs.~(\ref{eq:pk}) and (\ref{eq:ctlw}),
our qualitative descriptions in this and the previous subsection
should make intuitively clear that PK processes are nothing else but LWs,
where one chooses for the transition time PDF $T(\tau)$ an exponential
distribution.  This particular model has already been
introduced as a ``Brownian creeper" in Ref.~\cite{CAMYL15}, but there
the authors did not elucidate further the connection with the
mathematical formalism of PK processes.  
%Remarkably, this connection
%is also hinted at by the structural similarity between
%Eqs.~\eqref{eq:catt} and \eqref{LWFPE}, except for the integral
%operators that are a consequence of the power law scaling transition
%statistics of LWs.  

%In the following subsection we thus clarify the
%detailed connection between both processes by deriving equations
%determining their position PDFs.  This is achieved through the
%statistical analysis put forward by Fedotov and collaborators
%\cite{FTZ15,Fed16}, which was employed to derive the nonlinear
%diffusion equation~\eqref{LWFPE}.

\subsection{Establishing the connection between Poisson-Kac processes and L\'evy walks through the partial probability density wave functions representation.}
\label{subsec:ppdw}

The starting point of the statistical analysis formulated by Fedotov
and collaborators \cite{FTZ15,Fed16} is %, which we briefly review here, is
the following, more general, expression for the transition time
probability distribution $T(\tau)$,
\begin{equation}
T(\tau)= \lambda(\tau) \, \exp \left [ - \int_0^\tau \lambda(\theta) \, d \theta \right ]\:,
\label{eq:ttpdf}
\end{equation}
where $\lambda(\tau)$ denotes a generalised transition rate.  %Clearly,
By setting $\lambda(\tau)=\lambda=\mbox{const.}$, Eq.~\eqref{eq:ttpdf}
identifies the transition probability $T(\tau)$ with the familiar
Poissonian exponential distribution.  In contrast, by setting
\begin{equation}
\lambda(\tau)= \frac{\xi}{1+\tau}\:, 
\label{eq:lwtr}
\end{equation}
we obtain the power law tailed transition time probability
distribution Eq.~(\ref{eq:pltt}).  %Remarkably, 
This equation has a very
neat physical interpretation: It models a peculiar type of
persistence, where the probability of a transition decreases the
longer the random walker moves in one specific direction \cite{FTZ15}.
We further highlight that $\lambda(\tau)$ can alternatively be defined
by the equation $\lambda(\tau) = T(\tau)/\Lambda(\tau)$ \cite{CoMi65},
where $\Lambda(\tau)= \exp \left [ - \int_0^\tau \lambda(\theta) \, d
  \theta \right ]$ denotes the survival probability of the process.
Even if at first glance this is just a rewriting of
Eq.~\eqref{eq:ttpdf}, this definition is very advantageous in
practice, because it enables us to employ a huge toolbox of
statistical methods that have been developed in other branches of the
sciences for the estimation of the survival function $\Lambda$ from
empirical data \cite{Lee2003}.

Considering time dependent transition rates one can write down
balance equations for the two PPDW functions $p_\pm(x,\tau,t)$ (in
Refs.~\cite{FTZ15,Fed16} these quantities are called structural
probability density functions, denoted by $n_\pm(x,t,\tau)$). 
These represent the probability distributions of
finding a random walker at the position $x$ at time $t$ with
positive/negative ($\pm$) orientation of the velocity 
and transitional age $\tau$, 
with which we denote the time interval after the latest transition in the velocity direction. 
The evolution equations for $p_\pm(x,\tau,t)$ follow directly from their definition
and are given by \cite{FTZ15,Fed16}
\begin{eqnarray}
\frac{\partial p_{\pm}(x,\tau,t)}{\partial t} &  = &
- \frac{\partial p_{\pm}(x,\tau,t)}{\partial \tau} \mp 
b \,  \frac{\partial p_{\pm}(x,\tau,t)}{\partial x} - \lambda(\tau) \, p_{\pm}(x,\tau,t)
\:.\label{eq:ppdw}
\end{eqnarray}

In order to solve these PDEs, we need to equip them with suitable initial and boundary conditions. 
As regards the former, we need to specify 
the initial spatial probability distribution of the random walker, $p_0(x)$, 
the probabilities for each velocity direction, $\pi^0_{\pm}$, and 
the corresponding initial distributions of transitional ages, $\phi_{\pm}^0(\tau)$.  
This yields 
\begin{equation}
p_\pm(x,\tau,0)= \pi^0_{\pm} \, p_0(x) \, \phi_{\pm}^0(\tau) \:.
\label{eq:ic-gen}
\end{equation}
In the original refs.~\cite{FTZ15,Fed16},  
it is assumed that the walker, at the initial time, possesses a transitional age $\tau=0$ 
and uniformly distributed velocity directions. This means that, in Eq.~\eqref{eq:ic-gen}, 
$\pi^0_{\pm} =1/2$ and $\phi_{\pm}^0(\tau)=\delta(\tau)$, i.e.,
\begin{equation}
p_\pm(x,\tau,0)= \frac{1}{2} \, p_0(x) \, \delta (\tau)\:.
\label{eq:ic}
\end{equation}
We remark that our general initial conditions enable the investigation of more subtle aspects of these stochastic dynamics, such as ageing \cite{Giona2019age}.
As regards the latter, boundary conditions for the PPDW functions $p_\pm(x,t,\tau)$ are related to the details of the transition dynamics.
As an example, let us assume that at any transition time $\tau$ all walkers reverse the
velocity direction. In this case the boundary conditions at $\tau=0$ are given by
\begin{equation}
p_\pm(x,0,t)= \int_0^\infty \lambda(\tau^{\prime}) \, p_\mp(x,\tau^{\prime},t) \, \ud \tau^{\prime}\:.
\label{eq:bc}
\end{equation}
Under these assumptions, the resulting process is called a ``two-state model", 
as it consists of an alternating switching between two states \cite{ZuKl93a}.  
Other cases, like the ``velocity model", where particles choose their new direction randomly at any transition time, can also be modelled within this framework by generalising Eq.~(\ref{eq:bc}).
%\cmag{(see Supplementary Information, Section~???).}
%
%Finally, to be complete, we need to supplement the model with  
%regularity conditions at infinity with respect to both $\tau$ and $x$.  
%Specifically, we assume that $p_\pm(x,t,\tau)$, for
%any $x \in {\mathbb R}$ and $t >0$, decay faster than any polynomials
%for $\tau \rightarrow \infty$, and analogously for $|x|\rightarrow \infty$,
%for any $t,\tau \geq 0$,
%\begin{equation}
%\lim_{\tau \rightarrow \infty} \tau^q \, p_\pm(x,\tau,t)  =
%\lim_{|x| \rightarrow \infty} x^q \, p_\pm(x,\tau,t)=0
%\qquad \forall q=0,1,2,\dots \:.
%\label{eq:regul}
%\end{equation}
%

Having well defined the model 
\footnote{To be complete, we need also to supplement the model with
regularity conditions at infinity with respect to both $\tau$ and $x$.
Specifically, we assume that $p_\pm(x,t,\tau)$, for any $x \in
{\mathbb R}$ and $t >0$, decay faster than any polynomials for $\tau
\rightarrow \infty$, and analogously for $|x|\rightarrow \infty$, for
any $t,\tau \geq 0$,
\begin{equation}
\lim_{\tau \rightarrow \infty} \tau^q \, p_\pm(x,\tau,t)  =
\lim_{|x| \rightarrow \infty} x^q \, p_\pm(x,\tau,t)=0
\qquad \forall q=0,1,2,\dots \:.
\label{eq:regul}
\end{equation}
Eqs.~(\ref{eq:regul}) are  satisfied {\em a fortiori} if the initial
conditions admit compact support both in space and in $\tau$, 
owing to the finite velocity of propagation and to the physical meaning of $\tau$. 
In particular, for the initial conditions Eq.~(\ref{eq:ic}), 
and assuming, e.g., that $p_0(x)=0$ for $|x| > a$, we have 
$p_\pm(x,\tau,t)=0$ for $|x| > a + b \, t$ and likewise 
$p_\pm(x,\tau,t)=0$ for $\tau>t$.
Consequently, Eqs.~(\ref{eq:ppdw}) and
the integrals entering in Eqs.~(\ref{eq:bc}), 
(\ref{eq2_8}) and (\ref{eq:gpdw}), {\em ipso facto} are limited to the
closed interval $[0,t]$.}, 
we can now establish the connection between PK processes and LWs. 
For this purpose we define the auxiliary PPDW functions $P_\pm(x,t)$, 
i.e., the marginals of $p_\pm(x,\tau,t)$ with respect to the transitional age $\tau$,
\begin{equation}
P_\pm(x,t)= \int_0^\infty p_\pm(x,\tau^{\prime},t) \, \ud \tau^{\prime}\:.
\label{eq2_8}
\end{equation}
Integrating Eq.~(\ref{eq:ppdw}) with respect to $\tau$ while
enforcing the boundary conditions Eq.~(\ref{eq:bc}), 
we obtain the following evolution equations for $P_\pm(x,t)$, i.e., 
\begin{equation}
\frac{\partial P_\pm(x,t)}{\partial t}= \mp b \, \frac{\partial P_\pm(x,t)}{\partial x} \mp \int_0^\infty \lambda(\tau^{\prime}) \, \left [p_+(x,\tau^{\prime},t)-p_-(x,\tau^{\prime},t)
\right ] \, \ud \tau^{\prime}\:.
\label{eq:gpdw}
\end{equation}

In Refs.~\cite{FTZ15,Fed16}, 
Eqs.~\eqref{eq:gpdw} with the initial condition Eq.~\eqref{eq:ic}
are shown to generate LW dynamic. 
Interestingly, the authors also derive a closed fractional integro-differential evolution equation for the position statistics, 
$P(x,t)=P_+(x,t)+P_-(x,t)$. 
%Likewise, for genuine LWs the fractional integro-differential wave
%equation~(\ref{LWFPE}) can be obtained from Eq.~(\ref{eq:gpdw})
%\cite{Fed16}.  We remark that the convolution operators with respect
%to time yielding information about all the previous history of the
%process are a consequence of the time dependence modelled in the
%transition rate $\lambda(\tau)$ (see Eq.~(\ref{eq:lwtr})).
%Furthermore, 
However, this derivation 
%strictly holds for the particularly
%simple initial conditions Eq.~(\ref{eq:ic}) and 
cannot easily be extended to account for the more general initial
conditions Eq.~(\ref{eq:ic-gen}).  The structural stiffness of this
equation suggests that the spatial density $P(x,t)$ is not the most
natural and complete statistical description of this process.  In
contrast, only the PPDW functions $p_\pm(x,\tau,t)$ provide the
primitive statistical description of finite velocity processes, as
evidenced by the fact that, by including explicitly the transition
time $\tau$ as an additional independent coordinate, the corresponding
evolution equations~(\ref{eq:ppdw}) are Markovian and local in time.
This formulation therefore provides a big advantage for mathematical
analyses.

%Using the definition~(\ref{eq2_8}), 
We consider now the particular case of
$\lambda(\tau)=\lambda=\mbox{const}$; 
remarkably, this reproduces the simplest
two-state PK process first considered by Kac \cite{Kac74}
\begin{equation}
\frac{\partial P_\pm(x,t)}{\partial t} = \mp  b \, \frac{\partial P_\pm(x,t)}{\partial x} \mp\lambda \left [P_+(x,t)-P_-(x,t) \right ]\:.
\label{eq:pdwpk}
\end{equation}
%Therefore, in this case the PPDW marginal functions $P_\pm(x,t)$ fully
%characterise the statistical properties of the process.  
It is straightforward to derive from Eq.~(\ref{eq:pdwpk}) the Cattaneo
equation~(\ref{eq:catt}) for the distribution
$P(x,t)=P_+(x,t)+P_-(x,t)$.  
This argument demonstrates that classical
one-dimensional PK processes form a subset of LWs \cite{Fed16}.

The relation between Wiener processes, PK processes and LWs as
discussed above is our first main result, which is schematically
summarised in Fig.~\ref{fig:stokin}.  Conversely, one may now raise
the question whether we can exploit this connection in order to embed
LWs into a suitably generalised PK formalism and, correspondingly,
what novel diffusive features can be described within such a
generalised theory.  This problem is addressed in the following
Section~\ref{sec:generalised PK} yielding the new fourth outer layer
depicted in Fig.~\ref{fig:stokin}.

\begin{figure}[!t]
    \centering
    \includegraphics[width=180mm]{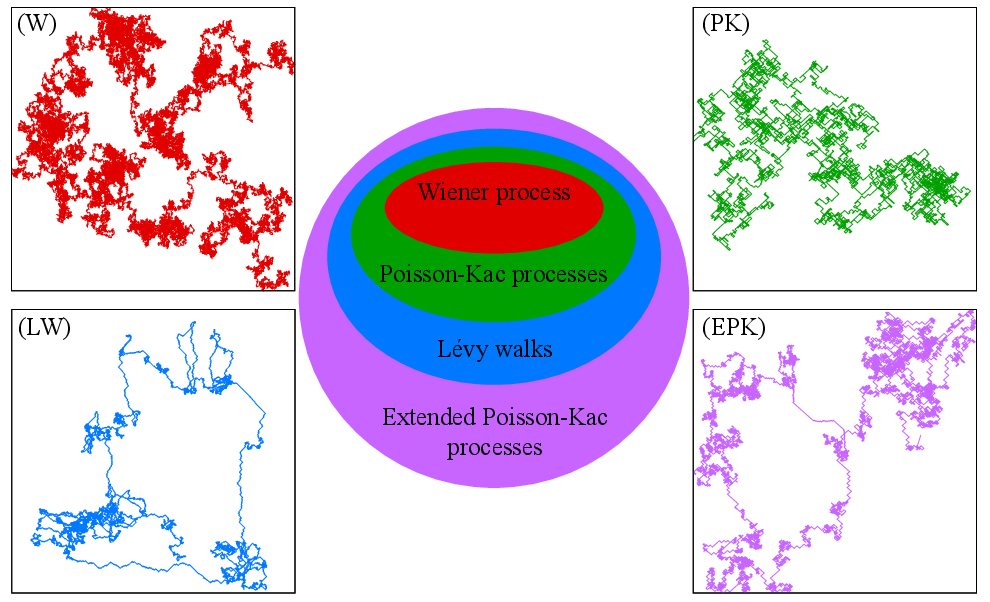} 
\caption{Schematic representation of the increasing level of
  generalisation in stochastic kinematics with finite propagation
  speed, from Wiener processes (W), to Poisson-Kac processes (PK), to
  L\'evy walks (LW), and finally to extended Poisson-Kac processes
  (EPK). The latter define the new class of stochastic models
  introduced in Section~\ref{sec:generalised PK}.  Overlapping
  Poisson-Kac processes, which are discussed specifically in
  Subsec.~\ref{sec:oPK}, are one particular instance of these
  processes.  Example orbits of realisations of these processes in two
  spatial dimensions are shown correspondingly.  For two-dimensional
  LWs, we simulate the two-state model (see
  Subsec.~\ref{subsec:ppdw}) with the transition rate
  Eq.~(\ref{eq:lwtr}) where $\xi=1.5$. For the class of EPK processes
  we choose a senescent LW (see Subsec.~\ref{subsec:senesc}) with
  $\xi=1$ and $\tau^0=1$.}
\label{fig:stokin}
\end{figure}

\subsection{Dissecting the structure of  Poisson-Kac processes}
\label{subsec:disspk}

Let us reconsider the process defined by Eq.~(\ref{eq:pk}).  In fully
general terms, this involves a set of state variables $\Sigma_X$,
which in our specific case only contains the process $X(t)$ itself 
(note that upper case letters refer to the stochastic processes, while
lower case to their realisation).  These state variables are defined
in some prescribed domain ${\mathcal D}_X \subseteq {\mathbb R}^n$, 
with $n$ being the total number of such variables.  In the case
considered here, ${\mathcal D}_X={\mathbb R}$.  The dynamics of these
state variables is controlled by a set of driving stochastic
processes, also called transitional parameters, $\Sigma_T$, which can
assume values in the set $D_T\subseteq {\mathbb R}^m$, 
with $m$ being the total number of such values.  The transitional parameters are
chosen such that their joint process with the state variables is
Markovian, while instead the state variables alone are non-Markovian.
For Eq.~(\ref{eq:pk}), in particular, the only transitional parameter
is the process $S(t)=(-1)^{\chi(t,\lambda)}$, which attains values in
$D_T=\{-1,1\}$.  In agreement with the condition above, we have shown
previously that, while $X(t)$ alone is non-Markovian, the couple
$(X(t),S(t))$ is Markovian instead.  The dynamics of the state
variables may also depend on a set of physical parameters $\Sigma_P$,
such as $b$ and $\lambda$ for the PK process Eq.~(\ref{eq:pk}),
generically defined in the domain ${\mathcal D}_P\in{\mathbb R}^p$,
with $p$ being the dimensionality of these parameters.  Finally, the
stochastic dynamic is given as a vector field $f:{\mathcal D}_X
\times {\mathcal D}_T \times {\mathcal D}_P \rightarrow {\mathcal
  D}_X$, expressing the temporal evolution of the state variables and
depending on the elements of the set $\Sigma_X \cup \Sigma_T \cup
\Sigma_P$.  Within this framework, it becomes clear how we can
formulate correctly the primitive statistical description of this
dynamic. This is achieved in terms of the PPDW functions of
the state variables, which are additionally parametrised by the
attainable values of the transitional parameters, owing to the
Markovian recombination mechanism that they provide.

According to this framework, in its essence the structure of a PK
process is constituted by the system $(\Sigma_X,\Sigma_T,\Sigma_P,f)$,
where $\Sigma_X= \{ X \}$, $\Sigma_T=\{ S \}$, $\Sigma_P=\{b,
\lambda\}$ and $f$ is specified by Eq.~(\ref{eq:pk}).  The
transitional parameter $S$ parametrises the statistical description of
the process, thus determining the system of PPDW functions $p(x,t;s)$,
$s \in D_T=\{-1,1\}$.  In our previous discussion, to simplify the
notation, we have identified $p(x,t;1)=P_+(x,t)$ and
$p(x,t,-1)=P_-(x,t)$.  The statistics of the process is then fully
determined by the Markovian evolution equations~(\ref{eq:pdwpk}) for
the PPDW functions.

\section{From L\'evy walks to extended Poisson-Kac processes}
\label{sec:generalised PK}

Motivated by the connection between LWs and PK processes just
established through the PPDW approach, in this section we formalise
the family of EPK processes, the new class of stochastic
processes with finite propagation speed that constitute the fourth
layer of generalisation in Fig.~\ref{fig:stokin}.  All the other
processes discussed so far can thus be recovered from them as special
cases.  We choose this terminology to distinguish them from the
generalised PK processes previously discussed in the literature
\cite{GBC17,Giona17a,GBC17b}.  In fact, these processes can be
obtained within our formalism.  

First, we define the concept of \textit{overlap}.  This consists in
introducing an additional coordinate to the classical description of
PK processes. This coordinate is a state variable that can be
interpreted as an internal time representing the transitional age of
the process, i.e., the time elapsed from the last velocity transition.
Simultaneously, this coordinate also plays the role of a transitional
parameter, as it directly controls the stochastic machinery of the
random velocity switches.  We call this coordinate overlapping, as it
belongs to both sets $\Sigma_X$ and $\Sigma_T$.  The overlapping
variable undergoes deterministic dynamics between different
transitions and discontinuous Markovian jumps at each transition
instants.  Remarkably, we demonstrate that this formalism can capture
a generic functional form for the transition time PDF, such as the
power law tailed Eq.~(\ref{eq:pltt}) characteristic of LW processes.
This coupling between the state variables and the driving stochastic
processes is therefore the key ingredient in order to embed LWs within
a generalised theory of PK processes.  We then derive the statistical
description of general overlapping PK processes in terms of partial
differential equations for their PPDW functions.  These processes are
however not the most general form of stochastic processes with finite
propagation velocities and transition rates.  For overlapping PK
processes, in fact, the Markovian jump dynamics of the overlapping and
state variables is assumed to be fully synchronised by definition.
Clearly, different processes can be obtained if we relax this
condition, e.g., by fully desynchronising the transitional dynamics of
overlapping and state variables.  In fact, we provide a recipe for how
the transitional synchronisation between these processes can be
modelled explicitly.  We denote as EPK processes the stochastic models
generated within this framework.

\subsection{Formulating a stochastic equation for generalised Poisson-Kac processes}
\label{subsec:nov}

With the knowledge of the PPDW approach %outlined in
(see Sec.~\ref{subsec:ppdw}) it is worthwhile to return to the basic
stochastic equation of motion (\ref{eq:pk}) defining the classical
one-dimensional PK process.  This equation yields PK dynamics by using
a {\em constant} transition rate $\lambda$ for the corresponding
Poisson counting process $\chi(t,\lambda)$. In contrast to this,
Eq.~(\ref{eq:ttpdf}), which was used to define the transition time PDF
$T(\tau)$ more generally, involves a {\em generalised} transition rate
$\lambda(\tau)$ that depends on the transition time $\tau$. This
simple observation of having intrinsically different transition rates
for PK processes and LWs suggests that LWs can be expressed in the
form of the suitably amended PK process
\begin{eqnarray}
\ud x(t) & = & b \, (-1)^{\chi(t,\lambda(\tau(t)))} \, \ud t \:,
\label{eq:apk1} \\ 
\ud \tau(t)& = & \ud t\:.
\label{eq:apk2}
\end{eqnarray}
Here $\chi(t,\lambda(\tau))$ represents a generalised Poisson process
whose transition rate $\lambda$ depends generically on the value
attained by the additional coordinate $\tau$, the transitional age,
which stands for the time elapsed after the last velocity transition.
In turn, $\tau$ is coupled to the physical time $t$ by
Eq.~(\ref{eq:apk2}). For example, using the time-dependent transition
rate Eq.~(\ref{eq:lwtr}) yields a generalised Poisson process
characterised by the power law transition time PDF
Eq.~(\ref{eq:pltt}) 
\footnote{A related model, called a fractional Poisson
process, was studied in the literature with $T(\tau)$ as a generalised
Mittag-Leffler function exhibiting power law tails
\cite{Lask03,SGM04}. We remark that in the following we relax the
condition of $\tau$ being positive to $\tau\in\mathbb{R}$.}.

Equations~(\ref{eq:apk1}) and~(\ref{eq:apk2}) are only valid in the
time interval between two velocity transitions. In order to extend
them over the entire history of the process, we need to supplement
them with %by suitable 
boundary conditions at the transition times. 
Intuitively, these boundary conditions must involve the
auxiliary variable $\tau$ and be discontinuous. In fact,
whenever a transition in $\chi(t,\lambda(\tau(t)))$ occurs, the
transitional age $\tau$ is reset to zero.  In contrast, because the
transition only changes the velocity direction, the stochastic process
$x(t)$ is continuous at the transition time.  In mathematical terms,
assuming that a transition occurs at the time instant $t^*$, we then
set
\begin{equation}
  x(t^*_+)=x(t_-^*) \: \quad \mbox{and} \quad \: \tau(t^*_+)=0 \:,
  \label{eq:apkbc}
\end{equation}
with the shorthand notation $f(t^*_{\pm})=\lim_{\epsilon\to0}
f(t^*\pm\epsilon)$ and $f$ any smooth or  {continuous} function.  For the
transition rate Eq.~(\ref{eq:lwtr}) the PK process defined by
Eqs.~(\ref{eq:apk1}) and (\ref{eq:apk2}) equipped with the boundary
conditions Eq.~(\ref{eq:apkbc}) generates a LW dynamics.  This is
demonstrated formally by calculating the evolution equations for the
PPDW functions of the process $x$, which can be shown to be equal to
Eqs.~(\ref{eq:ppdw}) (see Subsec.~\ref{subsec:LWrecovery}).  Within
this setting, a LW can therefore be interpreted as a form of
non-autonomous PK process depending explicitly on the internal time
coordinate $\tau$.  
%In this regard, we highlight that the dynamical
%rules Eqs.~(\ref{eq:apk1}) and (\ref{eq:apk2}) are highly
%discontinuous with respect to the inner temporal parametrisation
%expressed by $\tau$.

The observations above further suggest that LWs can be reformulated
within the theory of PK processes by defining the new state variable 
\begin{equation}
{\bf y}=
\left (
\begin{array}{c}
x \\
\tau
\end{array}
\right )\:.
\label{eq4_10}
\end{equation}
This formulation is analogous to the lift of the time coordinate that
is employed to transform a non-autonomous one-dimensional dynamical
system into an autonomous one in two dimensions \cite{Str18}. It thus
trivially leads to studying these processes in a space of dimension
higher than one, for which we will adapt the generalisation of the
theory of PK processes to higher dimensional state spaces described in
Refs.~\cite{GBC17,GBC17a,GBC17c}. {We note that this reformulated
  process belongs to the well-known class of renewal processes
  \cite{Cox1962}, which presents a big advantage.} Since our process
$x$ only admits two states with velocities $\pm b$, for the lifted
state variable ${\bf y}$ we define the two generalised velocity
vectors
\begin{equation}
{\bf b}(1)=
\left (
\begin{array}{c}
b \\
1
\end{array}
\right ) \; , \qquad
{\bf b}(-1)=
\left (
\begin{array}{c}
-b \\
1
\end{array}
\right )\:.
\label{eq4_11}
\end{equation}
%In order to model a LW process, 
%we further assume that $\lambda(\tau)$ is independent of the current stochastic state.
Using the setting Eqs.~(\ref{eq4_10}) and (\ref{eq4_11}), 
and assuming $\lambda$ to only depend on the transitional age variable $\tau$,
the equations of motion (\ref{eq:apk1}) and (\ref{eq:apk2})  can be compactly
expressed as
\begin{equation}
\ud {\bf y}(t)={\bf b}\left((-1)^{\chi(t,\lambda(\tau(t)))}\right) \, \ud t\:,
\label{eq:generalised PKv}
\end{equation}
equipped with the boundary conditions Eqs.~(\ref{eq:apkbc}) at each
time instant when a state transition occurs.  More general choices are
also possible.  For example, if we assume the dynamics of the
transitional age ${\mathcal T}(t)$ to be a stochastic process
possessing a Markovian transitional structure, the boundary condition
for $\tau(t^*_{+})$ becomes
\begin{eqnarray}
\tau(t^*_+) & = & \tau^\prime \qquad \mbox{with probability } 
\;  k(\tau^\prime,\tau(t^*_-))\;\:.
\label{eq4_15bis}
\end{eqnarray}
Evidently, we need to assume the following conditions on the transition probability  
$k(\tau^{\prime},\tau)$, i.e, 
\begin{equation}
k(\tau^\prime,\tau)\geq 0 \qquad \mbox{and} \qquad
\int_{-\infty}^\infty k(\tau^\prime,\tau) \, d \tau^\prime =1\:.
\label{eq4_13}
\end{equation}
The particular case of Eq.~(\ref{eq:apkbc}) corresponds to setting 
$k(\tau^\prime,\tau)= \delta(\tau^\prime)$.

The formulation provided by Eqs.~(\ref{eq4_10}), (\ref{eq4_11}) and
(\ref{eq:generalised PKv}) elucidates the following characteristic
features of the multivariate process ${\bf Y}(t)$: On the one hand, it
possesses an evident skew product structure, because we can formally
write ${\bf Y}(t)=(X(t;{\mathcal T}(t)),{\mathcal T}(t))$.  In fact,
while the transitional age process ${\mathcal T}(t)$ does not
incorporate the position process $X(t)$, the latter, in contrast,
depends explicitly on time $t$ and is simultaneously a nonlinear
functional of ${\mathcal T}(t)$ through the Poissonian transition rate
$\lambda$.  On the other hand, in Eq.~(\ref{eq:generalised PKv}) the
noise is manifestly governed by the state variable $\tau$.  This
coupling thus modulates the very fundamental stochastic structure of
the fluctuations, as is made evident by the fact that the transition
rate $\lambda(\tau)$ controls the correlation properties of the
resulting dynamics \footnote{We remark that all the stochastic
processes $X(t)$ possessing finite propagation velocity and bounded
$\lambda(\tau)$ described by Eq.~(\ref{eq:generalised PKv}) possess
Lipshitz trajectories.  Therefore, no issues arise with the definition
of stochastic integrals.}.  These properties reveal a striking change
of paradigm with respect to conventional PK processes, which is
determined by the overlap between the state variable ${\bf y}$ and the
transitional parameters controlling the randomisation dynamics as just
described.  This peculiar feature defines a new class of stochastic
processes with finite propagation velocity, called {\em overlapping
  PK} processes (OPK), that includes LWs as a special case.

\subsection{Overlapping Poisson-Kac processes}
\label{sec:oPK}

We now formalise the concept of overlap introduced previously by
specifying the formal structure of the multivariate stochastic process {\bf Y}(t) 
%, similarly to the discussion presented for classical PK
%processes in 
(see Subsec.~\ref{subsec:disspk}).  In fully general terms, we
define a PK process to be overlapping
if the following conditions hold true: 
(i) The sets of state variables, $\Sigma_X$, and of 
transitional parameters, $\Sigma_T$, possess a non-empty intersection,
\be
\Sigma_O = \Sigma_X \cap \Sigma_T \neq \emptyset \:.
\label{eqo_1}
\ee
(ii) The transition dynamics of the variables in $\Sigma_T$
depend exclusively on the dynamics of those 
in its set complementary to $\Sigma_O$, i.e., 
the set $\Sigma_T/\Sigma_O$,  
which contains all variables belonging to
$\Sigma_T$ but not to $\Sigma_O$. 
Furthermore, we assume the dynamics of these variables to be {Markovian.}
We acknowledge that non-Markovian dynamics for these variables can also be considered
but will not be discussed in this context. 
These two properties imply that the transitional
mechanism of an OPK process is essentially controlled by the Markovian transition
dynamics of the variables in $\Sigma_T/\Sigma_O$, 
while those in $\Sigma_O$ are characterised by 
a smooth evolution equation unless when a transition occurs, 
at which time instant they perform discontinuous jumps.
In this overlapped transition process,  all the physical parameters
characterising the Markovian dynamics of the variables in $\Sigma_T/\Sigma_O$
can be potentially modulated not only by the local state of the variables in $\Sigma_O$,
but also by that of the state variables belonging to $\Sigma_X$.

This is the basic mechanism characterising the evolution of a LW
process, as defined by Eqs.~(\ref{eq:apk1}) and (\ref{eq:apk2}) with
the boundary conditions Eq.~(\ref{eq:apkbc}).  For this stochastic
model we have $\Sigma_X=\{{\bf Y}\}=\{X,\mathcal{T}\}$ and
$\Sigma_T=\{S,\mathcal{T}\}$, with the generalised Poisson process
$S(t)=(-1)^{\chi(t,\lambda(\tau(t)))}$.  Consequently, we identify
$\Sigma_O=\{\mathcal{T}\}$ and $\Sigma_T/\Sigma_O=\{S\}$.  In
agreement with our previous argument, in a LW therefore the
transitional age process $\mathcal{T}$ exhibits a smooth temporal
dynamic (a linear growth in this specific case) except for randomly
distanced discontinuities occurring at all times when the transitional
parameter $S$ performs a transition (here specifically a sign flip).

With these definitions at our disposal, 
we can now derive the statistical characterisation of a general OPK process. 
To keep our formalism general, we assume $n$ spatial dimensions for the position process, 
${\bf X}(t)$, with domain ${\mathcal D}_X \subseteq {\mathbb R}^n$, 
and $m\leq n$ dimensions for the overlapped variables,  
$\boldsymbol{\mathcal T}(t)$, with domain ${\mathcal D}_\tau \subseteq {\mathbb R}^m$.  
Correspondingly, we set 
$\Sigma_X=\{{\bf Y} \}= \{{\bf X},\boldsymbol{\mathcal T}\}$. 
Stochasticity is generated in the model by defining a set of $n$-dimensional velocity vectors ${\bf b}(\boldsymbol{\alpha})$, which depend on a stochastic parameter
$\boldsymbol{\alpha} \in {\mathcal D}_\alpha \subseteq {\mathbb R}^d$ ($d=1,2,\dots,n$).
The set ${\mathcal D}_{\alpha}$ can be either discrete or continuous, 
thus providing us with several modelling opportunities for the underlying stochastic dynamics of the overlapping PK process. 
The stochastic temporal evolution of these variables is specified by introducing 
a Poisson field $\boldsymbol{\Xi}(t;\lambda,A)$ in ${\mathbb R}^d$, such that
\begin{equation}
\boldsymbol{\alpha}(t)=\boldsymbol{\Xi}(t;\lambda,A)\:,
\label{eq4_1}
\end{equation}
where $\boldsymbol{\Xi}(0;\lambda,A)=\boldsymbol{\alpha}_0 \in
{\mathcal D}_\alpha$.  
The Poisson field is a continuous stochastic
process attaining values in ${\mathcal D}_\alpha$ whose statistical
description satisfies a continuous Markov chain dynamics defined by
the transition rate function $\lambda \geq 0$ 
and by the transition probability kernel $A$. 
We specify the functional dependence of these parameters below;
further details on Poisson fields are given in Appendix~\ref{sec:pf}.
In this setting, we therefore define  
$\Sigma_T= \{\boldsymbol{\Xi}, \boldsymbol{\mathcal T}\}$,  
$\Sigma_O=\{\boldsymbol{\mathcal T}\} $ and 
$\Sigma_T/\Sigma_O=\{\boldsymbol{\Xi} \}$.

Finally, we assume for the lifted process ${\bf Y}(t)$ the following stochastic differential equation, 
\be
\ud {\bf y}(t) = \widetilde{\bf v}({\bf y}(t)) \, \ud t + \widetilde{\bf b}(\boldsymbol{\Xi}(t),{\bf y}(t)) \, \ud t\:,
\label{eqo_2}
\ee
where we introduced a deterministic biasing velocity field and a stochastic perturbation defined as, respectively,   
\be
\widetilde {\bf v}({\bf y})=
\left (
\begin{array}{c}
{\bf v}({\bf x},\boldsymbol{\tau}) \\
{\bf w}({\bf x},\boldsymbol{\tau})
\end{array}
\right )
\, ,
\qquad
\widetilde{\bf b}(\boldsymbol{\alpha},{\bf y})
=
\left (
\begin{array}{c}
{\bf b}(\boldsymbol{\alpha},{\bf x}, \boldsymbol{\tau}) \\
0
\end{array}
\right ) \:.
\label{eqo_3}
\ee
The stochastic velocity vectors
${\bf b} : {\mathcal D}_\alpha \times {\mathcal D}_X \times {\mathcal D}_\tau \mapsto {\mathbb R}^n$, 
parametrised with respect to the states $\boldsymbol{\alpha}$ of the Poisson field $\boldsymbol{\Xi}(t)$, 
are also further modulated by an explicit dependence on the model state variables.
If we neglect this dependence of the stochastic perturbation on the state variables and 
the deterministic field $\widetilde{\bf v}$, 
and we identify the Poisson field $\boldsymbol{\Xi}(t)$ with $S(t)$, 
Eq.~(\ref{eqo_3}) is fully analogous to Eq.~(\ref{eq:generalised PKv}). 
In addition, even the  constitutive properties of the Poisson field
 $\boldsymbol{\Xi}(t)$, i.e., 
 the transition rate $\lambda$ and the probability kernel $A$, 
 can be specified more generally as depending
on the variables belonging to the set $\Sigma_X \cup \Sigma_T$, i.e.,
\begin{eqnarray}
\lambda&=&\lambda({\bf x}, \boldsymbol{\tau},\boldsymbol{\alpha}) \:, \\ 
%&\lambda({\bf y},\boldsymbol{\alpha})
A& = & A({\bf x}, \boldsymbol{\alpha},\boldsymbol{\tau},\boldsymbol{\alpha}^\prime,\boldsymbol{\tau}^\prime) \:.
\label{eqo_4}
\end{eqnarray}
The local functional dependence of the kernel $A$ on the state
variables ${\bf x}$ preserves the validity of a locality
principle for the stochastic process ${\bf X}(t)$, i.e., non-local 
action at distance is not allowed in our model. 
This would not be preserved if $A$ also depended on ${\bf x}^\prime$; 
if such a dependence existed, it would in fact imply the possible
occurrence of discontinuous spatial jumps ${\bf x}^\prime \mapsto {\bf x}$.
At each transition time $t^*$ of the Poisson field $\boldsymbol{\Xi}$, we equip Eq.~\eqref{eqo_2} with the boundary condition 
\begin{eqnarray}
\boldsymbol{\tau}(t^*_+)  & = & \boldsymbol{\tau}^{\prime}  \qquad \mbox{with probability } 
\;  A({\bf x}(t^*_-), \boldsymbol{\alpha},\boldsymbol{\tau}(t^*_-),\boldsymbol{\alpha}^\prime,\boldsymbol{\tau}^\prime)\;\:.
\label{eq4_15bisp}
\end{eqnarray}
Eqs.~(\ref{eqo_2}) and (\ref{eqo_3}) explictly state that in any time
interval in which no transitions in the Poisson field
$\boldsymbol{\Xi}(t)$ occur, the dynamics of the overlapped variable
$\boldsymbol{\mathcal T}(t)$ follows a strictly deterministic
kinematics.  In agreement with our previous arguments, the overlapped
variables thus do not depend explicitly on the main transitional
parameter, here $\boldsymbol{\Xi}(t)$, but only implicitly %on it
through its transition dynamics.  Moreover, we remark that if $A({\bf
  x},
\boldsymbol{\alpha},\boldsymbol{\tau},\boldsymbol{\alpha}^\prime,\boldsymbol{\tau}^\prime)$
is different from zero for $\boldsymbol{\tau} \neq
\boldsymbol{\tau}^\prime$, at any transition instant of the stochastic
process $\boldsymbol{\Xi}(t)$ the overlapped variables
$\boldsymbol{\mathcal T}(t)$ may perform a discontinuous jump
$\boldsymbol{\tau}^\prime \mapsto \boldsymbol{\tau}$.  Consequently,
$\boldsymbol{\mathcal T}(t)$ can display nonlocal dynamics, which is
fully consistent with the locality principle of space-time
interactions, provided that the ${\boldsymbol \tau}$ variables do not
correspond to any space-time coordinate or physical field (otherwise
the principle of bounded propagation velocity would be %dramatically
violated) but solely internal non-geometrical variables of the system
(such as the transitional age).

The statistical characterisation of Eq.~(\ref{eqo_2}) is formally
identical to that of conventional generalised PK processes that has been derived in Refs.~\cite{GBC17,GBC17a,GBC17c}.  
In our case this is obtained in terms of the PPDW functions 
$p({\bf y},t,\boldsymbol{\alpha})=
p({\bf x},\boldsymbol{\tau},t,\boldsymbol{\alpha})$. 
Introducing the notation 
${\bf x} =(x_1,\dots,x_n)$, $\boldsymbol{\tau}=(\tau_1,\dots,\tau_m)$,
$\nabla_{\bf x}=(\partial_{x_1},\dots, \partial_{x_n})$ and 
$\nabla_{\boldsymbol{\tau}}=(\partial_{\tau_1},\dots,\partial_{\tau_m})$, 
and assuming the domains $\mathcal{D}_{\tau}$ and $\mathcal{D}_{\alpha}$ to be continuous, we obtain the evolution equation
\begin{align}
\frac{\partial p({\bf x},\boldsymbol{\tau},t,\boldsymbol{\alpha})}{\partial t} &=
- \nabla_{\bf x} \cdot \left [ {\bf v}({\bf x},\boldsymbol{\tau}) \, p({\bf x},\boldsymbol{\tau},t,\boldsymbol{\alpha}) \right ]
- \nabla_{\bf x} \cdot \left [ {\bf b}({\bf x}, \boldsymbol{\tau},\boldsymbol{\alpha}) \, p({\bf x},\boldsymbol{\tau},t,\boldsymbol{\alpha}) \right ] \nonumber 
 - \nabla_{\boldsymbol{\tau}} \cdot \left [ {\bf w}({\bf x},\boldsymbol{\tau}) \,
p({\bf x},\boldsymbol{\tau},t,\boldsymbol{\alpha}) \right ]
\nonumber \\
& \quad 
- \lambda({\bf x},\boldsymbol{\tau},\boldsymbol{\alpha}) \, 
 p({\bf x},\boldsymbol{\tau},t,\boldsymbol{\alpha})
 +  \int_{{\mathcal D}_\tau}  \left [
\int_{{\mathcal D}_\alpha} 
\lambda({\bf x},\boldsymbol{\tau}^\prime,\boldsymbol{\alpha}^\prime)
A({\bf x}, \boldsymbol{\alpha},\boldsymbol{\tau},\boldsymbol{\alpha}^\prime,\boldsymbol{\tau}^\prime) \, p({\bf x},\boldsymbol{\tau}^\prime,t,\boldsymbol{\alpha}^\prime) \, \ud \boldsymbol{\alpha}^\prime    \right ]
 \ud \boldsymbol{\tau}^\prime \:,
\label{eqo_5}
\end{align}
where
\begin{equation}
\int_{{\mathcal D}_\tau}  \left [ \int_{{\mathcal D}_\alpha}
A({\bf x}, \boldsymbol{\alpha},\boldsymbol{\tau},\boldsymbol{\alpha}^\prime,\boldsymbol{\tau}^\prime) \, \ud \boldsymbol{\alpha}  \right ]
\ud  \boldsymbol{\tau} = 1
\label{eqo_6}
\end{equation}
for any $\boldsymbol{\tau}^\prime \in {\mathcal D}_\tau$ and 
$\boldsymbol{\alpha}^\prime \in {\mathcal
  D}_\alpha$.  Likewise, if the set $\mathcal{D}_{\alpha}$ is
discrete, Eqs.~(\ref{eqo_5}) and (\ref{eqo_6}) still hold with the
corresponding integral terms suitably substituted by summations. We
note that, in this case, the function $p({\bf
  x},\boldsymbol{\tau},t,\boldsymbol{\alpha})$ is to be interpreted as
a probability (not a density) with respect to the stochastic variables
$\boldsymbol{\alpha}$.  It is straightforward to show that this
general framework can generate LW dynamic as a special case, see
Appendix~\ref{subsec:LWrecovery} for the derivation.

\subsection{Transitional asynchrony lines the route to extended Poisson-Kac processes}
\label{subset:markov}

We now develop our EPK theory in its most general form by introducing
the concept of {\em transitional asynchrony}.  Let us illustrate this
concept by reconsidering the dynamics of the variable
$\boldsymbol{\tau}$, which we have introduced for LWs to denote the
transitional age.  We now assume that $\boldsymbol{\tau}$ follows a
Markovian transition dynamics with rate $\mu({\bf
  x},\boldsymbol{\tau})$ and transition probability kernel $M({\bf
  x},\boldsymbol{\tau},\boldsymbol{\tau}^{\prime})$ (see
Appendix~\ref{sec:pf}).  By construction, this is a left stochastic
kernel, i.e., $M({\bf x},\boldsymbol{\tau},\boldsymbol{\tau}^{\prime})
\geq 0$, $\int_{{\mathcal D}_\tau} M({\bf
  x},\boldsymbol{\tau},\boldsymbol{\tau}^{\prime}) \, \ud
\boldsymbol{\tau}=1$ for all $\boldsymbol{\tau}^{\prime} \in {\mathcal
  D}_\tau$.  Potentially, a further deterministic evolution can be
superimposed to this Markovian transitional dynamics.  Here for
simplicity we keep the same one as in Eqs.~(\ref{eqo_2}) and
(\ref{eqo_3}).
%, but we neglect the dependence on the state variable ${\bf x}$; 
%thus, this is described by the function ${\bf w}(\boldsymbol{\tau})$.  
%Under these assumptions, the probability 
%$\mbox{Prob}[\{\boldsymbol{\tau}(t) \in
%  (\boldsymbol{\tau},\boldsymbol{\tau}+ \ud \boldsymbol{\tau}) \}]=
%\widehat{P}(\boldsymbol{\tau},t) \, \ud \boldsymbol{\tau}$
%($\ud \boldsymbol{\tau}$ denotes a $d$-dimensional infinitesimal volume in $\mathcal{D}_\tau$)
%is specified by the density function
%$\widehat{P}(\boldsymbol{\tau},t)$ 
%satisfying the temporal evolution equation
%\begin{equation}
%\frac{\partial \widehat{P}(\boldsymbol{\tau},t)}{\partial t}=
%- \nabla_{\boldsymbol{\tau}} \cdot \left [ {\bf w}(\boldsymbol{\tau}) \,
%\widehat{P}(\boldsymbol{\tau},t) \right ]
%-\mu(\boldsymbol{\tau}) \, \widehat{P}(\boldsymbol{\tau},t)
%+ \int_{{\mathcal D}_\tau} \mu(\boldsymbol{\tau}^{\prime}) \, M(\boldsymbol{\tau},\boldsymbol{\tau}^{\prime}) \, \widehat{P}(\boldsymbol{\tau}^{\prime},t) \, \ud \boldsymbol{\tau}^{\prime}
%\label{eq4_3b}\:.
%\end{equation}
The stochastic equation of motion for the state variable 
${\bf x}(t)$ is equal to that encapsulated in Eq.~(\ref{eqo_2}). 
%, i.e., 
%\begin{equation}
%\ud {\bf x}(t)= {\bf v}({\bf x}(t),\boldsymbol{\tau}(t)) \, \ud t + {\bf
%  b}(\boldsymbol{\Xi}(t),{\bf x}(t),\boldsymbol{\tau}(t)) \, dt\:.
%\label{eq4_4}
%\end{equation}
%cp.\ with Eq.~(\ref{eq:pk}). 
Its statistical description involves the PPDW functions $p({\bf x},t,\boldsymbol{\alpha},\boldsymbol{\tau})$, where now both $\boldsymbol{\alpha}$ and $\boldsymbol{\tau}$ are to be interpreted as stochastic parameters, 
which are solutions of the hyperbolic equations 
(as expressed in the form of first order equations with
respect to time $t$, position ${\bf x}$ and $\boldsymbol{\tau})$
\begin{align}
\frac{\partial p({\bf x},\boldsymbol{\tau},t,\boldsymbol{\alpha})}{\partial t} &=
- \nabla_{\bf x} \cdot \left [ {\bf v}({\bf x},\boldsymbol{\tau}) \, p({\bf x},\boldsymbol{\tau},t,\boldsymbol{\alpha}) \right ]
- \nabla_{\bf x} \cdot \left [ {\bf b}({\bf x}, \boldsymbol{\tau},\boldsymbol{\alpha}) \, p({\bf x},\boldsymbol{\tau},t,\boldsymbol{\alpha}) \right ] \nonumber 
 - \nabla_{\boldsymbol{\tau}} \cdot \left [ {\bf w}({\bf x},\boldsymbol{\tau}) \,
p({\bf x},\boldsymbol{\tau},t,\boldsymbol{\alpha}) \right ]
\nonumber \\
& \quad 
- \lambda({\bf x},\boldsymbol{\tau},\boldsymbol{\alpha}) \, 
 p({\bf x},\boldsymbol{\tau},t,\boldsymbol{\alpha})
+\int_{{\mathcal D}_\alpha} 
\lambda({\bf x},\boldsymbol{\tau},\boldsymbol{\alpha}^\prime)
A({\bf x}, \boldsymbol{\alpha},\boldsymbol{\tau},\boldsymbol{\alpha}^\prime,\boldsymbol{\tau}) \, p({\bf x},\boldsymbol{\tau},t,\boldsymbol{\alpha}^\prime) \, \ud \boldsymbol{\alpha}^\prime 
\nonumber \\
& \quad
- \mu({\bf x},\boldsymbol{\tau}) \, 
 p({\bf x},\boldsymbol{\tau},t,\boldsymbol{\alpha})
+\int_{{\mathcal D}_\tau} \mu({\bf x},\boldsymbol{\tau}^\prime)
M({\bf x},\boldsymbol{\tau},\boldsymbol{\tau}^\prime) \, p({\bf x},\boldsymbol{\tau}^{\prime},t,\boldsymbol{\alpha}) \,  \ud \boldsymbol{\tau}^\prime \:,
\label{eq4_5}
\end{align}

In formal terms, this generalised PK process is specified by the sets
of state and transitional variables $\Sigma_X=\{{\bf X}\}$ and
$\Sigma_T=\{\boldsymbol{\Xi},\boldsymbol{\mathcal{T}}\}$,
respectively.  In contrast, the overlapping PK process
Eq.~(\ref{eqo_2}) is defined by the sets $\Sigma_X=\{{\bf
  X},\boldsymbol{\mathcal{T}}\}$ and
$\Sigma_T=\{\boldsymbol{\Xi},\boldsymbol{\mathcal{T}}\}$.
Interestingly, these two different formal structures (in particular,
characterised by different transitional mechanisms) lead to different
statistical properties (compare Eqs.~(\ref{eqo_5})
and~(\ref{eq4_5})). Both are characterised by the interplay of the
main transitional parameters $\boldsymbol{\Xi}(t)$ and
$\boldsymbol{\mathcal{T}}(t)$ to determine the stochastic dynamic of
the position process ${\bf X}(t)$.  The difference between the two
formulations is a consequence of the different synchronisation between
%can be traced back to the mutual transitional independence 
the processes $\boldsymbol{\Xi}$ and $\boldsymbol{\mathcal{T}}$. 
%which highlights that the resulting processes describe two very different physical scenarios.
This concept is made evident by defining for each of the two transitional parameters 
the marginal transition time density, 
$T_{\boldsymbol{\Xi}}(t_1)$ and $T_{\boldsymbol{\mathcal{T}}}(t_2)$, respectively. 
Likewise, for the joint process $(\boldsymbol{\Xi}(t),\boldsymbol{\mathcal{T}}(t))$ we can specify the corresponding bivariate transition time density function
$T_{\boldsymbol{\Xi},\boldsymbol{\mathcal{T}}}(t_1,t_2)$.
Correspondingly, we can define the conditional transitional time density, 
%of $\tau_{\boldsymbol{\mathcal{T}}}$ conditional to $\tau_{\boldsymbol{\alpha}}$, 
%denoted as 
$T_{\boldsymbol{\mathcal{T}}|\boldsymbol{\Xi}}(t_2 \, | \, t_1)$, 
by the relation %(by simple probability theory) 
\begin{equation}
T_{\boldsymbol{\Xi},\boldsymbol{\mathcal{T}}}(t_1,t_2) = T_{\boldsymbol{\mathcal{T}}|\boldsymbol{\Xi}}(t_2 \, | \, t_1)
\, T_{\boldsymbol{\Xi}}(t_1) \:.
\label{eql_5}
\end{equation}
This quantity elucidates the different physics underlying the
generalised PK process defined by Eqs.~(\ref{eq4_5}) and the OPK
process Eq.~(\ref{eqo_5}). For the former
\begin{equation}
T_{\boldsymbol{\mathcal{T}}|\boldsymbol{\Xi}}(t_2 \, | \, t_1) = T_{\boldsymbol{\mathcal{T}}}(t_2) \:,
\end{equation}
meaning that the processes $\boldsymbol{\Xi}(t)$ and $\boldsymbol{\mathcal{T}}(t)$ are transitionally independent. Clearly, in this case the variable $\boldsymbol{\tau}$ loses its physical meaning of an elapsed time from the previous transition. 
%Markov processes, 
For the latter
\begin{equation}
T_{\boldsymbol{\mathcal{T}}|\boldsymbol{\Xi}}(t_2 \, | \, t_1) = \delta(t_2-t_1) \:, 
\end{equation}
i.e., the two processes are transitionally synchronised.
In this case, $\boldsymbol{\tau}$ is indeed an elapsed time, or transitional age.
%\begin{figure}[h]
%\begin{center}
%{\includegraphics[height=12cm]{generalised PK_review.eps}}
%\end{center}
%\caption{Qualitative difference in the realizations of
%the two stochastic processes $S(t)$ and $\Lambda(t)$ in 
%a overlapping PK process (panel a) and in a generalised PK process (panel b)
%parametrised with respect to $\lambda$. Lines (a) refers to
%$s(t)$, lines (b) to $\lambda(t)$.}
%\label{Figlam}
%\end{figure}
%This phenomenon is pictorially illustrated in figure \ref{Figlam}
%showing the qualitative evolution of a realization of $S(t)$ and
%$\Lambda(t)$ in the two cases.}

Remarkably, all the previous considerations can be applied as well to any model parameters. 
%, which can be Markovianised (in the sense defined above) and/or considered as overlap variables.   
%are neither  constant nor functions of the state variables, but
%In practice, this means that members of the model parameters set 
%$\Sigma_P$ become members of the set of transitional parameters $\Sigma_T$.
For simplicity, let us consider a one-dimensional PK process
(extending these arguments to the three-dimensional setting is
straightforward), 
which is specified formally by the sets of state variables, transitional
parameters and model parameters $\Sigma_X=\{X\}$, $\Sigma_T=\{S\}$ and
$\Sigma_P=\{b,\lambda\}$, respectively.  We then assume
$b=b_0\beta(t)$ and $\lambda=\lambda_0 \lambda(t)$.  A new family of
EPK processes can then be obtained by considering a subset of the
model parameters as transitional variables, i.e., $\Sigma_X=\{X\}$,
$\Sigma_T=\{S,\Lambda,B\}$, $\Sigma_P=\{b_0,\lambda_0\}$.  Similarly,
a new family of OPK processes can be obtained by considering them as
both state and transitional variables, i.e.,
$\Sigma_X=\{X,B,\Lambda\}$, $\Sigma_T=\{S,B,\Lambda\}$,
$\Sigma_P=\{b_0,\lambda_0\}$.  Each of these classes of finite
propagation velocity processes are characterised by different
transitional structures, thus leading to different statistical
properties.
%(two specific examples will be discussed later in Sections~\ref{subsec:markovb} and \ref{subsec:lam}).   
However, the argument above highlights that these two processes are particular cases of a wider class of models, where transitional asynchronies between the state variables and the main generator of microscopic stochasticity are encapsulated in the transitional time conditional density, for the example considered  
$T_{\{\beta,\Lambda\}|S}$. 

We denote these general models of stochastic dynamics with finite
propagation velocity as EPK processes.  The formulation of EPK theory
that includes OPK processes (and LWs among them) as special cases is
our second main result.  Figure~\ref{fig:stokin} pictorially describes
the inclusive relationship between the main classes of stochastic
kinematics formulated so far.  The common denominator between all
these stochastic models, except for Wiener processes (which are in
this sense a singular limit), is the assumption of a finite
propagation speed and finite transition rates.
The main difference resides in the statistics of
the transition times, which is exponential for PK processes, power law
tailed for LWs and fully generic for EPK processes; and in the
existence (or absence) of transitional asynchronies among the state
variables and the microscopic stochastic generator.
%This generates models with statistical properties hybrid of pure Markovianised and overlapping generalised PK processes.    
%These examples enable us to show the flexibility of generalised PK and overlapping PK schemes, and 
%and on the other hand, 
%to formalise the subtle but important difference between the two approaches.

This analysis elucidates the physical meaning and the broad range of
applications of our extension of conventional PK theory.
Transitionally independent PK models generically yield microscopic
processes subjected to external (environmental) fluctuations that
influence their local dynamics, but they can be considered independent of
the fluctuations in the local microscopic motion.
%This is reminiscent of the physical processes underlying several models of diffusing-diffusivity that have been discussed in recent years as they can generate Brownian yet non-Gaussian diffusion \cite{Wang:2012aa,Chechkin:2017aa}. In fact, later in subsection \ref{subsec:markovb} we will show that these approaches can be reproduced within our generalised theory, 
%which additionally provides a clear-cut physical interpretation of the superstatical effect. 
Conversely, OPK models capture complex microscopic 
fluctuations, the statistical description of which requires the introduction 
of inner transitionally synchronised degrees of freedom. 
This is for example the case of the transitional age $\tau$
%introduced in order to describe LW processes.
for LW processes. 
In between these two limiting cases, a spectrum of intermediate
situations can be defined \textit{ad hoc}, 
by specifying the transitional time conditional density between the state variables and the main transitional process. 
%(in the example above, $\{\beta,\Lambda\}$ and $S$ respectively). 
%$T_{\Lambda|S}(\tau_\lambda \, | \, \tau_s)$. 
%is neither impulsive nor
%coicides with $T_\Lambda(\tau_\lambda)$.

\section{Extended Poisson-Kac processes: case studies} %and perspectives}
\label{sec:ex}

We now discuss three specific examples of one-dimensional EPK processes.
First, %in Subsec.~\ref{subsec:senesc} 
we introduce a {\em transitional
  senescent} random walk, where the transitional dynamic depends
explicitly on the number of total transitions already occurred. 
We specifically study an EPK model where the age 
to which the walker is reset following a velocity transition 
is parametrised as an increasing function of the total number of transitions.
Second, %in Subsec.~\ref{subsec:markovb} 
we discuss an EPK model that
can reproduce ``Brownian yet non Gaussian" diffusion
\cite{Wang:2012aa}. 
This behaviour can be obtained by considering an EPK process 
where the walker velocity follows a Markovian jump dynamic 
transitionally independent from the corresponding dynamic of the Poisson field.
Differently from other phenomenological
approaches \cite{Chechkin:2017aa}, our model provides a clear
microscopic interpretation of this dynamics.
%  This approach is applied to a typical case study of 
%for which, the results recently obtained in \cite{Chechkin:2017aa}
%using a stochastic
%diffusion-diffusivity approach are  recovered in a simple way, by
%performing the Markovianisation of the velocity parameter.
Finally, %in Subsec.~\ref{subsec:lam} 
we formulate an EPK process with correlated transitional dynamic.
If these correlations are neglected, the model generates LW dynamic. 
If the correlations lead to increasing transition rates over time, 
the model yields a dynamic characterised by a sub- to superdiffusive crossover in the mean square
displacement.  
%To conclude, the relation between overlapping PK and Markovianised generalised PK,
%as two limiting ways of complexifying the fluctuations
%acting on PK models, is addressed in
%detail by taking the transition rate $\lambda$
% as a parameter.
The variety of diffusive dynamics that can be captured by EPK
processes highlights the modelling power of our theory.  We remark
that in this work we only discuss examples of OPK and transitionally
independent EPK processes.  The analysis of further EPK processes,
requiring the occurrence of more non-trivial multivariate distributions
of joint transition times (which is an intricate problem even for
finite Markov chains \cite{cmc1,cmc2} and associated counting
processes \cite{multicount}) will be developed in future
communications.
%\textcolor{red}{The final subsection ~\ref{subsec:persp} explores the
%application of EPK theory to classical and fundamentl problems
%of statistical physics.}

%\subsection{Senescent Walks as overlapping PK processes}
\subsection{Transitional Senescent Random Walks}
\label{subsec:senesc}

In 1961 Hayflick and Moorhead reported that cultured
 proliferating
human diploid cells stop cellular division after a limited number of
mytotic events \cite{Hayflick1961,Hayflick1965} showing   
 that this phenomenon is related to senescence, i.e. to
aging process occurring at a cellular level \cite{Harley1990,Krtolica2001}.
  Apart from its biological
and biochemical relevance, senescence is remarkable from a statistical
mechanical perspective, where it translates to the formulation of
random walk processes whose dynamic and transitional properties can
decay as the number of transitions increases.
In analogy with the terminology established in the biological context, 
we call this feature  {\em transitional senescence}. 
Correspondingly, we refer to transitionally senescent random walk processes implementing this feature.
%This is the essence of several aging models that have been discussed in the literature 
%, about which there exists a broad literature 
%\cite{}. 
A particular example has been discussed in the context of LWs in Ref.~\cite{Giona2019age}, 
%Specifically, 
where a progressive decrease of the walker velocity $b$
with the number of transitions has been shown to 
%been addressed in \cite{GDCCK19}
yield qualitative effects for the statistics of motion. 
Here we show how this feature can be easily accommodated within our general theory of EPK processes. 
%want to connect these
%phenomena with the general theory of overlapping PK processes 
%developed in the previous section.

%Formally, we say that a random walk is transitionally senescent if its 
%%corresponds to the fact that the
%dynamic properties and transitional mechanisms depend on the number of
%transition events occurred.  
Transitional senescence can be represented by the fact that
either the transition rate $\lambda$ and/or the walker velocity $b$
can become functions of the underlying stochastic counting process
$N(t)$, associated with the Markovian transitional structure of the
process.  The process $N(t)$ enumerates the transitions occurred up to the time
interval $[0,t)$.  As a pedagogical example, we consider first the case of a general transitionally senescent PK process.  
The system of transitional
parameters for this process is $\Sigma_T=\{S,N\}$.  Moreover, the
system of state variables is $\Sigma_X=\{X,N\}$, albeit the dynamics
of $N(t)$ is elementary.  In fact, $\ud N(t)/\ud t=0$ in any time
interval between two transitions, and $N(t) \mapsto N(t)+1$ at any
transition instant.  Because %$\Sigma_O=\{N\}$, meaning that the
counting process $N(t)$ is transitionally synchronised with $S(t)$,
this is for all intents and purposes an OPK process.  Its
statistical description involves the family of PPDW functions
$p_s^{(n)}(x,t)\equiv p(x,n,t,s)$ with $s=\pm$ and $n \in {\mathbb
Z}^{\geq0}$.
%which are parametrised with respect to %all the possible occurrences of 
%the process $S$ %and $N$, 
%i.e., $s=\pm$. %and $n \in {\mathbb Z}^{\geq}$.
%Since $N(t)$ attains values in $$,
%applying the same approach discussed for overlapping PK processes parametrised
%with respect to the transitional age, 
%its statistical description involves the family of PPDW functions 
To model in full generality the transitional senescence, we assume both the transition rate $\lambda_n$ and the velocity $b_n$ to be functions of the counting state $n$.
The evolution equations for the  associated OPK  process are thus expressed by
\begin{eqnarray}
\frac{\partial p_\pm^{(0)}(x,t)}{\partial t}
& = & \mp b_0 \, \frac{\partial p_\pm^{(0)}(x,t)}{\partial x}
-\lambda_0 \, p_\pm^{(0)}(x,t) \: ,
\label{eqs_1}
\\
\frac{\partial p_\pm^{(n)}(x,t)}{\partial t}
& = & \mp b_n \, \frac{\partial p_\pm^{(n)}(x,t)}{\partial x}
-\lambda_n \, p_\pm^{(n)}(x,t) + \lambda_{n-1} \, p_\mp^{(n-1)}(x,t) \: .
\nonumber
\end{eqnarray}
The generalisation to a senescent LW is now straightforward, 
as we just need to include the transitional age among the overlapping variables.
Therefore, $\Sigma_X=\{X,{\mathcal T}, N\}$, $\Sigma_T=\{S,{\mathcal  T},N \}$, 
and thus $\Sigma_O=\{{\mathcal T}, N\}$.
%Assuming constant velocity $b$,
The PPDW functions for this process are
$p_s^{(n)}(x,\tau,t)\equiv p(x,\tau,n,t,s)$,  
which, similar to Eq.~(\ref{eqs_1}), now read
\begin{eqnarray}
\frac{\partial p_\pm^{(0)}(x,\tau,t)}{\partial t}
 & = & \mp  b_0 \, \frac{\partial p_\pm^{(0)}(x,\tau,t)}{\partial x}
- \frac{\partial p_\pm^{(0)}(x,\tau,t)}{\partial \tau}
- \lambda_0(\tau) \, p_\pm^{(0)}(x,\tau,t) \:, \label{eqs_2} \\
\frac{\partial p_\pm^{(n)}(x,\tau,t)}{\partial t}
 & = & \mp b_n \, \frac{\partial p_\pm^{(n)}(x,\tau,t)}{\partial x}
- \frac{\partial p_\pm^{(n)}(x,\tau,t)}{\partial \tau}
- \lambda_n(\tau) \, p_\pm^{(n)}(x,\tau,t)
+ \int_0^\infty k_n(\tau,\tau^\prime) \, \lambda_{n-1}(\tau^\prime)
\, p_\mp^{(n-1)}(x,\tau^\prime,t) \, d \tau^\prime \: .
\nonumber
\label{eq:tslw}
\end{eqnarray}
Here the transitional senescence of the process is expressed by the
Markovian transition kernels $k_n(\tau,\tau^\prime)$, which are
assumed to depend on the counting state $n$. {The kernel
$k_n(\tau,\tau^\prime)$  may be an impulse Dirac delta function in $\tau$
or may admit a non-atomic support in $\tau$, consisting in an interval 
of values of $\tau$ for which $k_n(\tau,\tau^\prime)>0$.} 

Equations~(\ref{eqs_2}) represent the statistical characterisation of a general transitionally senescent LW. 
To illustrate how these processes can reveal novel dynamical features,
%Nevertheless, for %the sake of simplicity, 
we specify the senescing process such that, after any transition, 
the walker transitional age is not reset to zero but to a prescribed larger value.  
%can be modelled as if the effective age after
%any transition where not $\tau=0$ but a higher value.
In this respect, we introduce a diverging
sequence of non-negative numbers $\{ \tau_n^0 \}_{n=0}^\infty$ 
with $\tau_0^0=0$ and $\tau_n^0 < \tau_{n+1}^0$, such that
\begin{equation}
k_n(\tau,\tau^\prime)= \delta(\tau-\tau_n^0) \: , \qquad
\lambda_n(\tau)=\lambda(\tau) \: ,
\label{eqs_3}
\end{equation}
%$n=0,1,\dots$,
where $\lambda(\tau)$ is specified by Eq.~(\ref{eq:lwtr}). 
Clearly, each $\lambda_n(\tau)$ is defined in the age interval
$(\tau_n^0,\infty)$, 
which implies that the corresponding transition time densities also depend on $n$.
%, i.e., $T_n(\tau)$.
According to Eqs.~(\ref{eqs_3}), the age boundary conditions are
\begin{equation}
p_\pm^{(n)}(x,\tau_n^0,t)= \int_{\tau_{n-1}^0}^\infty
\lambda(\tau^{\prime}) \, p_\mp^{(n-1)}(x,\tau^{\prime},t) \, \ud \tau^{\prime} \:.
\label{eqs_4}
\end{equation}
We simulate numerically the stochastic process associated
with Eqs.~(\ref{eqs_2}) by further assuming constant
velocities $b_n=b$ and $\tau_n^0=(n-1) \, \tau^0$ (for $n>0$), where
$\tau^0$ is a constant positive parameter.  In Fig.~\ref{Figkl_2}(a)
we present the temporal evolution of the mean square displacement of
this dynamic, $\sigma_x^2(t)=\int_{-\infty}^\infty x^2 \, P(x,t) \,
\ud x$, obtained from stochastic simulations.  Here we have defined the
walker position distribution as
$P(x,t)=\sum_{n=0}^{\infty}\sum_{s=\pm} P_s^{(n)}(x,t)$ with the
marginal PPDW functions $P^{(n)}_s(x,t)=\int_{\tau_n^0}^{\infty}
p_s^{(n)}(x,\tau^{\prime},t)\ud \tau^{\prime}$.
We simulated $10^7$ trajectories, all initialised at the origin,
with $b=1$ and $\xi=1.5$ for different values of the parameter $\tau^0$. 
For $\tau^0=0$ we recover the conventional LW, in which case
$\sigma_x^2(t) \sim t^\gamma$ with $\gamma=(3-\xi)$.   
In contrast, for $\tau^0 \neq 0$
%, the scaling of $\sigma_x^2(t)$ deviates from this behaviour; in particular, 
we find the different long-term scaling 
$\sigma_x^2(t) \sim t^{\gamma_{\rm eff}}$, 
characterised by the effective exponent $\gamma_{\rm eff}>\gamma$.
This result is physically intuitive, because the transitional senescence 
induces a slowing-down in the transitional mechanism, 
which determines a more pronounced superdiffusive behaviour.
In Fig.~\ref{Figkl_2}(b) we present results for $\gamma_{\rm eff}$ for different values of the parameter $\xi$ predicted by the stochastic simulations.
%While the detailed analysis of a senescent LW is outside
%the scope of this article, it is worth
Remarkably, we observe that $\gamma_{\rm eff}>1$ 
%in the presence of this very simple process of senescence 
even for values $\xi>2$ for which the corresponding LW ($\tau^0=0$)
displays an Einsteinian scaling, $\gamma=1$.  For $0<\xi<1$ the
senescent LW exhibits ballistic diffusion, $\gamma_{\rm eff}=2$, similar
to its non-senescing counterpart.  To analytically predict how
$\gamma_{\text{eff}}$ depends on $\xi$ is an open problem left to
further studies.

\begin{figure}[!t]
\begin{center}
%{\includegraphics[height=6.5cm]{generalised PK_over_15.eps}}
{\includegraphics[width=18cm,keepaspectratio]{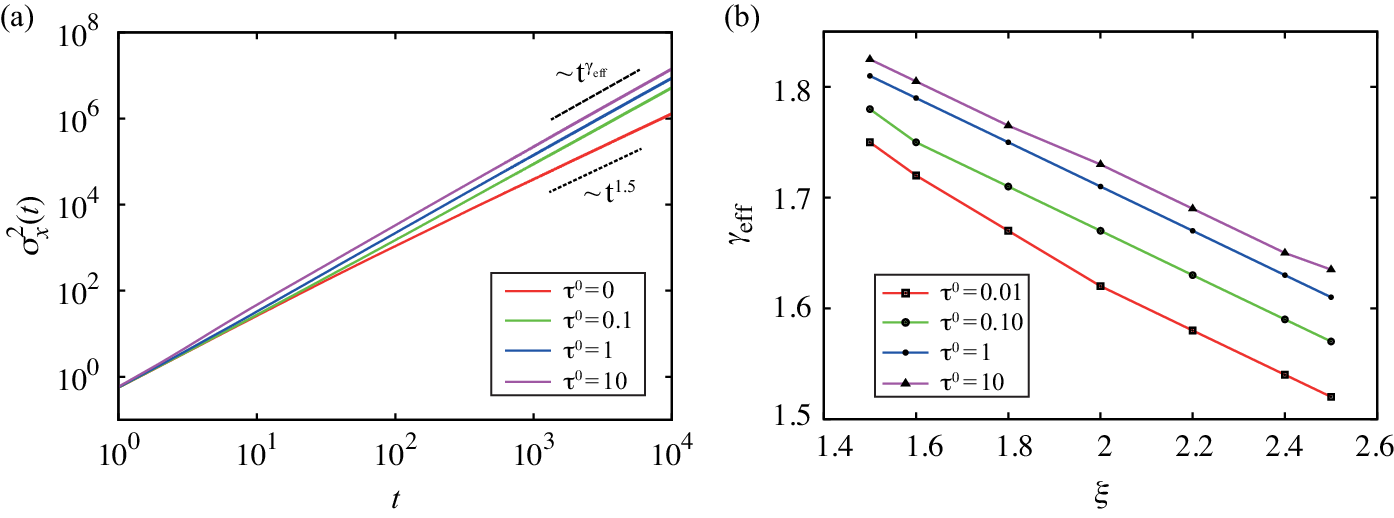}}
\end{center}
\caption{Superdiffusion for the model of a transitionally senescent
    LW, see Eq.~(\ref{eq:tslw}) and further specifications in the
    text. Here $\tau^0$ defines the senescence time. The special case
    $\tau^0=0$ corresponds to the conventional LW model without
    senescence. $\xi$ is the exponent of the power law distributed
    transition times, see Eqs.~(\ref{eq:pltt}), respectively
    (\ref{eq:lwtr}). For the speed of the walker we choose $b=1$.  (a)
    Temporal evolution of the mean square displacement $\sigma_x^2(t)$
    for $\xi=1.5$.  We see that senescence enhances superdiffusion
    beyond the conventional LW solution. (b) The scaling exponent
    $\gamma_{\rm eff}$ as a function of $\xi$ increases by increasing
    the senescence time $\tau^0=0$.}
\label{Figkl_2}
\end{figure}

%\subsection{Markovianisation of $b$}
%\subsection{Markovianisation of the walker velocity}
\subsection{Brownian yet non-Gaussian diffusive extended Poisson-Kac processes}
\label{subsec:markovb}

Brownian yet non-Gaussian diffusion is the hallmark of a specific
class of transport phenomena in out-of-equilibrium systems
\cite{Wang2009,Wang2012}.  It has recently been observed, among
others, for beads diffusing on lipid tubes and networks
\cite{Wang2009,Toyota2011,Valentine2001,e2014time,Samanta2016}, for
passive tracers immersed in active suspensions \cite{Leptos2009}, for 
heterogenous populations of moving nematodes \cite{Hapca2009}, and 
in the context of intracellular transport \cite{Witzel2019}.
The terminology refers generically to dynamics where the position mean
square displacement scales linearly for long times, while the position
statistics exhibits non-Gaussian tails. 
Clearly, this stochastic dynamic cannot be modelled by standard
Brownian motion. Hence, the formulation of suitable stochastic
processes that can capture this peculiar diffusive feature was subject
to numerous theoretical investigations in recent years.  According to
the observation that a Laplace distribution with linearly scaling
second moment can be derived from a superstatistical approach
\cite{Beck2003}, where Gaussian distributions are averaged over
Laplace distributed diffusion coefficients \cite{Hapca2009}, a family
of ``diffusing diffusivity" models has been proposed
\cite{Chubynsky2014,Chechkin:2017aa}. For these models the position
process is described by standard Brownian motion with a diffusion
coefficient performing a prescribed stochastic dynamic. We note that
microscopic derivations of this dynamic have recently been considered
in the context of active matter \cite{Kanazawa2020}.  Here we show
that EPK processes can naturally account for the hierarchical level of
fluctuations generating Brownian yet non-Gaussian diffusion by
allowing the walker speed to change over time according to a given
Markov chain dynamics.

For simplicity, we consider a PK process and assume $b(t)=b_0 \,
\beta(t)$, where $b_0>0$ is a constant parameter and $\beta(t)$ is a
stochastic process attaining values in $\mathcal{D}_\beta =
[0,b_{max}]$,
%$\mathcal{D}_\beta \subseteq {\mathbb R}^+$
which is characterised by a Markovian transition dynamics 
%If we denote as $P_b(\beta,t)$ its associated probability density at time  $t$,
%we can write down the temporal evolution equation
%\begin{equation}
%\frac{\partial P_b(\beta,t)}{\partial t}= - \mu \, P_b(\beta,t)
%+ \mu \, P^{\,\star}_b(\beta) \int_{\mathcal{D}_\beta} P_b(\beta^\prime,t) \,
%\ud \beta^\prime \: ,
%\label{eqkl_1}
%\end{equation}
%where $\mu$ is a constant transition rate and $P^{\,\star}_b(\beta)=
%\lim_{t \rightarrow \infty} P(\beta,t)$ the corresponding stationary
%distribution. For this we have assumed that the transition probability
%kernel $A(\beta,\beta^\prime)$ governing the redistribution amongst
%the velocity intensities yields the stationary distribution,
%$A(\beta,\beta^\prime)=P^{\,\star}_b(\beta)$.  
with a constant transition rate $\mu$ and 
the transition probability kernel $A(\beta,\beta^\prime)$ (see Appendix~\ref{sec:pf}). 
If we denote as $P_b(\beta,t)$ the probability density at time $t$ associated to this dynamic,
we assume the process to admit the stationary distribution  
$P^{\,\star}_b(\beta)= \lim_{t \rightarrow \infty} P(\beta,t)$. 
This requires the condition $A(\beta,\beta^\prime)=P^{\,\star}_b(\beta)$.
Consequently, we define the EPK process
\begin{equation}
\frac{d x(t)}{d t}= b_0 \, \beta(t) \, (-1)^{\chi(t,\lambda)} \:.
\label{eqkl_2}
\end{equation}
Under these assumptions, $\Sigma_X=\{X\}$, $\Sigma_T=\{S,\beta\}$ and
$\Sigma_P=\{b_0,\lambda\}$. 
%within the framework of our EPK theory
%outlined before.
%where $\beta(t)$ possesses the above described  Markovian dynamics.
%Eq.~(\ref{eqkl_2}), despite its formal simplicity,
%is the basic  archetype of superstatistical diffusing-diffusivity
%models, in which the diffusivity (in the present case the velocity
%intensity) is regarded as a stochastic process.

The statistical description of the EPK process Eq.~(\ref{eqkl_2}) involves
the PPDW functions $p_s(x,t,\beta)\equiv p(x,t,\beta,s)$. They are
parametrised with respect to $s= \pm$, corresponding to the
``microstochasticity'' in the local particle movements associated with
the Poissonian parity switching process, and with respect to
$\beta \in \mathcal{D}_{\beta}$, 
 corresponding to the ``superstatistical structure''
superimposed to the microscopic randomness \cite{Beck2003}. 
%Here $B$ is the set of attainable
%velocity intensities, modulo the constant prefactor $b_0$.
%As for any generalised PK process 
%The evolution equations for $p_\pm(x,t,\beta)$ 
%follow from the transition mechanism expressed
%by Eq.~(\ref{eqkl_1}), coupled with the Poissonian parity
%transitions entering Eq.~(\ref{eqkl_2}).
Given the transitional independence of the parameters $S$ and $\beta$, 
the evolution equations for $p_\pm(x,t,\beta)$ can be derived similarly to Eq.~\eqref{eq4_5}, i.e.,
%leads to the  following
%evolution equations for the PPDW functions,
\begin{align}
\frac{\partial p_\pm(x,t,\beta)}{\partial t} &=
 \mp b_0 \, \beta \, \frac{\partial p_\pm(x,t,\beta)}{\partial x}
\mp \lambda [ p_+(x,t,\beta)-p_-(x,t,\beta) ] -\mu \, p_\pm(x,t,\beta)
%\nonumber \\
%& \quad 
+ \mu \, P^{\,\star}_b(\beta) \, \int_{\mathcal{D}_{\beta}} p_\pm(x,t,\beta^\prime) \,
d \beta^\prime \:.
\label{eqkl_3}
\end{align}
We now assume initial equilibrium conditions with respect to the transitional
parameters $(s,\beta)$, and that all the particles are initially
located at $x=0$. This implies the initial condition
 $p_\pm(x,0,\beta)= P^{\,\star}_b(\beta)\, \delta(x)/2$.
The solution of Eq.~(\ref{eqkl_3}) with the above initial
conditions
 admits
a characteristic (and non trivial)
short-term behaviour, provided that
$\lambda \gg \mu$, i.e., that  a separation of time scales exists
between the two stochastic  contributions modulating the walker
dynamic Eq.~(\ref{eqkl_2}).
For short time scales, $t \ll 1/\mu$, the recombination among the
velocities is negligible, and consequently the short time solution
is simply the propagation of the initial condition via the
PK mechanism. 
%characterising the microscopic stochasticity in
%the velocity-direction switching.
Thus,
\begin{equation}
p_\pm(x,t,\beta) = \frac{P^{\,\star}_b(\beta)}{2} \left [ {\mathcal G}_{\pm,+}(x,t; b_0 \, \beta,\lambda)+ {\mathcal G}_{\pm, -}(x,t; b_0 \, \beta,\lambda) \right ]
\label{eqkl_4}
\end{equation}
where ${\mathcal G}_{s_1,s_2}(x,t; b_0 \,  \beta, \lambda)$ 
($s_1,s_2=\pm$) are the
entries of the tensorial Green functions 
%for the PK equation
%referred to the PPDW description of the process \cite{ming_giona}, 
for the PPDW equations of the PK process 
with velocity equal to $b_0 \, \beta$ and transition rate $\lambda$ \cite{ming_giona}.
If $\lambda$ is large enough, keeping fixed the ratio $b_0^2/2\lambda=D_0$,
the PK process approaches its Kac limit, which is the parabolic diffusion
equation. For each $\beta$, $[ {\mathcal G}_{\pm,+}(x,t; b_0\,\beta, \lambda)+ {\mathcal G}_{\pm, -}(x,t; b_0 \, \beta),\lambda] \rightarrow G_{D(\beta)}(x,t)$
%{\color{burntorange} Andrea: I modified this equation. Please check. In the previous formulation, it seems to there was an extra factor two in eq. (49).}
, where $G_{D(\beta)}(x,t)$
is the parabolic heat kernel for the value $D(\beta)=D_0 \, \beta^2$ of the
diffusivity. Thus, the overall marginal density $P(x,t)=\int_{\mathcal{D}_{\beta}} [p_+(x,t,\beta^{\prime})+p_-(x,t,\beta^{\prime})] \ud \beta^{\prime}$ approaches, in the Kac limit, the expression
\begin{equation}
P(x,t) = \int_{\mathcal{D}_{\beta}} \frac{P^{\,\star}_b(\beta)}{\sqrt{4 \, \pi \, D_0 \, \beta^2 \, t}}
\exp \left [- \frac{x^2}{4 \, D_0 \, \beta^2 \, t} \right ] \, \ud \beta^{\prime} \:.
\label{eqkl_5}
\end{equation}
%Consider a popular application of Eq.~(\ref{eqkl_5}):
Remarkably, if we assume 
%$\mathcal{D}_{\beta}=[0,\infty)$ and 
$P^{\,\star}_b(\beta)$ to be a generalised Gamma distribution, i.e., 
%\begin{equation}
$P^{\,\star}_b(\beta)=2 \, \kappa \, A \, \beta \, e^{-\kappa \beta^2}$
%\label{eqkl_7}
%\end{equation}
with $\kappa=D_0/D_*$ and 
the normalisation constant $A=(1-e^{-\kappa \, \beta_{\rm max}^2} )^{-1}$, 
%(see \cite{epaps}),
we recover at short time scales the Laplace distribution
\begin{equation}
P(x,t)=\frac{1}{\sqrt{4 \, D^* \, t}}
\exp \left( -\frac{|x|}{\sqrt{D^* \, t}}\right) \:,
\label{eqkl_6}
\end{equation}
%from Eq.~(\ref{eqkl_2}) - 
which has been  object of extensive investigations using  a variety of phenomenological diffusing diffusivity models \cite{Wang:2012aa,Hapca2009,Chechkin:2017aa,Chubynsky2014}. 
%{\color{burntorange} Andrea: citations missing} 
Our argument demonstrates that the process Eq.~(\ref{eqkl_2}) can be
regarded as the archetype of such models, with the advantage
that the superstatistical effect has a clear-cut physical
interpretation in terms of the Markovian recombination of the
microscopic velocities of the random walker.
%, in which the diffusivity (in the present case the velocity
%intensity) is regarded as a stochastic process.}

In Fig.~\ref{Figkl_4} we validate our theoretical predictions on the short time behaviour
of our model through stochastic simulations of Eq.~(\ref{eqkl_2}). 
We set $D_0=1$, $D^*=1$, $\mu=1/2$ and $b_{max}=\infty$. 
We use two different values of the transition rate, $\lambda=\{10,\,10^2\}$. 
%while the value of $\mu$ equals $1/2$, 
We run $10^7$ independent trajectories, each initially located at $x=0$.
In agreement with our theoretical considerations, we observe 
an excellent agreement between the simulation data and the
Laplace distribution Eq.~(\ref{eqkl_6}) for $\lambda=10^2$
(panel (a)) up to $t \leq 2=1/\mu$.
For longer times, the approach towards the long-term asymptotics
starts to appear, driven  by the recombination dynamics
associated with the transition mechanism  of the stochastic process $\beta(t)$.
For $\lambda=10$ (panel (b)), the early short-time behaviour, specifically
the data at $t=0.2$, shows a significant deviation from
Eq.~(\ref{eqkl_6}).
For this $\lambda$ and at this time scale, the recombination mechanism
of the velocity switching process $S(t)$ is not fast enough to
allow the PK dynamics to be accurately approximated by its
parabolic Kac limit.

\begin{figure}[!t]
\begin{center}
%  {\includegraphics[width=18cm,keepaspectratio]{dd_lam.pdf}}
  {\includegraphics[width=18cm,keepaspectratio]{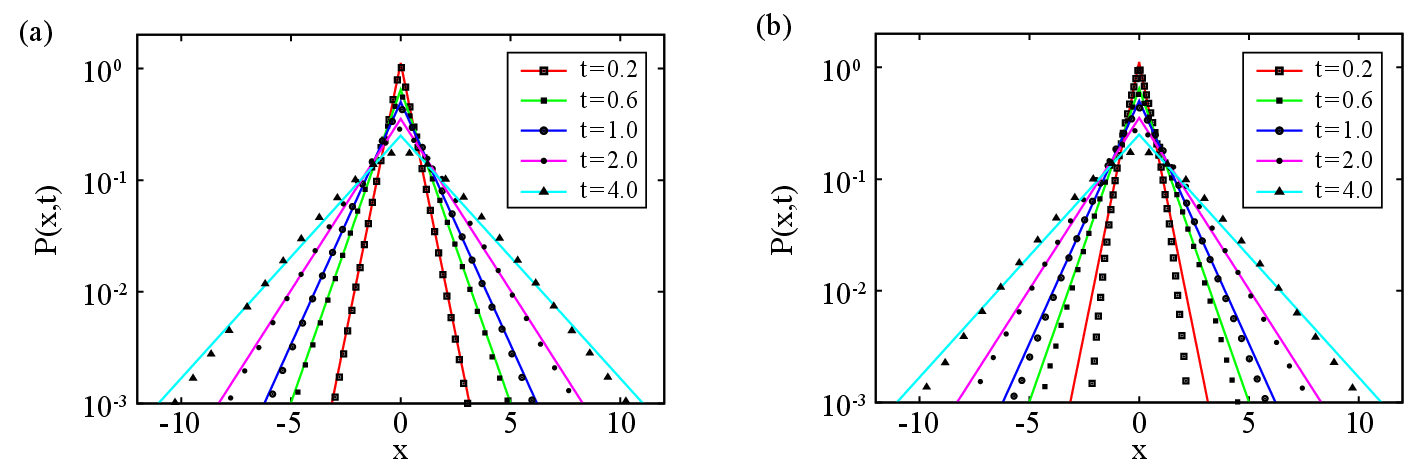}}
\end{center}
\caption{Short time diffusive properties of the EPK model
    Eq.~(\ref{eqkl_2}).  All model and simulation parameters are given
    in the text.  Shown is the position probability density function
    $P(x,t)$ for the transition rates $\lambda=\{10,10^2\}$ (panels
    (b) and (a) respectively).  Symbols correspond to the results of
    stochastic simulations at increasing values of time $t$ as given
    in the legends. Lines represent the Laplace distribution
    Eq.~(\ref{eqkl_6}).}
\label{Figkl_4}
\end{figure}

The asymptotic (long-term) behaviour  of Eq.~(\ref{eqkl_2}) corresponds to
the Kac limit of Eq.~(\ref{eqkl_3}). 
%for $\lambda \rightarrow \infty$,
%keeping fixed  the value $b_0^2/2 \lambda=D_0$.
In this limit, $p_\pm (x,t,\beta) \simeq P(x,t) \, P^{\,\star}_b(\beta)/2$
and, following identical calculations developed in \cite{GBC17,GBC17a,GBC17c},
%{\color{burntorange}Andrea: We should provide them in the Supplementary also.}
one recovers the parabolic diffusion equation $\partial_t P(x,t)
= D_{\rm eff} \, \partial_x^2 P(x,t)$, with an effective
diffusivity $D_{\rm eff}=D_0 \, \langle \beta^2 \rangle/(1- e^{-\kappa \, \beta_{\rm max}^2})$
where 
$\langle \beta^2 \rangle= \int_{\mathcal{D}_{\beta}} \beta^{\prime\,2} \, P^{\,\star}_b(\beta^{\prime}) \, \ud \beta^{\prime}$. Correspondingly, the
long-time asymptotics of Eq.~(\ref{eqkl_3})
is expressed by  the Gaussian heat kernel
\begin{equation}
P(x,t)= \frac{1}{\sqrt{4 \, \pi \, D_{\rm eff} \, t}} \, \exp \left
( - \frac{x^2}{4 \, D_{\rm eff} \, t} \right ) \:.
\label{eqkl_8}
\end{equation}
Figure~\ref{Figkl_5} confirms these predictions numerically.
%(panel a) depicts the long-term statistical
%properties  of Eq.~(\ref{eqkl_2})
The simulation protocol and model parameters used are the same as for
%for the same values of the parameters used for the simulations shown in 
Fig.~\ref{Figkl_4}. 
We only present the case 
%Simulations refer to 
$\lambda=10$, since the long-term asymptotic behaviour
%approaches that of the corresponding
%Kac-limit Eq.~(\ref{eqkl_8}), 
is the same for any value of $\lambda$.
In this case, $D_0=1$,
$\langle \beta^2 \rangle=1$, so that $D_{\rm eff}=1$.
In panel (a) the agreement between our prediction Eq.~(\ref{eqkl_8}) and the simulation data is excellent. In
panel (b) we show that the scaling of the mean square displacement $\sigma_x^2(t)$ is linear in time over all the time scales considered. 
We note that for finite $b_{max}$ a ballistic scaling for  the mean square displacement
$\sigma_x^2(t) \sim t^2$ for times $t \leq 1/\lambda$ 
is also observed due to the bounded propagation speed. 
This demonstrates that our model Eq.~(\ref{eqkl_2}) can successfully reproduce Brownian yet non-Gaussian diffusive behaviour.

\begin{figure}[!t]
\begin{center}
%  {\includegraphics[width=18cm,keepaspectratio]{dd_longt.pdf}}
  {\includegraphics[width=18cm,keepaspectratio]{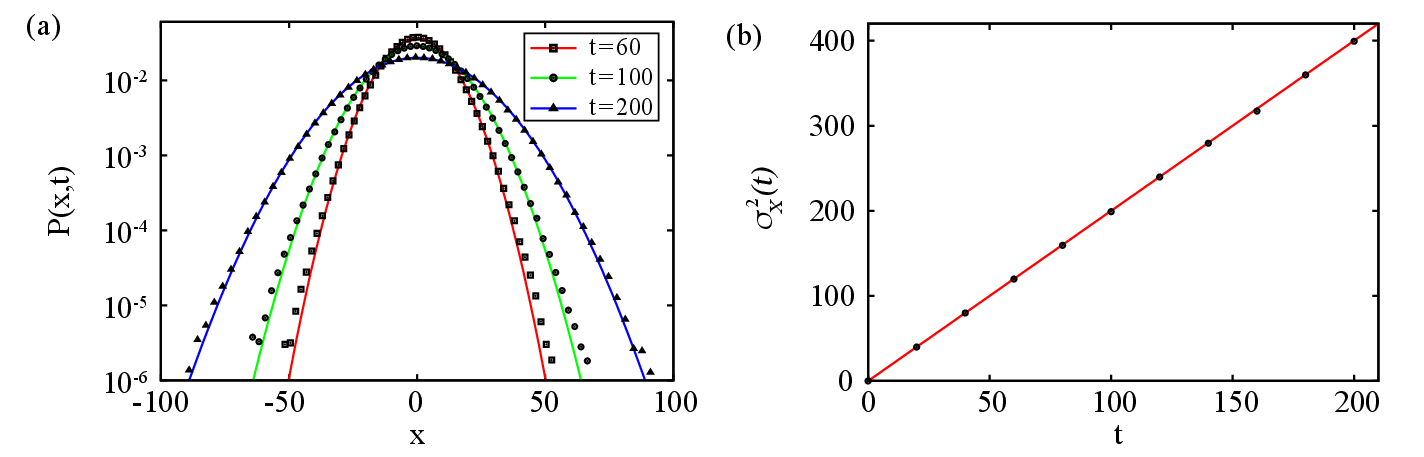}}
\end{center}
\caption{Long time diffusive properties of the EPK model
  Eq.~(\ref{eqkl_2}). Here we consider only the transition rate
  $\lambda=10$, all the other model and simulation parameters are the
  same as in Fig.~\ref{Figkl_4}, see also the text. Symbols correspond
  to the results of stochastic simulations.  (a) Position probability
  density function $P(x,t)$ at different times $t$ as given in the
  legend.  Lines represent the Gaussian long time limit
  Eq.~(\ref{eqkl_8}). (b) Mean square displacement $\sigma_x^2(t)$.
  The solid line corresponds to the Einsteinian scaling
  $\sigma_x^2(t)= 2 D_{\rm eff} \, t$, with $D_{\rm eff}=1$.}
\label{Figkl_5}
\end{figure}

\subsection{Subdiffusive L\'evy Walks}
\label{subsec:lam}

In their original formulation based on CTRWs, LWs have been shown to
capture ballistic, normal and superdiffusive behaviour, according to
the scaling properties of their transition time density distribution
\cite{ZDK15}.  Other diffusional features, such as particularly
subdiffusion, could only be achieved in a generalised version of LW
dynamics, where a power-law kinematic relation between the
displacement of the walker and the transition time was imposed \cite{AlRa18}.
%Namely, when we set (recall our discussion in Sec.~\ref{sec:fvp})
%$\phi(\Delta,\tau)=\delta(|\Delta|-b\tau^{\nu})T(\tau)/2$ (with $\nu
%\neq \xi$). A subdiffusive scaling of the mean square displacement for
%this generalised model has been shown to exist for a certain range of
%the parameters $\xi$, $\nu$ in Ref.~\cite{AlRa18}.  
We remark that
the occurrence of long-term subdiffusive scaling in stochastic
processes possessing finite propagation velocity has also already been
obtained for symmetric random walks on fractals \cite{Ben2000} or
generalised PK processes in pre-fractal media \cite{GBC16b}.
Motivated by these results, in this section we show that 
%within our theoretical framework 
we can formulate an EPK process
%with much more flexible diffusive dynamics 
that can capture short-term subdiffusion
solely as the result of microscopic correlations among its transition rates.

%We now discuss the Markovianisation of the transition rate $\lambda$. 
We consider an EPK process where the transition rate of the Poissonian switching process $S$ is described by the stochastic process $\Lambda(t)$ attaining values in the bounded interval $\mathcal{D}_{\lambda}=[0,\lambda_{\rm max}]$.
Specifically, we assume $\Lambda$ to generate a Markov chain dynamics,
with transition rate $\mu(\lambda)$ and probability transition
kernel $\mathcal{M}(\lambda,\lambda^\prime)$ (see Appendix~\ref{sec:pf}).
%The temporal evolution of the PDF of this process, $P(\lambda,t)$
%(here the averaging is over all independent realisations of the Markov chain dynamics), 
%is described by the following equation,
%\begin{equation}
%\frac{\partial P(\lambda,t)}{\partial t}= - \mu(\lambda) \,
%P(\lambda,t) + \int_0^{\lambda_{\rm max}} \lambda^\prime
%\mathcal{M}(\lambda,\lambda^\prime) P(\lambda',t)\, \ud
%\lambda^\prime \:.
%\label{eql_1}
%\end{equation}
The corresponding extended PK process can then be defined as  
\begin{equation}
\frac{d x(t)}{d t}= b \, (-1)^{\chi(t,\lambda(t))} \: .
\label{eql_0}
\end{equation}
%where $\lambda(t)$ denotes one specific realization of the process $\Lambda(t)$. 
%Similarly to the previous model discussed, this process superimposes a level of stochasticity manifest in the transition rate dynamic to the microscopic local stochasticity
%induced by the switching process $S(t)=(-1)^{\chi(t,\lambda(t))}$.
%The Markovianisation process  introduced in the previous two subsections implies that 
%Both $S(t)$ and $\Lambda(t)$ possess independent transition dynamics (which
%does not means in principle that they are independent random
%processes). 
The PPDW functions for this process, 
$p_s(x,t,\lambda)\equiv p(x,t,\lambda,s)$, 
are parametrised with respect to all the transitional parameters, here 
$s=\pm$ and $\lambda\in\mathcal{D}_{\lambda}$. 
The transitional parameters are transitionally independent as in the previous example. 
The temporal evolution equations can then be obtained similarly to Eq.~\eqref{eq4_5}, i.e.,
%This property enables us to derive straightforwardly
%the evolution equation for the ppdw 
%of the associated generalised PK process that are solutions of the
%These functions satisfy the balance equations
\begin{align}
\frac{\partial p_\pm(x,t,\lambda)}{\partial t} &= \mp
b \, \frac{\partial p_\pm(x,t,\lambda)}{\partial x} \mp \lambda \, [
p_+(x,t,\lambda)-p_-(x,t,\lambda)] - \mu(\lambda) \, p_\pm(x,t,\lambda)
\nonumber \\
& \quad + 
\int_0^{\lambda_{\rm max}} \mu(\lambda^\prime) \, 
\mathcal{M}(\lambda,\lambda^\prime)
\, p_\pm(x,t,\lambda^\prime) \, \ud \lambda^\prime \:.
\label{eql_3}
\end{align}
Solutions of these equations are uniquely determined by the initial condition $p_\pm(x,\lambda,0)=p_\pm^0(x,\lambda)$. 
%This is different from OPK processes, Eqs.~(\ref{eqo_5}), where an additional boundary condition was required to handle the discontinuity in the transitional age dynamic.   

First, we demonstrate that the process Eq.~(\ref{eql_0}) generates a
dynamic that shares the long-time statistical characteristics of
the conventional LW.  Let us assume
%that $A_\lambda(\lambda,\lambda^\prime)$
% does not depend on the initial state by solely on the final one,
$\mathcal{M}(\lambda,\lambda^\prime)= \pi^*(\lambda)$, 
where $\pi^*(\lambda)$ is the equilibrium  density function of the transition rate process.
%For this case, Eq.~(\ref{eql_4}) simplifies as
%\begin{equation}
%\frac{\partial p_\pm(x\lambda,t)}{\partial t} =  \mp b \, \frac{\partial p_\pm(x,\lambda,t)}{\partial x} -\lambda \, p_\pm(x,\lambda,t)
% +  \pi_\lambda(\lambda) \, \int \lambda^\prime \, p_\mp(x,\lambda^\prime,t)
%\, d \lambda^\prime
%\label{eql_8}
%\end{equation}
%Since $\pi_\lambda(\lambda)$ is the density
%function for the transition rate and, 
%the conditional transition time density  
Under these assumptions, the transition time density for this process is given by  
\begin{equation}
T(\tau) 
%= \int_{\mathcal{D}_{\lambda}} T(\tau \, | \, \lambda) \, \pi_\lambda(\lambda)
%\, d \lambda 
= \int_{\mathcal{D}_{\lambda}} \lambda \, e^{-\lambda \, \tau} \, \pi^*(\lambda) \, \ud \lambda \: .
\label{eql_9}
\end{equation}
%where $T(\tau \, | \, \lambda)= \lambda \, e^{-\lambda \, \tau}$,
%it follows that the effective transition time density $T_{\rm eff}(\tau)$
%associated with the ogPK model Eq.~(\ref{eql_8}) is given by
This equation follows by recalling that, once we fix $\lambda$, the time $\tau$ elapsed  before the next transition is a random variable sampled from an exponential distribution with mean $\lambda$. 
We now specify the equilibrium density as    
%By simply considering a density $\pi_\lambda(\lambda)$ producing,
%via Eq.~(\ref{eql_9}),
%the effective transition time density $T_{\rm eff}(\tau)$ of a
%LW, an alternative way to generate this class of processes out of gPK
%theory is provided.
%Since the values of $\lambda$ for classical LW are upper-bounded,
%without loss of generality, 
%let us suppose that $\lambda_{\rm max}=1$,
%so that $\pi_\lambda(\lambda)$ is defined in the unit interval.
%The well-known diffusive anomalies  characterising LW (superdiffusive and ballistic regime)
%can be recovered by assuming for $\pi_\lambda(\lambda)$ the
%expression
\begin{equation}
\pi^*(\lambda) = (1+ \alpha) \, \lambda^\alpha
%\qquad \lambda \in [0,1]
\label{eql_10}
\end{equation}
with $\alpha > -1$ and $\lambda_{max}=1$. In this case, 
%(* we should include some  material here as supplementary material *)
$T(\tau) \sim \tau^{-(\alpha+2)}$ for large $\tau$. 
Therefore, this process reproduces qualitatively all the characteristic long-term diffusive features of the conventional LW as defined in Section~\ref{sec:fvp}, 
%characterised by a transition-time
%density $T(\tau) \sim \tau^{-(\xi+1)}$ are recovered 
provided we set $\xi=\alpha+1$. 
In particular, we can show that 
%Indicating with $\sigma_x^2(t)$ the
%mean square displacement at time $t$, it follows that:
(i) for $-1 < \alpha <0$ the process is ballistic,
i.e., $\sigma_x^2(t) \sim t^2$; (ii) for $0 <\alpha < 1$ the process
is superdiffusive,
%\begin{equation}
$\sigma_x^2(t) \sim t^{2-\alpha}$; and 
%\label{eql_11}
%\end{equation}
(iii) for $\alpha >1$, the process exhibits a linear Einsteinian scaling,
$\sigma_x^2(t) \sim t$.
Furthermore, we can show that (iv) the invariant function 
$p^*(z) =  \sigma_x(t) \, P(x,t) |_{x= z \, \sigma_x(t)}$,
with the probability density function for the process 
$P(x,t)= \int_0^1 \left [p_+(x,t,\lambda)+p_-(x,t,\lambda) \right ] \, \ud \lambda$,
is the same as that of the conventional  LW.  
We verified all these results in numerical simulations; see Figs.~\ref{Figlw3}(a) and (b) for some representative examples.
We note that the different transition time densities for the process Eq.~(\ref{eql_0}) and the LW Eq.~(\ref{eq:apk2}) only affect their short-time statistical properties. 

Remarkably, the formulation Eq.~(\ref{eql_0}) of LW dynamics enables 
the explicit modelling of highly non-trivial transitional correlations 
through the full transition kernel $\mathcal{M}(\lambda,\lambda^\prime)$.  
%The mechanism of subdiffusive motion in a LW should be therefore
%sought in the lack of convergence of the transitional rate distribution
In particular, we assume that the transition rate process $\Lambda$ %Eq.~(\ref{eql_1})
does not possess an invariant density. This is ensured by the condition 
\begin{equation}
\int_{\mathcal{D}_{\lambda}} \lambda \, \mathcal{M}(\lambda,\lambda^\prime) \, \ud \lambda > \lambda^\prime \: . 
\label{eql_12}
\end{equation}
In physical terms, this condition induces a progressive shift over time 
%induced by $A_\lambda(\lambda,\lambda^\prime)$ 
towards higher and higher values of the transition rate.  
%>\lambda^\prime$, occurring if
%the condition
%is fulfilled.
%To illustrate the phenomenon, consider an example.
As a specific example, we assume 
%that the particle ensemble possesses
%at $t=0$ the same value of $\lambda$, say $\lambda=1$.
%and define $A_\lambda(\lambda,\lambda^\prime)$ as
$\mathcal{M}(\lambda,\lambda^\prime)= 
(\lambda^\prime)^\nu/(a_2+a_1)$ for
 $\lambda \in [\lambda^\prime-a_1/(\lambda^\prime)^\nu, \lambda^\prime+a_2/(\lambda^\prime)^\nu]$
and zero otherwise,
where $a_1,\, a_2>0$ ($a_2 > a_1$) and $\nu>0$ are constant. 
The shift is clearly determined by the fact that $\lambda$ at each transition is sampled uniformly from an asymmetric interval 
%Despite the apparent complexity of this representation the idea
%behind it is  very simple.
%Given the actual value $\lambda^\prime$, the value of $\lambda$
%after the transition is uniformly chosen in the interval
$[\lambda^\prime  -c(\lambda^\prime) a_1,\lambda^\prime+c(\lambda^\prime) a_2]$, where $c(\lambda^\prime)=(\lambda^\prime)^{-\nu}$ decreases progressively. 
This function is introduced to slow down the shift that would, otherwise, rapidly stop the motion.
%depends on $\lambda^\prime$.
%Since $a_2>a_1$, say $a_1=0.4$ and $a_2=0.45$,
%given the actual value $\lambda^\prime$, the value of $\lambda$
%after the transition  is in average higher that $\lambda^\prime$,
%fulfilling the requirement Eq.~(\ref{eq11}).
This shift towards higher values of $\lambda$ determines a progressive decrease of the local diffusivity, leading potentially to subdiffusive behaviour.
%For this reason,  we set $c=1/(\lambda^\prime)^\nu$,
%with $\nu=5$.
This is verified in Fig.~\ref{Figlw3}(c), where we plot $\sigma_x^2$ for this
process obtained from numerical simulations. We run $10^5$ independent trajectories. Starting from a ballistic scaling for short times, as typical of all processes possessing finite propagation velocity, the mean square displacement for $t >10^2$ exhibits an anomalous long-time scaling with subdiffusive exponent $\gamma=0.8$.
We note that this scaling is observed over more than four decades, $t \in [10^2,10^6]$.
%{\color{burntorange}Andrea: I would show here also the pdf of the process. Except for the ballistic peaks, what kind of distribution do we see? Compare with L\'evy. And plot as second panel in Fig. 5.}

\begin{figure}
  \includegraphics[width=18cm]{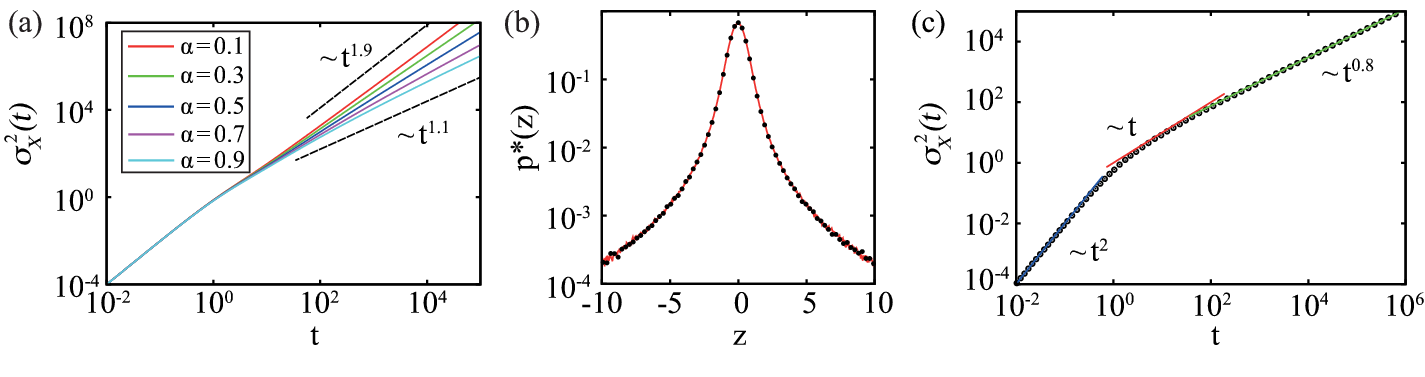}
\caption{Statistical characterisation of the EPK model Eq.~(\ref{eql_3}) for equilibrated transition rate dynamics
%, i.e., $\mathcal{M}(\lambda,\lambda^\prime)=\pi^*(\lambda)$ with $\pi^*(\lambda)$ as in Eq.~(\ref{eql_10}),
(panels (a) and (b)) and for the general non-equilibrated case (panel
  (c)) (see main text for details).  (a) Mean square displacement
  $\sigma_x^2(t)$ for $\alpha=\{0.1,\,0.3,\,0.5,\,0.7,\,0.9\}$, where
  $\alpha$ is the exponent of the power law determining the
  equilibrium density of the transition rates, see Eq.~(\ref{eql_10}).
%vs $t$ for the GPK-formulated LW
%model defined by Eq.~(\ref{eql_3a})
Lines are generated by numerical simulations of the stochastic dynamic.
%Panel (a): the arrow indicates increasing values of $\alpha=0.1,\,0.3,\,0.5,\,0.7,\,0.9$. %Lines (a) and (b) represent  the
Dashed lines indicate the scaling predictions 
% $\sigma_x^2(t) \sim t^{1.9}$ and $\sigma_x^2(t) \sim t^{1.1}$,  respectively,
%associated with
for the extremal values of $\alpha$ in the range considered.  (b)
Invariant long time density $p^*(z)$ ($z=x/\sigma_x(t)$) at
$\alpha=0.5$ (symbols), compared with the distribution of its
corresponding LW (solid lines).
%with $\xi=1.5$.
(c) Mean square displacement $\sigma_x^2(t)$ for the same EPK model in
non-equilibrated conditions. Three different diffusive regimes are
indicated by straight lines.
%described in the main
%text. 
Markers represent the results of numerical simulations.
Solid lines indicate the different scaling regimes exhibited by this dynamic.
%scalings $\sigma_x^2(t) \sim t^2$, $\sigma_x^2(t) \sim t$,
%and $\sigma_x^2(t) \sim t^\gamma$,
%with $\gamma=0.8$.
}
\label{Figlw3}
\end{figure}

\section{Conclusions and perspectives}
\label{sec:concl}

Stochastic processes form a cornerstone of our mathematical
description of physical reality. They enable the modelling of a wide
variety of transport phenomena in the natural and social sciences,
such as the random movements of cells, bacteria and viruses, the
fluctuations of climate and the volatility of financial markets
\cite{BCKV16,KRS08,MeKl04}. Typical stochastic models considered,
however, fail to ensure finite velocities thus violating Einstein's
theory of special relativity.  While these models still capture the
correct statistics of motion on sufficiently long time scales, their
representation of the real world is thus intrinsically
defective. Partially they also lead to mathematical problems like
diverging moments for the probability distributions of a random
walker, with direct implications for physical observables obtained
from such models. To solve this deep conceptual problem, stochastic
processes with finite propagation speed have been
introduced. Paradigmatic examples are PK processes
\cite{Tay21,Gold51,Kac74} and LWs
\cite{SKW82,GNZ85,ShlKl85,KBS87,Shles87,GZR88} yielding normal and
anomalous diffusion, respectively.  Despite their joint feature of
finite propagation speeds, however, these two fundamental classes of
stochastic processes have so far coexisted without exploring any
cross-links between them.

Inspired by the novel formulation of LW dynamics proposed by Fedotov
and collaborators \cite{FTZ15,Fed16}, in this article we explored the
connection between LWs and PK processes by showing that the latter
models can be understood as a particular case of the former
ones. Clarifying the relation between these two dynamics, by including
Wiener processes as a special case, yielded our first main
result. This is represented in Fig.~\ref{fig:stokin} by the first
three inner circles. In turn, this observation suggested the most
natural stochastic differential equations describing LW path dynamics,
Eqs.~(\ref{eq:apk1}) and (\ref{eq:apk2}), which are obtained from suitably
generalising the formalism of PK processes. This formulation neatly
results from the definition of a LW process and, in this sense,
greatly differs from other phenomenological models published in the
literature that rely on subordination techniques
\cite{Eule2012,Wang2019} or fractional derivatives
\cite{Magdziarz2012}, 
as the statistical characterization involves
first-order evolution equations in time and space, 
whose mathematical structure resembles the linear Boltzmann equation
\cite{cercignani}. Owing  to this analogy and to the analogy
between the evolution equations for the partial densities
and the mathematics of radiative transfer \cite{chandrasekhar}, 
the {mathematical approaches} 
developed in these two fields can be consistently transferred to
the study of EPK processes \cite{cercignani,chandrasekhar}.
  With a reverse-engineering approach,
we then used the cross-link between these processes to formulate a
very general theoretical framework for stochastic models with finite
propagation speed, which we called EPK theory. This theory contains
LWs as a special case, as is depicted again in Fig.~\ref{fig:stokin}
by the fourth most outer circle. This is our second main result.

Motivated by experimental applications we then demonstrated by three
explicit, practical examples the potential and the modelling power
offered by our novel theory. 
We showed that EPK processes can capture
senescing phenomena, where the mechanism for velocity changes depends
explicitly on on the number of transitions occurred.
From EPK theory we also obtained a microscopic
interpretation of the intriguing and very actively explored
{transport phenomenon associated
with  Brownian  yet non-Gaussian diffusion
%diffusive phenomenon of Brownian yet non-Gaussian diffusion
\cite{Wang:2012aa,Chechkin:2017aa}.}  Finally, we demonstrated that LWs
may not only be superdiffusive but also subdiffusive, depending on
more subtle microscopic details of the LW dynamic as captured by EPK
theory.
These novel diffusional features (anomalies) are ultimately 
obtained by exploiting the internal coupling between state variables 
and transitional parameters characteristic of EPK processes.

In this paper, we outlined the general framework of EPK theory 
%The examples discussed in this article we focused on pure random motion 
in unbounded domains and in the absence of biasing fields
(potential and/or pressure-driven velocity fields). 
%In point of fact, EPK processes and their statistical characterization can be transferred
The extension of our theory to transport problems in 
bounded settings can be achieved straightforwardly 
by applying the boundary conditions
(absorbing, reflecting, or of mixed nature) already developed for hyperbolic
transport problems involving PK processes and LWs \cite{brasiello,gels,Giona2019age}. 
%Similarly,
The effects of external biasing fields can be included in two ways: 
The first one is to consider the EPK counterparts of the classical Ornstein-Uhlenbeck model,
as discussed in \cite{GBC17c} for simple PK processes. 
The second one is %Alternatively,
to consider the effect of the external potential on the transitional properties of the EPK model, 
%a kinematic EPK process (here kinematic means that only particle
%position is considered as a state variable, as throughout this article)
by allowing for a dependence of the transition rates on the positional state variables.
%Within this framework 
A number of important open problems and
consequences still remain to be addressed, %which 
especially concerning the mathematical features of EPK models 
and their experimental applications. 
%We therefore conclude by providing an
%outlook to what we consider to be the most important ones.  

From the
perspective of stochastic theory, the first intriguing open problem is
the formulation of a spectral theory of finite propagation speed
processes (most notably LWs among them).  We have preliminary evidence
that the constraint imposed by the finite velocity may manifest itself
as a lower bound in the real part of the spectrum.  If confirmed, it
would be interesting to investigate whether this is a universal
property of all stochastic processes possessing finite propagation
velocity.  In this direction, a second top-priority problem is to
develop a homogenisation theory for EPK processes.  This is
fundamental, as it could enable to calculate the hierarchy of moments
and correlation functions associated with these stochastic processes.
These are essential quantities to explore theoretically the ergodic
and aging properties of a given class of stochastic processes.  In
addition, these results are also relevant from the perspective of
experimental applications.  Indeed, both moments and correlations are
observables that can often be measured with great accuracy in
experiments. Therefore, by comparing experimental data with model
predictions of these observables, they could be employed to discern
the most suitable finite propagation speed stochastic model for a
given empirical diffusion dynamics.  To complement this argument, our
considerations further suggest that experimentalists should make an
effort to measure the probability distribution of diffusive
observables (such as position and/or velocity) with sufficiently large
statistics to assess reliably the tails of these distributions.  While
stochastic processes with or without finite propagation speed might
exhibit the same long-term asymptotic scaling of moments and/or
correlations, the presence of ballistic peaks in the tails of these
distributions is a hallmark that the diffusion observed empirically
possesses finite velocities.  Finally, while in this work we have
focused on the single-particle case, another relevant research
direction is the study of the emerging collective behaviour in
ensembles of EPK particles potentially interacting among
themselves. Along these lines, and by introducing persistence in terms
of correlations, one may crosslink finite velocity stochastic theory
with the very recent field of active matter
\cite{Toner2005,Schweitzer2007,Ramaswamy2010,Vicsek2012,Marchetti2013}.
Given the additional biologically-motivated features that can be
included in EPK theory, it will be interesting to systematically
investigate what EPK models can further reproduce in terms of
biologically significant dynamics.

We conclude by highlighting that EPK processes are not only
of fundamental importance as models of transport phenomena, 
but also that they can be applied to basic and fundamental problems 
of statistical physics, providing a novel view of the latter. 
%The fundamental  signature of 
%EPK processes is not only the bounded
%propagation velocity but also the boundedness of the transition
%rates. The latter property permits to describe the dynamics
The boundedness of the transition rates characteristic of EPK processes
allows to describe the dynamics
of processes controlled by elementary events occurring
at random transition times. This is the case of the quantum
phenomenology associated with the interaction between molecules
and photons. In this case, the elementary processes involve emission
and absorption of energy quanta \cite{einstein1917}. In this
class  falls also the stochastic representation
of quantum systems interacting with an external environment
\cite{petruccione}. At sufficiently high (ambient) temperatures 
these processes can be described by a Poissonian statistics
of the transition times \cite{einstein1917,petruccione}.
In the limit of very low temperatures, reachable e.g. by means of
laser cooling techniques, the transitional statistics becomes
more complex and correlated \cite{lasercooling} providing
a natural application for EPK processes.

Another important field of application is the hydrodynamics 
of colloidal systems whenever particle/fluid interactions are accounted for
in detail. Physically, this implies the transition from
the Einstein-Langevin picture of Brownian motion (Stokes regime), 
%%where the fluid responds instantaneously to particle motion (Stokes regime), thus
%%giving rise solely to the Stokesian friction, 
to the realm where fluid inertia is no longer negligible %considered,
(time-dependent Stokes regime) 
%%, determining the occurrence of
%%the Basset force and of the dynamic added-mass effect 
\cite{stokes1851,landau}.
These effects, originally discovered by Stokes in 1851 
%in his study of the  pendulum motion  in a liquid 
\cite{stokes1851}, determine  the  occurrence of long-term power-law tails
in the velocity autocorrelation functions,
recently observed experimentally \cite{bm1,bm2,bm3} for Brownian particles in a liquid. 
%The development of 
Preliminary results on a velocity-based
representation of fluctuation-dissipation theorem %addressed in \cite{gionafd}
indicates that EPK processes possessing a continuous velocity distribution,
can be amongst the simplest candidate for representing the thermal/hydrodynamic
fluctuations controlling Brownian and colloidal particle motion
in inertial fluids both in the free space and in confined geometries.

{\bf Acknowledgements:} RK thanks the TU Berlin and particularly his
host, Prof.~S.H.L.Klapp, for funding in the form of a Mercator
Visiting Professorship, financed by the Deutsche
Forschungsgemeinschaft (DFG, German Research Foundation) —
Projektnummer~163436311 — SFB~910.  He also acknowledges an External
Fellowship from the London Mathematical Laboratory.

\appendix 

\section{Continuous-time random walks and their relation to Lévy walks}\label{sec:ctrw}

A one-dimensional continuous-time random walk is a stochastic process 
that describes the dynamic of a random walker on the real line that can   
wait for a random time $\tau$ at its current position 
before jumping over a random distance $\Delta$ either
to the left or to the right. 
The pair of jump length and waiting time, $(\Delta,\tau)$, is sampled from a prescribed joint PDF
$\phi(\Delta,\tau)$.  The jump length distribution is
$w(\Delta)=\int_0^{+\infty} \phi(\Delta,\tau) \ud\tau$, where
$\Delta\in\mathbb{R}$; likewise, the waiting time distribution is
$T(\tau)=\int_{-\infty}^{+\infty} \phi(\Delta,\tau) \ud\Delta$.  Hence,
a CTRW dynamic can be more naturally defined with respect to an
operational time $n \in {\mathbb N}$, which simply counts the
jump events occurred.  The particle position $x_n$ at the physical
time $t_n$ is then given by the equations of motion
\begin{align}
x_{n+1}&=x_n+\Delta_n \: , & t_{n+1}=t_n+\tau_n \: , 
\label{eq:ctrw}
\end{align}
with the pairs $(\Delta_h,\tau_h)$ and $(\Delta_k,\tau_k)$ being
independent of each other for $h \neq k$.  Traditionally, jump lengths
and waiting times are further assumed to be independent, meaning that
we can factorise the joint distribution as
$\phi(\Delta,\tau)=w(\Delta) \, T(\tau)$.  Under this assumption, the
CTRW dynamic resembles that of the Wiener process Eq.~\eqref{eq:wiener} 
(which can be recovered in fact by assuming $w(\Delta)$ and $T(\tau)$ to be Gaussians \cite{MeKl00}), 
in the sense that the walker can perform
potentially unbounded random displacements depending on the prescribed
jump length distribution $w(\Delta)$. 
%waiting times and jump lengths \cite{MeKl00}, 
This is manifest particularly
in the L\'evy flight model \cite{Man82,Hughes1981,Shlesinger1995},
where the probabilities of sampling very large displacements are
enhanced by assuming a power law tailed jump length distribution.

In order to model a LW, we impose on Eqs.~\eqref{eq:ctrw} the
constraint of a finite and constant propagation speed $b$
\cite{SKW82,GNZ85,ShlKl85,KBS87,Shles87,GZR88,ZuKl93a}. This implies
introducing a relation between $\Delta_n$ and $\tau_n$ in the form of
\begin{equation}
\Delta_n= s_n\, b \, \tau_n, 
\end{equation}
where $s_n$ are random variables attaining values
$\pm 1$ with equal probability.  We highlight that this is not just a
technicality, but it implies a complete change of perspective. Within
this different interpretation, the waiting time $\tau$ in the original
CTRW formulation becomes in the LW the transition time for a velocity
change in one direction.  In the special case of a two-state model the
direction of motion is inverted at each transition event
\cite{ZuKl93a}, $\Delta_n= b\,s_0 \, (-1)^n \, \tau_n$, where $s_0=\pm
1$ with equal probability determines the initial direction of
motion.

\section{Poisson fields: definition and evolution equation}\label{sec:pf}

A $d$-dimensional Poisson field is a continuous stochastic
process attaining values in a set, ${\mathcal D}_\alpha$, 
whose statistical description satisfies a continuous Markov chain dynamic defined by
the transition rate function $\lambda(\boldsymbol{\alpha}) \geq 0$,
and by the transition probability kernel
$A(\boldsymbol{\alpha},\boldsymbol{\alpha}^{\prime})$.  
The symbols $\boldsymbol{\alpha},\boldsymbol{\alpha}^{\prime}$ denote $d$-dimensional vectors in ${\mathcal D}_\alpha$. 
Let us now
assume ${\mathcal D}_\alpha$ to be continuous.  Thus, the probability
$\mbox{Prob}[\{\boldsymbol{\Xi}(t) \in
  (\boldsymbol{\alpha},\boldsymbol{\alpha}+ \ud \boldsymbol{\alpha})
  \}]= \widehat{P}(\boldsymbol{\alpha},t) \, \ud \boldsymbol{\alpha}$
($\ud \boldsymbol{\alpha}$ denotes a $d$-dimensional infinitesimal
volume in $\mathcal{D}_\alpha$) is determined by the density function
$\widehat{P}(\boldsymbol{\alpha},t)$ that satisfies the linear
evolution equation
\begin{equation}
\frac{\partial \widehat{P}(\boldsymbol{\alpha},t)}{\partial t}=
-\lambda(\boldsymbol{\alpha}) \, \widehat{P}(\boldsymbol{\alpha},t)
+ \int_{{\mathcal D}_\alpha} \lambda(\boldsymbol{\alpha}^{\prime}) \, A(\boldsymbol{\alpha},\boldsymbol{\alpha}^{\prime}) \, \widehat{P}(\boldsymbol{\alpha}^{\prime},t) \, \ud \boldsymbol{\alpha}^{\prime}
\label{eq4_3}\:.
\end{equation}
The transition
probability kernel $A(\boldsymbol{\alpha},\boldsymbol{\alpha}^{\prime})$ is a
left stochastic kernel, i.e.,
$A(\boldsymbol{\alpha},\boldsymbol{\alpha}^{\prime}) \geq 0$, $\int_{{\mathcal
    D}_\alpha} A(\boldsymbol{\alpha},\boldsymbol{\alpha}^{\prime}) \, \ud
\boldsymbol{\alpha}=1$ for all $\boldsymbol{\alpha}^{\prime} \in {\mathcal
  D}_\alpha$.
  If ${\mathcal D}_\alpha$ is discrete instead, the statistical description of the Poisson field $\boldsymbol{\Xi}(t)$ is obtained directly in terms of the probability 
$\mathcal{P}(\boldsymbol{\alpha},t)=\mbox{Prob}[\{\boldsymbol{\Xi}(t) =
  \boldsymbol{\alpha} \in {\mathcal D}_\alpha \}]$, which satisfies an equation similar to Eq.~(\ref{eq4_3}) with the integral term substituted by the summation 
 $\sum_{\boldsymbol{\alpha^{\prime}}\in {\mathcal D}_\alpha} \lambda(\boldsymbol{\alpha}^{\prime}) \, A(\boldsymbol{\alpha},\boldsymbol{\alpha}^{\prime}) \, \mathcal{P}(\boldsymbol{\alpha}^{\prime},t)$.
 Naturally, the condition on the transition kernel becomes 
$\sum_{\boldsymbol{\alpha}\in {\mathcal D}_\alpha}  A(\boldsymbol{\alpha},\boldsymbol{\alpha}^{\prime})=1$ for every $\boldsymbol{\alpha}^\prime \in {\mathcal D}_\alpha$.
We remark that any continuous Markov chain transition dynamic satisfies an equation similar to Eq.~\eqref{eq4_3} with given rate and kernel functions.   

%
%Its position PDF, $P(x,t)\equiv\prec\delta(x-X(t))\succ$, where $\prec
%\cdot \succ$ denotes averaging over independent realisations of the
%Poisson process $\chi$, obeys the Cattaneo equation \cite{Kac74}
%\be
%\frac{1}{2 \, \lambda} \frac{\partial^2 P(x,t)}{\partial t^2}
%+ \frac{\partial P(x,t)}{\partial t} = D \frac{\partial^2 P(x,t)}
%{\partial x^2}\:.
%\label{eq:catt}
%\ee
%
%Interestingly, only later
%a closed temporal evolution equation for the position statistics of a
%LW, analogous to the Cattaneo equation~\eqref{eq:catt} for PK
%processes and to the diffusion equation for Wiener processes, has been
%derived (see details below) by Fedotov \cite{Fed16},
%\begin{multline}
%\dersecpar{}{t} P(x,t) + 
%\frac{1}{2} \left[ \derpar{}{t} - b \derpar{}{x} \right] \int_0^t K(t^{\prime}) P(x- b\, t^{\prime},t-t^{\prime}) \diff{t^{\prime}} + 
%\frac{1}{2} \left[ \derpar{}{t} + b \derpar{}{x} \right]  \int_0^t K(t^{\prime}) P(x+b\, t^{\prime},t-t^{\prime}) \diff{t^{\prime}} = 
%b^2 \dersecpar{}{x} P(x,t) \: ,  
%\label{LWFPE}
%\end{multline}
%where the memory kernel is defined in temporal Laplace transform as 
%$K(s)=s\, T(s)/(1-T(s))$. 
%The integral operators appearing in this equation, also known as ``fractional substantial derivatives" \cite{sokolov2003,Friedrich2006,cairoli2015}, are another evident manifestation of the spatio-temporal coupling characteristic of the LW dynamic. 

\section{L\'evy walks as specific overlapping Poisson-Kac
    processes}\label{subsec:LWrecovery}

For a one-dimensional LW (see Subsec.~\ref{subsec:ppdw}) our
general framework boils down to $n=m=1$, ${\bf x}=x$,
$\boldsymbol{\tau}=\tau$, and $\boldsymbol{\alpha}= \alpha$, which
attains only the discrete values ${\mathcal
  D}_\alpha=\{\pm\}$. Moreover, ${\bf v}(x,\tau)=0$,
$\lambda=\lambda(\tau)$, ${\bf w}(x,\tau)=1$ and ${\bf b}(x,\tau,\alpha)=
\text{b}(\alpha)$, with $\text{b}(\pm)=\pm b$. %, $b(-1)=-b$.  
Given the atomic nature of the transitional parameter $S(t)$, we can define
%=(-1)^{\chi(t,\lambda(\tau))}$ belonging to $\Sigma_T/\Sigma_O$, one has
\begin{eqnarray}
p(x,\tau,t,\alpha) & = & \sum_{s=\pm} p_s(x,\tau,t) \, \delta(\alpha-s) \:, \\
%\nonumber \\
%\text{b}(\alpha)& = & \alpha \, b  \:,\nonumber \\
A(x,\tau,\alpha,\tau^\prime,\alpha^\prime) & =  & \sum_{s,q=\pm} 
\widehat{A}_{sq}(\tau,\tau^\prime) \, \delta(\alpha-s) \,
\delta(\alpha^\prime-q) \:,
\label{eqo_7}
\end{eqnarray}
in which the $2 \times 2$ transition kernel $A$ does not depend on the state variable $x$.
In order to recover  a LW it is sufficient to consider
the factorisation
\begin{equation}
\widehat{A}_{sq}(\tau,\tau') = A_{sq} \, k(\tau,\tau')\:,
\label{eq4_12}
\end{equation}
where ${\bf A}=(A_{sq})_{s,q=\pm}$ is a left stochastic matrix and
%$k(\tau,\tau')\geq 0$ is the kernel introduced in Subsec.~\ref{subsec:ppdw}. 
%As already observed, 
%$k(\tau,\tau')$ in the specific case of LWs 
%does not depend on the age $\tau^\prime$ at the transition, but 
%it is chosen such that the transitional age is reset to zero. Therefore,  
%solely on $\tau$,
%\begin{equation}
$k(\tau,\tau^\prime)=\delta(\tau)$ (see Subsec.~\ref{subsec:ppdw}). 
%\label{eq4_12a}
%\end{equation}
%Furthermore, we consider 
For the matrix ${\bf A}$, we consider the simplest form
defining a two-state process, i.e., 
%$A_{sq}=1-\delta_{sq}$
%with the Kronecker delta function $\delta_{sq}$.
\begin{equation} 
{\bf A}= \left (
\begin{array}{cc}
0 & 1 \\
1 & 0 
\end{array}
\right )\:.
\label{eq4_14}
\end{equation}
Implementing all these simplifications in Eq.~(\ref{eqo_5}), 
we obtain the equations for the PPDW functions $p_s(x,\tau,t)$,  
%, $h=\pm$ become
\be
\frac{\partial p_s(x,\tau,t)}{\partial t} =  - s \, b \, \frac{\partial p_s(x,\tau,t)}{\partial x} - \frac{\partial p_s(x,\tau,t)}{\partial \tau}
- \lambda(\tau) \, p_s(x,\tau,t)
+  \delta(\tau) \, \int_0^\infty \lambda(\tau^\prime) \, p_{\pi(s)}(x,\tau^\prime,t) \, \ud \tau^\prime \:,
\label{eq4_15a}
\ee
where $\pi(s)$ represents the index %, $h=\pm$
flipping operation $\pi(\pm)=\mp$, 
which accounts for the structure of the Markovian transition matrix ${\bf A}$. 
%defined in Eq.~(\ref{eq4_14}).
For any $\tau>0$, the term $\propto \delta(\tau)$ is null, such that 
Eq.~(\ref{eq4_15a}) reduces to the Eq.~(\ref{eq:ppdw}) previously derived for LW processes (see Subsec.~\ref{subsec:ppdw}). 
%\begin{eqnarray}
%\frac{\partial p_+(x,\tau,t)}{\partial t}= -b \,  \frac{\partial p_+(x,\tau,t)}
%{\partial x} -  \frac{\partial p_+(x,\tau,t)}{\partial \tau} -\lambda(\tau) \, p_+(x,\tau,t) \nonumber \\
%\frac{\partial p_-(x,\tau,t)}{\partial t}= b \, \frac{\partial p_-(x,\tau,t)}
%{\partial x}  - \frac{\partial p_-(x,\tau,t)}{\partial \tau} - \lambda(\tau) \, p_-(x,\tau,t)
%\label{eq4_20}
%\end{eqnarray}
%
The impulsive forcing term $\propto \delta(\tau)$ 
%entering the last term of 
in Eq.~(\ref{eq4_15a}) only affects the behaviour of
$p_s(x,\tau,t)$ near $\tau=0$. The resolution
of this impulsive discontinuity  can be achieved by integrating
Eq.~(\ref{eq4_15a}) in the transitional age $\tau$ over an interval $[-\varepsilon,\varepsilon]$, 
with arbitrary $\varepsilon \rightarrow 0$.  
This interval contains $\tau=0$ as its internal point. 
%, letting $\varepsilon \rightarrow 0$. 
In doing this, we are implicitly extending the domain of   
%the only conceptual extension is to regards 
the PPDW functions $p_s(x,\tau,t)$ to negative $\tau$.  
% transitional ages 
%defined for $\tau \in [-\varepsilon,\infty)$, i.e. also
%for negative $\tau$, letting $\varepsilon \rightarrow \infty$.
%
Performing the integration, %in $\tau$, %over $[-\varepsilon,\varepsilon]$ 
the  $O(1)$-contributions stem solely from the convective term along the age-abscissa, 
$\partial p_s/\partial \tau$, 
and from the last impulsive recombination term $\propto \delta(\tau)$, 
while all other remaining terms are $O(\varepsilon)$. 
Thus, they are negligible in the limit $\epsilon\to0$. 
In this way, Eq.~(\ref{eq4_15a}) yields
\begin{equation}
-\int_{-\varepsilon}^\varepsilon \frac{\partial p_s(x,\tau,t)}{\partial \tau} \, \ud \tau + 
\int_0^\infty \lambda(\tau^\prime) \, p_{\pi(s)}(x,\tau^\prime,t) \, \ud \tau^\prime 
%+ O(\varepsilon) 
=0 \:.
\label{eq4_20a}
\end{equation}
Since $\lim_{\varepsilon \rightarrow 0} p_s(x,\varepsilon,t)=p_s(x,0,t)$
and $\lim_{\varepsilon \rightarrow 0^+} p_s(x,-\varepsilon,t)=0$, 
Eq.~(\ref{eq4_20a}) reduces to
%\begin{equation}
%p_h(x,0,t) = \int_0^\infty \lambda(\tau^\prime) \,p_{\pi(h)}(x,\tau^\prime,t) \, d \tau^\prime \:,
%\label{eq4_21}
%\end{equation}
%which is equivalent to 
the boundary condition Eq.~(\ref{eq:bc}). 
%
%Eq.~(\ref{eq4_20})  coincides with Eqs.~(\ref{eq:ppdw}),  and 
%Eq.~(\ref{eq4_21}) with 
%Eq.~(\ref{eq:bc})  by
%trivially identifying $p_\pm(x,\tau,t)$ with $p_\pm(x,t,\tau)$.
%
In point of fact, Eq.~(\ref{eq4_15a}) represents in a compact form
both the evolution equations~(\ref{eq:ppdw})
%(\ref{eq4_20}) or
and the age boundary condition~(\ref{eq:bc}).  This equation
possesses the typical integro-differential structure of a generalised
PK process in the presence of an impulsive recombination in $\tau$, in
which the dynamics for $\tau \in (0,\infty)$ gives rise to
Eqs.~(\ref{eq:ppdw}), whereas the impulsive discontinuity at $\tau=0$
can be resolved by the boundary condition Eq.~(\ref{eq:bc}).

We highlight that within our generalised theory LWs can be defined
straightforwardly in higher spatial dimensions.  In contrast, only few
models in two spatial dimensions with \textit{ad hoc} assumptions on
the transitional dynamics have been obtained by using the conventional
formalism of CTRWs \cite{ZFDB16}.  As a matter of fact, all these
models can be naturally recovered as special cases of our
Eq.~(\ref{eqo_5}), as follows: First, in the \textit{product model},
we assume that the random walker performs completely independent
one-dimensional LWs in both the $x$- and $y$-direction.  Therefore,
within our formalism, the PPDW functions can be factorised as
$p(\textbf{x},\boldsymbol{\tau},t,\boldsymbol{\alpha})=p(x,\tau_x,t,\alpha_x)p(y,\tau_y,t,\alpha_y)$,
where we introduced transitional age variables $\tau_x$ and $\tau_y$,
and independent stochastic parametrisations $\alpha_x$ and $\alpha_y$
in each direction.  If we assume that all other characteristic
parameters of these processes are the same as those of the
one-dimensional LW discussed above, each PPDW functions then satisfy
the same Eq.~\eqref{eq4_15a} in its corresponding spatial direction.
%Within our formalism, therefore, we set 
%$\boldsymbol{\tau}=(\tau_x,\tau_y)$, $\boldsymbol{\alpha}=(\alpha_x,\alpha_y)$ with 
%$\mathcal{D}_{\alpha_x},\mathcal{D}_{\alpha_y}=\{\pm\} $, and
%$\textbf{b}(\textbf{x},\boldsymbol{\tau},\boldsymbol{\alpha})=\textbf{b}(\boldsymbol{\alpha})=(\text{b}(\alpha_x),\text{b}(\alpha_y))$. 
Second, in the \textit{XY model}, we assume that the random walker can only move along the x- or y-axis in either direction. In this case, we therefore set   
$\boldsymbol{\tau}=\tau$, $\boldsymbol{\alpha}=\alpha$ with 
$\mathcal{D}_{\alpha}=\{0,1,2,3\}$, and 
$\textbf{b}(\textbf{x},\tau,\alpha)=\textbf{b}(\alpha)=b(\cos{(\alpha\pi/2)},\sin{(\alpha\pi/2)})$. 
As the stochastic parameter can only attain discrete values, we can make assumptions similar to Eq.~(\ref{eqo_7}) and~(\ref{eq4_12}) with the only difference being in the definition of the Markovian transitional matrix; in this case, this is defined as  
%$A_{ij}=(1-\delta_{ij})/3 $.     
\begin{equation} 
{\bf A}= \frac{1}{3}
\left (
\begin{array}{cccc}
0 & 1 & 1 & 1 \\
1 & 0 & 1 & 1 \\
1 & 1 & 0 & 1 \\
1 & 1 & 1 & 0
\end{array}
\right )\:.
%\label{eq4_15}
\end{equation}
Finally, in the \textit{uniform model}, we assume that the random
walker at each transition can choose any random new direction. 
%In this case, 
We therefore set $\boldsymbol{\tau}=\tau$,
$\boldsymbol{\alpha}=\alpha$ with $\mathcal{D}_{\alpha}=[0,2\pi)$, and
  $\textbf{b}(\textbf{x},\tau,\alpha)=\textbf{b}(\alpha)=b(\cos{(\alpha)},\sin{(\alpha)})$.
  As the stochastic parameter here attains values in a continuous set,
  Eq.~(\ref{eqo_5}) applies directly with the transitional matrix
  %defined as
  \begin{equation}
  A(\textbf{x},\alpha,\tau,\alpha^{\prime},\tau^{\prime})=\frac{\delta(\tau)}{2\pi}.
  \end{equation}
  %These considerations already indicate the power and flexiblity
  %of our novel theoretical framework.
 
%\bibliography{summ31}
  
%\bibliography{epkrefs}

\begin{thebibliography}{111}%
\makeatletter
\providecommand \@ifxundefined [1]{%
 \@ifx{#1\undefined}
}%
\providecommand \@ifnum [1]{%
 \ifnum #1\expandafter \@firstoftwo
 \else \expandafter \@secondoftwo
 \fi
}%
\providecommand \@ifx [1]{%
 \ifx #1\expandafter \@firstoftwo
 \else \expandafter \@secondoftwo
 \fi
}%
\providecommand \natexlab [1]{#1}%
\providecommand \enquote  [1]{``#1''}%
\providecommand \bibnamefont  [1]{#1}%
\providecommand \bibfnamefont [1]{#1}%
\providecommand \citenamefont [1]{#1}%
\providecommand \href@noop [0]{\@secondoftwo}%
\providecommand \href [0]{\begingroup \@sanitize@url \@href}%
\providecommand \@href[1]{\@@startlink{#1}\@@href}%
\providecommand \@@href[1]{\endgroup#1\@@endlink}%
\providecommand \@sanitize@url [0]{\catcode `\\12\catcode `\$12\catcode
  `\&12\catcode `\#12\catcode `\^12\catcode `\_12\catcode `\%12\relax}%
\providecommand \@@startlink[1]{}%
\providecommand \@@endlink[0]{}%
\providecommand \url  [0]{\begingroup\@sanitize@url \@url }%
\providecommand \@url [1]{\endgroup\@href {#1}{\urlprefix }}%
\providecommand \urlprefix  [0]{URL }%
\providecommand \Eprint [0]{\href }%
\providecommand \doibase [0]{http://dx.doi.org/}%
\providecommand \selectlanguage [0]{\@gobble}%
\providecommand \bibinfo  [0]{\@secondoftwo}%
\providecommand \bibfield  [0]{\@secondoftwo}%
\providecommand \translation [1]{[#1]}%
\providecommand \BibitemOpen [0]{}%
\providecommand \bibitemStop [0]{}%
\providecommand \bibitemNoStop [0]{.\EOS\space}%
\providecommand \EOS [0]{\spacefactor3000\relax}%
\providecommand \BibitemShut  [1]{\csname bibitem#1\endcsname}%
\let\auto@bib@innerbib\@empty
%</preamble>
\bibitem [{\citenamefont {Bunde}\ \emph {et~al.}(2018)\citenamefont {Bunde},
  \citenamefont {Caro}, \citenamefont {Kaerger},\ and\ \citenamefont
  {Vogl}}]{BCKV16}%
  \BibitemOpen
  \bibinfo {editor} {\bibfnamefont {A.}~\bibnamefont {Bunde}}, \bibinfo
  {editor} {\bibfnamefont {J.}~\bibnamefont {Caro}}, \bibinfo {editor}
  {\bibfnamefont {J.}~\bibnamefont {Kaerger}}, \ and\ \bibinfo {editor}
  {\bibfnamefont {G.}~\bibnamefont {Vogl}},\ eds.,\ \href@noop {} {\emph
  {\bibinfo {title} {Diffusive Spreading in Nature, Technology and Society}}}\
  (\bibinfo  {publisher} {Springer},\ \bibinfo {address} {Berlin},\ \bibinfo
  {year} {2018})\BibitemShut {NoStop}%
\bibitem [{\citenamefont {van Kampen}(1992)}]{vK}%
  \BibitemOpen
  \bibfield  {author} {\bibinfo {author} {\bibfnamefont {N.}~\bibnamefont {van
  Kampen}},\ }\href@noop {} {\emph {\bibinfo {title} {Stochastic processes in
  physics and chemistry}}}\ (\bibinfo  {publisher} {North Holland},\ \bibinfo
  {address} {Amsterdam},\ \bibinfo {year} {1992})\BibitemShut {NoStop}%
\bibitem [{\citenamefont {Risken}(1996)}]{Risk}%
  \BibitemOpen
  \bibfield  {author} {\bibinfo {author} {\bibfnamefont {H.}~\bibnamefont
  {Risken}},\ }\href@noop {} {\emph {\bibinfo {title} {The {F}okker-{P}lanck
  Equation}}},\ \bibinfo {edition} {2nd}\ ed.\ (\bibinfo  {publisher}
  {Springer},\ \bibinfo {address} {Berlin},\ \bibinfo {year}
  {1996})\BibitemShut {NoStop}%
\bibitem [{\citenamefont {Gardiner}(2009)}]{Gard09}%
  \BibitemOpen
  \bibfield  {author} {\bibinfo {author} {\bibfnamefont {C.}~\bibnamefont
  {Gardiner}},\ }\href@noop {} {\emph {\bibinfo {title} {Stochastic Methods: A
  Handbook for the Natural and Social Sciences}}},\ \bibinfo {edition} {4th}\
  ed.\ (\bibinfo  {publisher} {Springer},\ \bibinfo {address} {Berlin},\
  \bibinfo {year} {2009})\BibitemShut {NoStop}%
\bibitem [{\citenamefont {Zwanzig}(2001)}]{Zwan01}%
  \BibitemOpen
  \bibfield  {author} {\bibinfo {author} {\bibfnamefont {R.}~\bibnamefont
  {Zwanzig}},\ }\href@noop {} {\emph {\bibinfo {title} {Nonequilibrium
  statistical mechanics}}}\ (\bibinfo  {publisher} {Oxford University Press},\
  \bibinfo {address} {Oxford},\ \bibinfo {year} {2001})\BibitemShut {NoStop}%
\bibitem [{\citenamefont {Langevin}(1908)}]{Lang08}%
  \BibitemOpen
  \bibfield  {author} {\bibinfo {author} {\bibfnamefont {P.}~\bibnamefont
  {Langevin}},\ }\bibfield  {title} {\enquote {\bibinfo {title} {Sur la
  th\'eorie du mouvement Brownien},}\ }\href@noop {} {\bibfield  {journal}
  {\bibinfo  {journal} {C.R. Acad. Sci. (Paris)}\ }\textbf {\bibinfo {volume}
  {146}},\ \bibinfo {pages} {530--533} (\bibinfo {year} {1908})},\ \bibinfo
  {note} {see also the English translation in Am. J. Phys. {\bf 65}, 1079
  (1997)}\BibitemShut {NoStop}%
\bibitem [{\citenamefont {Coffey}\ \emph {et~al.}(2004)\citenamefont {Coffey},
  \citenamefont {Kalmykov},\ and\ \citenamefont {Waldron}}]{CKW04}%
  \BibitemOpen
  \bibfield  {author} {\bibinfo {author} {\bibfnamefont {W.}~\bibnamefont
  {Coffey}}, \bibinfo {author} {\bibfnamefont {Y.~P.}\ \bibnamefont {Kalmykov}},
  \ and\ \bibinfo {author} {\bibfnamefont {J.~T.}\ \bibnamefont {Waldron}},\
  }\href@noop {} {\emph {\bibinfo {title} {The Langevin Equation}}}\ (\bibinfo
  {publisher} {World Scientific},\ \bibinfo {address} {Singapore},\ \bibinfo
  {year} {2004})\BibitemShut {NoStop}%
\bibitem [{\citenamefont {Gnedenko}\ and\ \citenamefont
  {Kolmogorov}(1954)}]{Gnedenko1954}%
  \BibitemOpen
  \bibfield  {author} {\bibinfo {author} {\bibfnamefont {B.~V.}\ \bibnamefont
  {Gnedenko}}\ and\ \bibinfo {author} {\bibfnamefont {A.~N.}\ \bibnamefont
  {Kolmogorov}},\ }\href@noop {} {\emph {\bibinfo {title} {{Limit distributions
  for sums of independent random variables}}}}\ (\bibinfo  {publisher}
  {Addison-Wesley, Cambridge, United States},\ \bibinfo {year}
  {1954})\BibitemShut {NoStop}%
\bibitem [{\citenamefont {Cattaneo}(1948)}]{Catt48}%
  \BibitemOpen
  \bibfield  {author} {\bibinfo {author} {\bibfnamefont {C.}~\bibnamefont
  {Cattaneo}},\ }\bibfield  {title} {\enquote {\bibinfo {title} {Sulla
  conduzione del calore},}\ }\href@noop {} {\bibfield  {journal} {\bibinfo
  {journal} {Atti Semin. Mat. Fis. Univ. Modena}\ }\textbf {\bibinfo {volume}
  {3}},\ \bibinfo {pages} {83--101} (\bibinfo {year} {1948})}\BibitemShut
  {NoStop}%
\bibitem [{\citenamefont {Dunkel}\ and\ \citenamefont
  {H{\"a}nggi}(2009)}]{DuHa09}%
  \BibitemOpen
  \bibfield  {author} {\bibinfo {author} {\bibfnamefont {J.}~\bibnamefont
  {Dunkel}}\ and\ \bibinfo {author} {\bibfnamefont {P.}~\bibnamefont
  {H{\"a}nggi}},\ }\bibfield  {title} {\enquote {\bibinfo {title} {Relativistic
  Brownian motion},}\ }\href@noop {} {\bibfield  {journal} {\bibinfo  {journal}
  {Phys. Rep.}\ }\textbf {\bibinfo {volume} {471}},\ \bibinfo {pages} {1 -- 73}
    (\bibinfo {year} {2009})}\BibitemShut {NoStop}%
\bibitem [{\citenamefont {Rezzolla}\ and\ \citenamefont
  {Zanotti}(2013)}]{ReZa13}%
  \BibitemOpen
  \bibfield  {author} {\bibinfo {author} {\bibfnamefont {L.}~\bibnamefont
  {Rezzolla}}\ and\ \bibinfo {author} {\bibfnamefont {O.}~\bibnamefont
  {Zanotti}},\ }\href@noop {} {\emph {\bibinfo {title} {Relativistic
  Hydrodynamics}}}\ (\bibinfo  {publisher} {Oxford University Press},\ \bibinfo
  {address} {Oxford},\ \bibinfo {year} {2013})\BibitemShut {NoStop}%
\bibitem [{\citenamefont {Blum}\ \emph {et~al.}(2006)\citenamefont {Blum},
  \citenamefont {Bruns}, \citenamefont {Rademacher}, \citenamefont {Voss},
  \citenamefont {Willenberg},\ and\ \citenamefont {Krause}}]{Blum2006}%
  \BibitemOpen
  \bibfield  {author} {\bibinfo {author} {\bibfnamefont {J.}\
  \bibnamefont {Blum}}, \bibinfo {author} {\bibfnamefont {S.}\ \bibnamefont
  {Bruns}}, \bibinfo {author} {\bibfnamefont {D.}\ \bibnamefont
  {Rademacher}}, \bibinfo {author} {\bibfnamefont {A.}\ \bibnamefont
  {Voss}}, \bibinfo {author} {\bibfnamefont {B.}\ \bibnamefont
  {Willenberg}}, \ and\ \bibinfo {author} {\bibfnamefont {M.}\ \bibnamefont
  {Krause}},\ }\bibfield  {title} {\enquote {\bibinfo {title} {Measurement of
  the translational and rotational Brownian motion of individual particles in a
  rarefied gas},}\ }\href@noop {} {\bibfield  {journal} {\bibinfo  {journal}
  {Phys. Rev. Lett.}\ }\textbf {\bibinfo {volume} {97}},\ \bibinfo {pages}
  {230601} (\bibinfo {year} {2006})}\BibitemShut {NoStop}%
\bibitem [{\citenamefont {Yoo}\ \emph {et~al.}(1990)\citenamefont {Yoo},
  \citenamefont {Liu},\ and\ \citenamefont {Alfano}}]{YLA90}%
  \BibitemOpen
  \bibfield  {author} {\bibinfo {author} {\bibfnamefont {K.~M.}\ \bibnamefont
  {Yoo}}, \bibinfo {author} {\bibfnamefont {Feng}\ \bibnamefont {Liu}}, \ and\
  \bibinfo {author} {\bibfnamefont {R.~R.}\ \bibnamefont {Alfano}},\ }\bibfield
   {title} {\enquote {\bibinfo {title} {When does the diffusion approximation
  fail to describe photon transport in random media?}}\ }\href@noop {}
  {\bibfield  {journal} {\bibinfo  {journal} {Phys. Rev. Lett.}\ }\textbf
  {\bibinfo {volume} {64}},\ \bibinfo {pages} {2647--2650} (\bibinfo {year}
  {1990})}\BibitemShut {NoStop}%
\bibitem [{\citenamefont {Chang}\ \emph {et~al.}(2008)\citenamefont {Chang},
  \citenamefont {Okawa}, \citenamefont {Garcia}, \citenamefont {Majumdar},\
  and\ \citenamefont {Zettl}}]{COGMZ08}%
  \BibitemOpen
  \bibfield  {author} {\bibinfo {author} {\bibfnamefont {C.~W.}\ \bibnamefont
  {Chang}}, \bibinfo {author} {\bibfnamefont {D.}~\bibnamefont {Okawa}},
  \bibinfo {author} {\bibfnamefont {H.}~\bibnamefont {Garcia}}, \bibinfo
  {author} {\bibfnamefont {A.}~\bibnamefont {Majumdar}}, \ and\ \bibinfo
  {author} {\bibfnamefont {A.}~\bibnamefont {Zettl}},\ }\bibfield  {title}
  {\enquote {\bibinfo {title} {Breakdown of Fourier's law in nanotube thermal
  conductors},}\ }\href@noop {} {\bibfield  {journal} {\bibinfo  {journal}
  {Phys. Rev. Lett.}\ }\textbf {\bibinfo {volume} {101}},\ \bibinfo {pages}
  {075903} (\bibinfo {year} {2008})}\BibitemShut {NoStop}%
\bibitem [{\citenamefont {Sellitto}\ \emph {et~al.}(2016)\citenamefont
  {Sellitto}, \citenamefont {Cimmelli},\ and\ \citenamefont {Jou}}]{SCJ16}%
  \BibitemOpen
  \bibfield  {author} {\bibinfo {author} {\bibfnamefont {A.}~\bibnamefont
  {Sellitto}}, \bibinfo {author} {\bibfnamefont {V.~A.}\ \bibnamefont
  {Cimmelli}}, \ and\ \bibinfo {author} {\bibfnamefont {D.}~\bibnamefont
  {Jou}},\ }\href@noop {} {\emph {\bibinfo {title} {Mesoscopic Theories of Heat
  Transport in Nanosystems}}}\ (\bibinfo  {publisher} {Springer},\ \bibinfo
  {address} {Berlin},\ \bibinfo {year} {2016})\BibitemShut {NoStop}%
\bibitem [{\citenamefont {Joseph}\ and\ \citenamefont
  {Preziosi}(1989)}]{JoPr89}%
  \BibitemOpen
  \bibfield  {author} {\bibinfo {author} {\bibfnamefont {D.~D.}\ \bibnamefont
  {Joseph}}\ and\ \bibinfo {author} {\bibfnamefont {L.}\ \bibnamefont
  {Preziosi}},\ }\bibfield  {title} {\enquote {\bibinfo {title} {Heat waves},}\
  }\href@noop {} {\bibfield  {journal} {\bibinfo  {journal} {Rev. Mod. Phys.}\
  }\textbf {\bibinfo {volume} {61}},\ \bibinfo {pages} {41--73} (\bibinfo
  {year} {1989})}\BibitemShut {NoStop}%
\bibitem [{\citenamefont {Metzler}\ and\ \citenamefont
  {Klafter}(2000)}]{MeKl00}%
  \BibitemOpen
  \bibfield  {author} {\bibinfo {author} {\bibfnamefont {R.}~\bibnamefont
  {Metzler}}\ and\ \bibinfo {author} {\bibfnamefont {J.}~\bibnamefont
  {Klafter}},\ }\bibfield  {title} {\enquote {\bibinfo {title} {The random
  walk's guide to anomalous diffusion: A fractional dynamics approach},}\
  }\href@noop {} {\bibfield  {journal} {\bibinfo  {journal} {Phys. Rep.}\
  }\textbf {\bibinfo {volume} {339}},\ \bibinfo {pages} {1--77} (\bibinfo
  {year} {2000})}\BibitemShut {NoStop}%
\bibitem [{\citenamefont {Klages}\ \emph {et~al.}(2008)\citenamefont {Klages},
  \citenamefont {Radons},\ and\ \citenamefont {Sokolov}}]{KRS08}%
  \BibitemOpen
  \bibinfo {editor} {\bibfnamefont {R.}~\bibnamefont {Klages}}, \bibinfo
  {editor} {\bibfnamefont {G.}~\bibnamefont {Radons}}, \ and\ \bibinfo {editor}
  {\bibfnamefont {I.~M.}\ \bibnamefont {Sokolov}},\ eds.,\ \href@noop {} {\emph
  {\bibinfo {title} {Anomalous transport: Foundations and Applications}}}\
  (\bibinfo  {publisher} {Wiley-VCH},\ \bibinfo {address} {Berlin},\ \bibinfo
  {year} {2008})\BibitemShut {NoStop}%
\bibitem [{\citenamefont {Shlesinger}\ \emph {et~al.}(1993)\citenamefont
  {Shlesinger}, \citenamefont {Zaslavsky},\ and\ \citenamefont
  {Klafter}}]{SZK93}%
  \BibitemOpen
  \bibfield  {author} {\bibinfo {author} {\bibfnamefont {M.~F.}\ \bibnamefont
  {Shlesinger}}, \bibinfo {author} {\bibfnamefont {G.~M.}\ \bibnamefont
  {Zaslavsky}}, \ and\ \bibinfo {author} {\bibfnamefont {J.}~\bibnamefont
  {Klafter}},\ }\bibfield  {title} {\enquote {\bibinfo {title} {Strange
  kinetics},}\ }\href@noop {} {\bibfield  {journal} {\bibinfo  {journal}
  {Nature}\ }\textbf {\bibinfo {volume} {363}},\ \bibinfo {pages} {31--37}
  (\bibinfo {year} {1993})}\BibitemShut {NoStop}%
\bibitem [{\citenamefont {Klafter}\ \emph {et~al.}(1996)\citenamefont
  {Klafter}, \citenamefont {Shlesinger},\ and\ \citenamefont
  {Zumofen}}]{KSZ96}%
  \BibitemOpen
  \bibfield  {author} {\bibinfo {author} {\bibfnamefont {J.}~\bibnamefont
  {Klafter}}, \bibinfo {author} {\bibfnamefont {M.~F.}\ \bibnamefont
  {Shlesinger}}, \ and\ \bibinfo {author} {\bibfnamefont {G.}~\bibnamefont
  {Zumofen}},\ }\bibfield  {title} {\enquote {\bibinfo {title} {Beyond
  {B}rownian motion},}\ }\href@noop {} {\bibfield  {journal} {\bibinfo
  {journal} {Phys. Today}\ }\textbf {\bibinfo {volume} {49}},\ \bibinfo {pages}
  {33--39} (\bibinfo {year} {1996})}\BibitemShut {NoStop}%
\bibitem [{\citenamefont {Metzler}\ \emph {et~al.}(2014)\citenamefont
  {Metzler}, \citenamefont {Jeon}, \citenamefont {Cherstvy},\ and\
  \citenamefont {Barkai}}]{MJCB14}%
  \BibitemOpen
  \bibfield  {author} {\bibinfo {author} {\bibfnamefont {R.}\ \bibnamefont
  {Metzler}}, \bibinfo {author} {\bibfnamefont {J.-H.}\ \bibnamefont
  {J.}}, \bibinfo {author} {\bibfnamefont {A.~G.}\ \bibnamefont
  {Cherstvy}}, \ and\ \bibinfo {author} {\bibfnamefont {E.}\ \bibnamefont
  {Barkai}},\ }\bibfield  {title} {\enquote {\bibinfo {title} {Anomalous
  diffusion models and their properties: non-stationarity{,} non-ergodicity{,}
  and ageing at the centenary of single particle tracking},}\ }\href@noop {}
  {\bibfield  {journal} {\bibinfo  {journal} {Phys. Chem. Chem. Phys.}\
  }\textbf {\bibinfo {volume} {16}},\ \bibinfo {pages} {24128--24164} (\bibinfo
  {year} {2014})}\BibitemShut {NoStop}%
\bibitem [{\citenamefont {Zaburdaev}\ \emph {et~al.}(2015)\citenamefont
  {Zaburdaev}, \citenamefont {Denisov},\ and\ \citenamefont {Klafter}}]{ZDK15}%
  \BibitemOpen
  \bibfield  {author} {\bibinfo {author} {\bibfnamefont {V.}~\bibnamefont
  {Zaburdaev}}, \bibinfo {author} {\bibfnamefont {S.}~\bibnamefont {Denisov}},
  \ and\ \bibinfo {author} {\bibfnamefont {J.}~\bibnamefont {Klafter}},\
  }\bibfield  {title} {\enquote {\bibinfo {title} {L\'{e}vy walks},}\
  }\href@noop {} {\bibfield  {journal} {\bibinfo  {journal} {Rev. Mod. Phys.}\
  }\textbf {\bibinfo {volume} {87}},\ \bibinfo {pages} {483--529} (\bibinfo
  {year} {2015})}\BibitemShut {NoStop}%
\bibitem [{\citenamefont {{Metzler}}\ and\ \citenamefont
  {{Klafter}}(2004)}]{MeKl04}%
  \BibitemOpen
  \bibfield  {author} {\bibinfo {author} {\bibfnamefont {{R.}}~\bibnamefont
  {{Metzler}}}\ and\ \bibinfo {author} {\bibfnamefont {{J.}}~\bibnamefont
  {{Klafter}}},\ }\bibfield  {title} {\enquote {\bibinfo {title} {The
  restaurant at the end of the random walk: Recent developments in the
  description of anomalous transport by fractional dynamics},}\ }\href@noop {}
  {\bibfield  {journal} {\bibinfo  {journal} {J. Phys. A: Math. Gen.}\ }\textbf
  {\bibinfo {volume} {37}},\ \bibinfo {pages} {R161--R208} (\bibinfo {year}
  {2004})}\BibitemShut {NoStop}%
\bibitem [{\citenamefont {Viswanathan}\ \emph {et~al.}(2011)\citenamefont
  {Viswanathan}, \citenamefont {{da Luz}}, \citenamefont {Raposo},\ and\
  \citenamefont {Stanley}}]{VLRS11}%
  \BibitemOpen
  \bibfield  {author} {\bibinfo {author} {\bibfnamefont {G.~M}\ \bibnamefont
  {Viswanathan}}, \bibinfo {author} {\bibfnamefont {M.~G.~E.}\ \bibnamefont {{da
  Luz}}}, \bibinfo {author} {\bibfnamefont {E.~P.}\ \bibnamefont {Raposo}}, \
  and\ \bibinfo {author} {\bibfnamefont {H.~E.}\ \bibnamefont {Stanley}},\
  }\href@noop {} {\emph {\bibinfo {title} {The physics of foraging}}}\
  (\bibinfo  {publisher} {Cambridge University Press},\ \bibinfo {address}
  {Cambridge},\ \bibinfo {year} {2011})\BibitemShut {NoStop}%
\bibitem [{\citenamefont {Mandelbrot}(1982)}]{Man82}%
  \BibitemOpen
  \bibfield  {author} {\bibinfo {author} {\bibfnamefont {B.~B.}\ \bibnamefont
  {Mandelbrot}},\ }\href@noop {} {\emph {\bibinfo {title} {The fractal geometry
  of nature}}}\ (\bibinfo  {publisher} {W.H. Freeman and Company},\ \bibinfo
  {address} {San Francisco},\ \bibinfo {year} {1982})\BibitemShut {NoStop}%
\bibitem [{\citenamefont {Klages}(2016)}]{Kla16}%
  \BibitemOpen
  \bibfield  {author} {\bibinfo {author} {\bibfnamefont {R.}~\bibnamefont
  {Klages}},\ }\href@noop {} {\enquote {\bibinfo {title} {Search for food of
  birds, fish and insects},}\ } (\bibinfo {year} {2016}),\ \bibinfo {note}
  {book chapter in: A.Bunde, J.Caro, J.Kaerger, G.Vogl (Eds.), Diffusive
  Spreading in Nature, Technology and Society. (Springer, Berlin)}\BibitemShut
  {NoStop}%
\bibitem [{\citenamefont {Shlesinger}\ \emph {et~al.}(1982)\citenamefont
  {Shlesinger}, \citenamefont {Klafter},\ and\ \citenamefont {Wong}}]{SKW82}%
  \BibitemOpen
  \bibfield  {author} {\bibinfo {author} {\bibfnamefont {M.~F.}\ \bibnamefont
  {Shlesinger}}, \bibinfo {author} {\bibfnamefont {J.}~\bibnamefont {Klafter}},
  \ and\ \bibinfo {author} {\bibfnamefont {Y.~M.}\ \bibnamefont {Wong}},\
  }\bibfield  {title} {\enquote {\bibinfo {title} {{Random walks with infinite
  spatial and temporal moments}},}\ }\href@noop {} {\bibfield  {journal}
  {\bibinfo  {journal} {J. Stat. Phys.}\ }\textbf {\bibinfo {volume} {27}},\
  \bibinfo {pages} {499--512} (\bibinfo {year} {1982})}\BibitemShut {NoStop}%
\bibitem [{\citenamefont {Geisel}\ \emph {et~al.}(1985)\citenamefont {Geisel},
  \citenamefont {Nierwetberg},\ and\ \citenamefont {Zacherl}}]{GNZ85}%
  \BibitemOpen
  \bibfield  {author} {\bibinfo {author} {\bibfnamefont {T.}~\bibnamefont
  {Geisel}}, \bibinfo {author} {\bibfnamefont {J.}~\bibnamefont {Nierwetberg}},
  \ and\ \bibinfo {author} {\bibfnamefont {A.}~\bibnamefont {Zacherl}},\
  }\bibfield  {title} {\enquote {\bibinfo {title} {Accelerated diffusion in
  {J}osephson junctions and related chaotic systems},}\ }\href@noop {}
  {\bibfield  {journal} {\bibinfo  {journal} {Phys. Rev. Lett.}\ }\textbf
  {\bibinfo {volume} {54}},\ \bibinfo {pages} {616--619} (\bibinfo {year}
  {1985})}\BibitemShut {NoStop}%
\bibitem [{\citenamefont {{Shlesinger}}\ and\ \citenamefont
  {{Klafter}}(1985)}]{ShlKl85}%
  \BibitemOpen
  \bibfield  {author} {\bibinfo {author} {\bibfnamefont {{M.~F.}}\ \bibnamefont
  {{Shlesinger}}}\ and\ \bibinfo {author} {\bibfnamefont {{J.}}~\bibnamefont
  {{Klafter}}},\ }\bibfield  {title} {\enquote {\bibinfo {title} {Accelerated
  diffusion in {J}osephson-junctions and related chaotic systems - comment},}\
  }\href@noop {} {\bibfield  {journal} {\bibinfo  {journal} {Phys. Rev. Lett.}\
  }\textbf {\bibinfo {volume} {54}},\ \bibinfo {pages} {2551} (\bibinfo {year}
  {1985})}\BibitemShut {NoStop}%
\bibitem [{\citenamefont {Klafter}\ \emph {et~al.}(1987)\citenamefont
  {Klafter}, \citenamefont {Blumen},\ and\ \citenamefont {Shlesinger}}]{KBS87}%
  \BibitemOpen
  \bibfield  {author} {\bibinfo {author} {\bibfnamefont {J.}~\bibnamefont
  {Klafter}}, \bibinfo {author} {\bibfnamefont {A.}~\bibnamefont {Blumen}}, \
  and\ \bibinfo {author} {\bibfnamefont {M.~F.}\ \bibnamefont {Shlesinger}},\
  }\bibfield  {title} {\enquote {\bibinfo {title} {{Stochastic pathway to
  anomalous diffusion}},}\ }\href@noop {} {\bibfield  {journal} {\bibinfo
  {journal} {Phys. Rev. A}\ }\textbf {\bibinfo {volume} {35}},\ \bibinfo
  {pages} {3081--3085} (\bibinfo {year} {1987})}\BibitemShut {NoStop}%
\bibitem [{\citenamefont {Shlesinger}\ \emph {et~al.}(1987)\citenamefont
  {Shlesinger}, \citenamefont {West},\ and\ \citenamefont {Klafter}}]{Shles87}%
  \BibitemOpen
  \bibfield  {author} {\bibinfo {author} {\bibfnamefont {M.~F.}\ \bibnamefont
  {Shlesinger}}, \bibinfo {author} {\bibfnamefont {B.~J.}\ \bibnamefont {West}},
  \ and\ \bibinfo {author} {\bibfnamefont {J.}~\bibnamefont {Klafter}},\
  }\bibfield  {title} {\enquote {\bibinfo {title} {{L{\'e}vy dynamics of
  enhanced diffusion: Application to turbulence}},}\ }\href@noop {} {\bibfield
  {journal} {\bibinfo  {journal} {Phys. Rev. Lett.}\ }\textbf {\bibinfo
  {volume} {58}},\ \bibinfo {pages} {1100--1103} (\bibinfo {year}
  {1987})}\BibitemShut {NoStop}%
  \bibitem [{\citenamefont {Geisel}\ \emph {et~al.}(1988)\citenamefont {Geisel},
  \citenamefont {Zacherl},\ and\ \citenamefont {Radons}}]{GZR88}%
  \BibitemOpen
  \bibfield  {author} {\bibinfo {author} {\bibfnamefont {T.}~\bibnamefont
  {Geisel}}, \bibinfo {author} {\bibfnamefont {A.}~\bibnamefont {Zacherl}}, \
  and\ \bibinfo {author} {\bibfnamefont {G.}~\bibnamefont {Radons}},\
  }\bibfield  {title} {\enquote {\bibinfo {title} {{Chaotic diffusion and 1/f-noise
          of particles in two-dimensional solids}},}\ }\href@noop {}
  {\bibfield  {journal} {\bibinfo  {journal} {Z. Phys. B}\
  }\textbf {\bibinfo {volume} {71}},\ \bibinfo {pages} {117--127} (\bibinfo
  {year} {1988})}\BibitemShut {NoStop}%
\bibitem [{\citenamefont {{Zumofen}}\ and\ \citenamefont
  {{Klafter}}(1993)}]{ZuKl93a}%
  \BibitemOpen
  \bibfield  {author} {\bibinfo {author} {\bibfnamefont {{G.}}~\bibnamefont
  {{Zumofen}}}\ and\ \bibinfo {author} {\bibfnamefont {{J.}}~\bibnamefont
  {{Klafter}}},\ }\bibfield  {title} {\enquote {\bibinfo {title}
  {Scale-invariant motion in intermittent chaotic systems},}\ }\href@noop {}
  {\bibfield  {journal} {\bibinfo  {journal} {Phys. Rev. E}\ }\textbf {\bibinfo
  {volume} {47}},\ \bibinfo {pages} {851--863} (\bibinfo {year}
  {1993})}\BibitemShut {NoStop}%
\bibitem [{\citenamefont {Shlesinger}\ (1979)}]{shlesinger79}%
  \BibitemOpen
  \bibfield  {author} {\bibinfo {author} {\bibfnamefont {{M.~F.}}\ \bibnamefont
  {{Shlesinger}}},\ }\bibfield  {title} {\enquote {\bibinfo {title} {Correlation effects on frequency dependent conductivity: 
  application to superionic conductors},}\ }\href@noop {} {\bibfield  {journal} {\bibinfo  {journal} {Solid State Commun.}\
  }\textbf {\bibinfo {volume} {32}},\ \bibinfo {pages} {1207--1210} (\bibinfo {year}
  {1979})}\BibitemShut {NoStop}%
\bibitem [{\citenamefont {Becker-Kern}\ \emph {et~al.}(2004)\citenamefont
  {Becker-Kern}, \citenamefont {Meerschaert},\ and\ \citenamefont
  {Scheffler}}]{BKMS04}%
  \BibitemOpen
  \bibfield  {author} {\bibinfo {author} {\bibfnamefont {P.}~\bibnamefont
  {Becker-Kern}}, \bibinfo {author} {\bibfnamefont {M.~M.}\ \bibnamefont
  {Meerschaert}}, \ and\ \bibinfo {author} {\bibfnamefont {H.~-P.}\ \bibnamefont
  {Scheffler}},\ }\bibfield  {title} {\enquote {\bibinfo {title} {{Limit
  theorems for coupled continuous time random walks}},}\ }\href@noop {}
  {\bibfield  {journal} {\bibinfo  {journal} {Ann. Prob.}\ }\textbf {\bibinfo
  {volume} {32}},\ \bibinfo {pages} {730--756} (\bibinfo {year}
  {2004})}\BibitemShut {NoStop}%
\bibitem [{\citenamefont {Rebenshtok}\ \emph {et~al.}(2014)\citenamefont
  {Rebenshtok}, \citenamefont {Denisov}, \citenamefont {H\"anggi},\ and\
  \citenamefont {Barkai}}]{RDHB14}%
  \BibitemOpen
  \bibfield  {author} {\bibinfo {author} {\bibfnamefont {A.}~\bibnamefont
  {Rebenshtok}}, \bibinfo {author} {\bibfnamefont {S.}~\bibnamefont {Denisov}},
  \bibinfo {author} {\bibfnamefont {P.}~\bibnamefont {H\"anggi}}, \ and\
  \bibinfo {author} {\bibfnamefont {E.}~\bibnamefont {Barkai}},\ }\bibfield
  {title} {\enquote {\bibinfo {title} {Non-normalizable densities in strong
  anomalous diffusion: Beyond the central limit theorem},}\ }\href@noop {}
  {\bibfield  {journal} {\bibinfo  {journal} {Phys. Rev. Lett.}\ }\textbf
  {\bibinfo {volume} {112}},\ \bibinfo {pages} {110601} (\bibinfo {year}
  {2014})}\BibitemShut {NoStop}%
\bibitem [{\citenamefont {Fedotov}(2016)}]{Fed16}%
  \BibitemOpen
  \bibfield  {author} {\bibinfo {author} {\bibfnamefont {S.}\ \bibnamefont
  {Fedotov}},\ }\bibfield  {title} {\enquote {\bibinfo {title} {Single
  integrodifferential wave equation for a L\'evy walk},}\ }\href@noop {}
  {\bibfield  {journal} {\bibinfo  {journal} {Phys. Rev. E}\ }\textbf {\bibinfo
  {volume} {93}},\ \bibinfo {pages} {020101} (\bibinfo {year}
  {2016})}\BibitemShut {NoStop}%
\bibitem [{\citenamefont {Magdziarz}\ and\ \citenamefont
  {Zorawik}(2016)}]{MaZo16}%
  \BibitemOpen
  \bibfield  {author} {\bibinfo {author} {\bibfnamefont {M.}\ \bibnamefont
  {Magdziarz}}\ and\ \bibinfo {author} {\bibfnamefont {T.}\ \bibnamefont
  {Zorawik}},\ }\bibfield  {title} {\enquote {\bibinfo {title} {Explicit
  densities of multidimensional ballistic L\'evy walks},}\ }\href@noop {}
  {\bibfield  {journal} {\bibinfo  {journal} {Phys. Rev. E}\ }\textbf {\bibinfo
  {volume} {94}},\ \bibinfo {pages} {022130} (\bibinfo {year}
  {2016})}\BibitemShut {NoStop}%
\bibitem [{\citenamefont {Palyulin}\ \emph {et~al.}(2019)\citenamefont
  {Palyulin}, \citenamefont {Blackburn}, \citenamefont {Lomholt}, \citenamefont
  {Watkins}, \citenamefont {Metzler}, \citenamefont {Klages},\ and\
  \citenamefont {Chechkin}}]{PBL19}%
  \BibitemOpen
  \bibfield  {author} {\bibinfo {author} {\bibfnamefont {V.~V.}\
  \bibnamefont {Palyulin}}, \bibinfo {author} {\bibfnamefont {G.}\
  \bibnamefont {Blackburn}}, \bibinfo {author} {\bibfnamefont {M.~A.}\
  \bibnamefont {Lomholt}}, \bibinfo {author} {\bibfnamefont {N.~W.}\
  \bibnamefont {Watkins}}, \bibinfo {author} {\bibfnamefont {R.}\
  \bibnamefont {Metzler}}, \bibinfo {author} {\bibfnamefont {R.}\
  \bibnamefont {Klages}}, \ and\ \bibinfo {author} {\bibfnamefont {A.~V.}\
  \bibnamefont {Chechkin}},\ }\bibfield  {title} {\enquote {\bibinfo {title}
  {First passage and first hitting times of L{\'{e}}vy flights and L{\'{e}}vy
  walks},}\ }\href@noop {} {\bibfield  {journal} {\bibinfo  {journal} {New J.
  Phys.}\ }\textbf {\bibinfo {volume} {21}},\ \bibinfo {pages} {103028}
  (\bibinfo {year} {2019})}\BibitemShut {NoStop}%
\bibitem [{\citenamefont {Reynolds}(2018)}]{Reyn18}%
  \BibitemOpen
  \bibfield  {author} {\bibinfo {author} {\bibfnamefont {A.~M.}\ \bibnamefont
  {Reynolds}},\ }\bibfield  {title} {\enquote {\bibinfo {title} {Current status
  and future directions of {L}{\'e}vy walk research},}\ }\href@noop {}
  {\bibfield  {journal} {\bibinfo  {journal} {Biol. Open}\ }\textbf {\bibinfo
  {volume} {7}},\ \bibinfo {pages} {03016} (\bibinfo {year}
  {2018})}\BibitemShut {NoStop}%
\bibitem [{\citenamefont {Taylor}(1921)}]{Tay21}%
  \BibitemOpen
  \bibfield  {author} {\bibinfo {author} {\bibfnamefont {G.~I.}\ \bibnamefont
  {Taylor}},\ }\bibfield  {title} {\enquote {\bibinfo {title} {Diffusion by
  continuous movements},}\ }\href@noop {} {\bibfield  {journal} {\bibinfo
  {journal} {Proc. London Math Soc.}\ }\textbf {\bibinfo {volume} {20}},\
  \bibinfo {pages} {196--212} (\bibinfo {year} {1921})}\BibitemShut {NoStop}%
\bibitem [{\citenamefont {Goldstein}(1951)}]{Gold51}%
  \BibitemOpen
  \bibfield  {author} {\bibinfo {author} {\bibfnamefont {S.}~\bibnamefont
  {Goldstein}},\ }\bibfield  {title} {\enquote {\bibinfo {title} {On diffusion
  by discontinuous movements, and on the telegraph equation},}\ }\href@noop {}
  {\bibfield  {journal} {\bibinfo  {journal} {Q. J. Mech. Appl. Math.}\
  }\textbf {\bibinfo {volume} {4}},\ \bibinfo {pages} {129--156} (\bibinfo
  {year} {1951})}\BibitemShut {NoStop}%
\bibitem [{\citenamefont {Kac}(1974)}]{Kac74}%
  \BibitemOpen
  \bibfield  {author} {\bibinfo {author} {\bibfnamefont {M.}~\bibnamefont
  {Kac}},\ }\bibfield  {title} {\enquote {\bibinfo {title} {A stochastic model
  related to the telegrapher's equation},}\ }\href@noop {} {\bibfield
  {journal} {\bibinfo  {journal} {Rocky Mount. J. Math.}\ }\textbf {\bibinfo
  {volume} {4}},\ \bibinfo {pages} {497--509} (\bibinfo {year}
  {1974})}\BibitemShut {NoStop}%
\bibitem [{Note1()}]{Note1}%
  \BibitemOpen
  \bibinfo {note} {We note that any attempt to extend this equation to higher
  spatial dimensions fails to ensure the positivity of the corresponding PDF
  \cite {KoeBe98,GBC17}.}\BibitemShut {Stop}%
\bibitem [{\citenamefont {Cattaneo}(1958)}]{Catt58}%
  \BibitemOpen
  \bibfield  {author} {\bibinfo {author} {\bibfnamefont {C.}~\bibnamefont
  {Cattaneo}},\ }\bibfield  {title} {\enquote {\bibinfo {title} {Sur une forme
  de l'equation de la chaleur eliminant le paradoxe d'une propagation
  instantanee},}\ }\href@noop {} {\bibfield  {journal} {\bibinfo  {journal}
  {C. R. Acad. Sc. Paris}\ }\textbf {\bibinfo {volume} {247}},\ \bibinfo
  {pages} {431--433} (\bibinfo {year} {1958})}\BibitemShut {NoStop}%
\bibitem [{\citenamefont {Bena}(2006)}]{Bena06}%
  \BibitemOpen
  \bibfield  {author} {\bibinfo {author} {\bibfnamefont {I.}~\bibnamefont
  {Bena}},\ }\bibfield  {title} {\enquote {\bibinfo {title} {{Dichotomous
  Markov noise: exact results for out-of-equilibrium systems}},}\ }\href@noop
  {} {\bibfield  {journal} {\bibinfo  {journal} {Int. J. Mod. Phys. B}\
  }\textbf {\bibinfo {volume} {20}},\ \bibinfo {pages} {2825--2888} (\bibinfo
  {year} {2006})}\BibitemShut {NoStop}%
\bibitem [{\citenamefont {Weiss}(2007)}]{Weiss07}%
  \BibitemOpen
  \bibfield  {author} {\bibinfo {author} {\bibfnamefont {G.~H.}\ \bibnamefont
  {Weiss}},\ }\bibfield  {title} {\enquote {\bibinfo {title} {{Some
  applications of persistent random walks and the telegrapher ' s equation}},}\
  }\href@noop {} {\bibfield  {journal} {\bibinfo  {journal} {Physica A}\
  }\textbf {\bibinfo {volume} {311}},\ \bibinfo {pages} {381--410} (\bibinfo
  {year} {2007})}\BibitemShut {NoStop}%
\bibitem [{\citenamefont {Kolesnik}(2008)}]{Kol08}%
  \BibitemOpen
  \bibfield  {author} {\bibinfo {author} {\bibfnamefont {A.~D.}\ \bibnamefont
  {Kolesnik}},\ }\bibfield  {title} {\enquote {\bibinfo {title} {Random motions
  at finite speed in higher dimensions},}\ }\href@noop {} {\bibfield  {journal}
  {\bibinfo  {journal} {J. Stat. Phys.}\ }\textbf {\bibinfo {volume} {131}},\
  \bibinfo {pages} {1039--1065} (\bibinfo {year} {2008})}\BibitemShut {NoStop}%
\bibitem [{\citenamefont {Kolesnik}\ and\ \citenamefont
  {Pinsky}(2011)}]{Kol11}%
  \BibitemOpen
  \bibfield  {author} {\bibinfo {author} {\bibfnamefont {A.~D.}\ \bibnamefont
  {Kolesnik}}\ and\ \bibinfo {author} {\bibfnamefont {M.~A.}\ \bibnamefont
  {Pinsky}},\ }\bibfield  {title} {\enquote {\bibinfo {title} {Random
  evolutions are driven by the hyperparabolic operators},}\ }\href@noop {}
  {\bibfield  {journal} {\bibinfo  {journal} {J. Stat. Phys.}\ }\textbf
  {\bibinfo {volume} {142}},\ \bibinfo {pages} {828--846} (\bibinfo {year}
  {2011})}\BibitemShut {NoStop}%
\bibitem [{\citenamefont {Gaveau}\ \emph {et~al.}(1984)\citenamefont {Gaveau},
  \citenamefont {Jacobson}, \citenamefont {Kac},\ and\ \citenamefont
  {Schulman}}]{GJKS84}%
  \BibitemOpen
  \bibfield  {author} {\bibinfo {author} {\bibfnamefont {B.}~\bibnamefont
  {Gaveau}}, \bibinfo {author} {\bibfnamefont {T.}~\bibnamefont {Jacobson}},
  \bibinfo {author} {\bibfnamefont {M.}~\bibnamefont {Kac}}, \ and\ \bibinfo
  {author} {\bibfnamefont {L.~S.}\ \bibnamefont {Schulman}},\ }\bibfield
  {title} {\enquote {\bibinfo {title} {Relativistic extension of the analogy
  between quantum mechanics and Brownian motion},}\ }\href@noop {} {\bibfield
  {journal} {\bibinfo  {journal} {Phys. Rev. Lett.}\ }\textbf {\bibinfo
  {volume} {53}},\ \bibinfo {pages} {419--422} (\bibinfo {year}
  {1984})}\BibitemShut {NoStop}%
\bibitem [{\citenamefont {Rosenau}(1993)}]{Ros93}%
  \BibitemOpen
  \bibfield  {author} {\bibinfo {author} {\bibfnamefont {P.}~\bibnamefont
  {Rosenau}},\ }\bibfield  {title} {\enquote {\bibinfo {title} {Random walker
  and the telegrapher's equation: A paradigm of a generalized hydrodynamics},}\
  }\href@noop {} {\bibfield  {journal} {\bibinfo  {journal} {Phys. Rev. E}\
  }\textbf {\bibinfo {volume} {48}},\ \bibinfo {pages} {R655--R657} (\bibinfo
  {year} {1993})}\BibitemShut {NoStop}%
\bibitem [{\citenamefont {M{\"u}ller}\ and\ \citenamefont
  {Ruggeri}(1993)}]{MuRu93}%
  \BibitemOpen
  \bibfield  {author} {\bibinfo {author} {\bibfnamefont {I.}~\bibnamefont
  {M{\"u}ller}}\ and\ \bibinfo {author} {\bibfnamefont {T.}~\bibnamefont
  {Ruggeri}},\ }\href@noop {} {\emph {\bibinfo {title} {Extended
  Thermodynamics}}}\ (\bibinfo  {publisher} {Springer},\ \bibinfo {address}
  {New York},\ \bibinfo {year} {1993})\BibitemShut {NoStop}%
\bibitem [{\citenamefont {Giona}\ \emph
  {et~al.}(2017{\natexlab{a}})\citenamefont {Giona}, \citenamefont
  {Brasiello},\ and\ \citenamefont {Crescitelli}}]{GBC17}%
  \BibitemOpen
  \bibfield  {author} {\bibinfo {author} {\bibfnamefont {M.}\
  \bibnamefont {Giona}}, \bibinfo {author} {\bibfnamefont {A.}\
  \bibnamefont {Brasiello}}, \ and\ \bibinfo {author} {\bibfnamefont
  {S.}\ \bibnamefont {Crescitelli}},\ }\bibfield  {title} {\enquote
  {\bibinfo {title} {Stochastic foundations of undulatory transport phenomena:
  generalized Poisson-Kac processes --- part i basic theory},}\ }\href@noop {}
  {\bibfield  {journal} {\bibinfo  {journal} {J. Phys. A: Math. Theor.}\
  }\textbf {\bibinfo {volume} {50}},\ \bibinfo {pages} {335002/1--43} (\bibinfo
  {year} {2017}{\natexlab{a}})}\BibitemShut {NoStop}%
\bibitem [{\citenamefont {Giona}(2017)}]{Giona17a}%
  \BibitemOpen
  \bibfield  {author} {\bibinfo {author} {\bibfnamefont {M.}~\bibnamefont
  {Giona}},\ }\bibfield  {title} {\enquote {\bibinfo {title} {Variational
  principles and Lagrangian functions for stochastic processes and their
  dissipative statistical descriptions},}\ }\href@noop {} {\bibfield  {journal}
  {\bibinfo  {journal} {Physica A}\ }\textbf {\bibinfo {volume} {473}},\
  \bibinfo {pages} {561--577} (\bibinfo {year} {2017})}\BibitemShut {NoStop}%
\bibitem [{\citenamefont {Giona}\ \emph
  {et~al.}(2017{\natexlab{b}})\citenamefont {Giona}, \citenamefont
  {Brasiello},\ and\ \citenamefont {Crescitelli}}]{GBC17b}%
  \BibitemOpen
  \bibfield  {author} {\bibinfo {author} {\bibfnamefont {M.}~\bibnamefont
  {Giona}}, \bibinfo {author} {\bibfnamefont {A.}~\bibnamefont {Brasiello}}, \
  and\ \bibinfo {author} {\bibfnamefont {S.}~\bibnamefont {Crescitelli}},\
  }\bibfield  {title} {\enquote {\bibinfo {title} {Kac limit and thermodynamic
  characterization of stochastic dynamics driven by Poisson-Kac
  fluctuations},}\ }\href@noop {} {\bibfield  {journal} {\bibinfo  {journal}
  {Eur. Phys. J. Spec. Top.}\ }\textbf {\bibinfo {volume} {226}},\ \bibinfo
  {pages} {2299--2310} (\bibinfo {year} {2017}{\natexlab{b}})}\BibitemShut
  {NoStop}%  
\bibitem [{\citenamefont {Einstein}(1917)}]{einstein1917}%
  \BibitemOpen
  \bibfield  {author} {\bibinfo {author} {\bibfnamefont {A.}~\bibnamefont
  {Einstein}},\ }\bibfield  {title} {\enquote {\bibinfo {title} {Zur Quantentheorie der Strahlung},}\ 
  }\href@noop {} {\bibfield  {journal} {\bibinfo  {journal}
  {Phys. Zeitschrift}\ }\textbf {\bibinfo {volume} {18}},\ \bibinfo
  {pages} {121--128} (\bibinfo {year} {1917})}\BibitemShut
  {NoStop}%
\bibitem [{\citenamefont {H.-P. Breuer}(2002)\citenamefont {Breuer},\ and \ \citenamefont {Petruccione}}]{petruccione}%
  \BibitemOpen
  \bibfield  {author} {\bibinfo {author} {\bibfnamefont {H.~-P.}~\bibnamefont {Breuer}},\ and \
  \bibinfo {author} {\bibfnamefont {F.}~\bibnamefont {Petruccione}}, \
   }\href@noop {} {\emph {\bibinfo {title} {The theory of open quantum systems}}}\ (\bibinfo  {publisher} {Clarendon Press},\ \bibinfo
  {address} {Oxford},\ \bibinfo {year} {2002})\BibitemShut {NoStop}%
\bibitem [{\citenamefont {Weiss}(1994)}]{Weiss94}%
  \BibitemOpen
  \bibfield  {author} {\bibinfo {author} {\bibfnamefont {G.~H.}\ \bibnamefont
  {Weiss}},\ }\href@noop {} {\emph {\bibinfo {title} {Aspects and applications
  of the random walk}}}\ (\bibinfo  {publisher} {North-Holland},\ \bibinfo
  {address} {Amsterdam},\ \bibinfo {year} {1994})\BibitemShut {NoStop}%
\bibitem [{\citenamefont {Giona}(2018)}]{Giona18}%
  \BibitemOpen
  \bibfield  {author} {\bibinfo {author} {\bibfnamefont {M.}~\bibnamefont
  {Giona}},\ }\bibfield  {title} {\enquote {\bibinfo {title} {Lattice random
  walk: an old problem with a future ahead},}\ }\href@noop {} {\bibfield
  {journal} {\bibinfo  {journal} {Phys. Scr.}\ }\textbf {\bibinfo {volume}
  {93}},\ \bibinfo {pages} {095201} (\bibinfo {year} {2018})}\BibitemShut
  {NoStop}%
\bibitem [{\citenamefont {Fedotov}\ \emph {et~al.}(2015)\citenamefont
  {Fedotov}, \citenamefont {Tan},\ and\ \citenamefont {Zubarev}}]{FTZ15}%
  \BibitemOpen
  \bibfield  {author} {\bibinfo {author} {\bibfnamefont {S.}\ \bibnamefont
  {Fedotov}}, \bibinfo {author} {\bibfnamefont {A.}\ \bibnamefont {Tan}}, \
  and\ \bibinfo {author} {\bibfnamefont {A.}\ \bibnamefont {Zubarev}},\
  }\bibfield  {title} {\enquote {\bibinfo {title} {{Persistent random walk of
  cells involving anomalous effects and random death}},}\ }\href@noop {}
  {\bibfield  {journal} {\bibinfo  {journal} {Phys. Rev. E}\ }\textbf {\bibinfo
  {volume} {91}},\ \bibinfo {pages} {1--10} (\bibinfo {year}
    {2015})}\BibitemShut {NoStop}%
\bibitem [{\citenamefont {Giona}\ \emph
  {et~al.}(2019{\natexlab{a}})\citenamefont {Giona}, \citenamefont
  {D'Ovidio}, \citenamefont {Cocco}, \citenamefont {Cairoli},\ and\
  \citenamefont {Klages}}]{Giona2019age}%
  \BibitemOpen
  \bibfield  {author} {\bibinfo {author} {\bibfnamefont {M.}\
  \bibnamefont {Giona}}, \bibinfo {author} {\bibfnamefont {M.}\ \bibnamefont
  {D'Ovidio}}, \bibinfo {author} {\bibfnamefont {Davide}\ \bibnamefont
  {Cocco}}, \bibinfo {author} {\bibfnamefont {A.}\ \bibnamefont {Cairoli}},
  \ and\ \bibinfo {author} {\bibfnamefont {R.}\ \bibnamefont {Klages}},\
  }\bibfield  {title} {\enquote {\bibinfo {title} {Age representation of
  L{\'e}vy walks: partial density waves, relaxation and first passage time
  statistics},}\ }\href@noop {} {\bibfield  {journal} {\bibinfo  {journal} {J.
  Phys. A}\ }\textbf {\bibinfo {volume} {52}},\ \bibinfo {pages} {384001}
  (\bibinfo {year} {2019}{\natexlab{a}})}\BibitemShut {NoStop}%
\bibitem [{\citenamefont {Chen}\ \emph {et~al.}(2015)\citenamefont {Chen},
  \citenamefont {Wang},\ and\ \citenamefont {Granick}}]{CWG15}%
  \BibitemOpen
  \bibfield  {author} {\bibinfo {author} {\bibfnamefont {K.}~\bibnamefont
  {Chen}}, \bibinfo {author} {\bibfnamefont {B.}~\bibnamefont {Wang}}, \ and\
  \bibinfo {author} {\bibfnamefont {S.}~\bibnamefont {Granick}},\ }\bibfield
  {title} {\enquote {\bibinfo {title} {Memoryless self-reinforcing
  directionality in endosomal active transport within living cells},}\
  }\href@noop {} {\bibfield  {journal} {\bibinfo  {journal} {Nat. Mater.}\
  }\textbf {\bibinfo {volume} {14}},\ \bibinfo {pages} {589--593} (\bibinfo
  {year} {2015})}\BibitemShut {NoStop}%
\bibitem [{\citenamefont {Song}\ \emph {et~al.}(2018)\citenamefont {Song},
  \citenamefont {Moon}, \citenamefont {Jeon},\ and\ \citenamefont
  {Park}}]{SMJP18}%
  \BibitemOpen
  \bibfield  {author} {\bibinfo {author} {\bibfnamefont {M.~S.}\
  \bibnamefont {Song}}, \bibinfo {author} {\bibfnamefont {H.~C.}\
  \bibnamefont {Moon}}, \bibinfo {author} {\bibfnamefont {J.~H.}\
  \bibnamefont {Jeon}}, \ and\ \bibinfo {author} {\bibfnamefont {H.~Y.}\
  \bibnamefont {Park}},\ }\bibfield  {title} {\enquote {\bibinfo {title}
  {{Neuronal messenger ribonucleoprotein transport follows an aging L{\'{e}}vy
  walk}},}\ }\href@noop {} {\bibfield  {journal} {\bibinfo  {journal} {Nat.
  Commun.}\ }\textbf {\bibinfo {volume} {9}},\ \bibinfo {pages} {1--8}
  (\bibinfo {year} {2018})}\BibitemShut {NoStop}%
\bibitem [{\citenamefont {Korabel}\ \emph {et~al.}(2018)\citenamefont
  {Korabel}, \citenamefont {Waigh}, \citenamefont {Fedotov},\ and\
  \citenamefont {Allan}}]{KWFA18}%
  \BibitemOpen
  \bibfield  {author} {\bibinfo {author} {\bibfnamefont {N.}\
  \bibnamefont {Korabel}}, \bibinfo {author} {\bibfnamefont {T.~A.}\
  \bibnamefont {Waigh}}, \bibinfo {author} {\bibfnamefont {S.}\
  \bibnamefont {Fedotov}}, \ and\ \bibinfo {author} {\bibfnamefont {V.~J.}\
  \bibnamefont {Allan}},\ }\bibfield  {title} {\enquote {\bibinfo {title}
  {Non-Markovian intracellular transport with sub-diffusion and run-length
  dependent detachment rate},}\ }\href@noop {} {\bibfield  {journal} {\bibinfo
  {journal} {PLoS ONE}\ }\textbf {\bibinfo {volume} {13}},\ \bibinfo {pages}
  {e0207436/1--23} (\bibinfo {year} {2018})}\BibitemShut {NoStop}%
\bibitem [{\citenamefont {Wang}\ \emph
  {et~al.}(2012{\natexlab{a}})\citenamefont {Wang}, \citenamefont {K.},
  \citenamefont {Bae},\ and\ \citenamefont {Granick}}]{Wang:2012aa}%
  \BibitemOpen
  \bibfield  {author} {\bibinfo {author} {\bibfnamefont {B.}~\bibnamefont
  {Wang}}, \bibinfo {author} {\bibfnamefont {J.}\ \bibnamefont {Kuo}},
  \bibinfo {author} {\bibfnamefont {S.~C.}\ \bibnamefont {Bae}}, \ and\
  \bibinfo {author} {\bibfnamefont {S.}\ \bibnamefont {Granick}},\
  }\bibfield  {title} {\enquote {\bibinfo {title} {When {Brownian} diffusion is
  not {Gaussian}},}\ }\href@noop {} {\bibfield  {journal} {\bibinfo  {journal}
  {Nature Mater.}\ }\textbf {\bibinfo {volume} {11}},\ \bibinfo {pages}
  {481--485} (\bibinfo {year} {2012}{\natexlab{a}})}\BibitemShut {NoStop}%
\bibitem [{\citenamefont {Chechkin}\ \emph {et~al.}(2017)\citenamefont
  {Chechkin}, \citenamefont {Seno}, \citenamefont {Metzler},\ and\
  \citenamefont {Sokolov}}]{Chechkin:2017aa}%
  \BibitemOpen
  \bibfield  {author} {\bibinfo {author} {\bibfnamefont {A.~V.}\
  \bibnamefont {Chechkin}}, \bibinfo {author} {\bibfnamefont {F.}\
  \bibnamefont {Seno}}, \bibinfo {author} {\bibfnamefont {R.}\ \bibnamefont
  {Metzler}}, \ and\ \bibinfo {author} {\bibfnamefont {I.~M.}\ \bibnamefont
  {Sokolov}},\ }\bibfield  {title} {\enquote {\bibinfo {title} {Brownian yet
  non-{Gaussian} diffusion: From superstatistics to subordination of diffusing
  diffusivities},}\ }\href@noop {} {\bibfield  {journal} {\bibinfo  {journal}
  {Phys. Rev. X}\ }\textbf {\bibinfo {volume} {7}},\ \bibinfo {pages} {021002}
  (\bibinfo {year} {2017})}\BibitemShut {NoStop}%
\bibitem [{\citenamefont {Pinsky}(1991)}]{Pins91}%
  \BibitemOpen
  \bibfield  {author} {\bibinfo {author} {\bibfnamefont {M.~A}\ \bibnamefont
  {Pinsky}},\ }\href@noop {} {\emph {\bibinfo {title} {Lectures on Random
  Evolution}}}\ (\bibinfo  {publisher} {World Scientific},\ \bibinfo {address}
  {Singapore},\ \bibinfo {year} {1991})\BibitemShut {NoStop}%
\bibitem [{\citenamefont {Zaburdaev}\ \emph
  {et~al.}(2016{\natexlab{a}})\citenamefont {Zaburdaev}, \citenamefont
  {Fouxon}, \citenamefont {Denisov},\ and\ \citenamefont {Barkai}}]{ZFDB16}%
  \BibitemOpen
  \bibfield  {author} {\bibinfo {author} {\bibfnamefont {V.}~\bibnamefont
  {Zaburdaev}}, \bibinfo {author} {\bibfnamefont {I.}~\bibnamefont {Fouxon}},
  \bibinfo {author} {\bibfnamefont {S.}~\bibnamefont {Denisov}}, \ and\
  \bibinfo {author} {\bibfnamefont {E.}~\bibnamefont {Barkai}},\ }\bibfield
  {title} {\enquote {\bibinfo {title} {Superdiffusive dispersals impart the
  geometry of underlying random walks},}\ }\href@noop {} {\bibfield  {journal}
  {\bibinfo  {journal} {Phys. Rev. Lett.}\ }\textbf {\bibinfo {volume} {117}},\
  \bibinfo {pages} {270601} (\bibinfo {year} {2016}{\natexlab{a}})}\BibitemShut
  {NoStop}%
\bibitem [{\citenamefont {Albers}\ and\ \citenamefont
  {Radons}(2018{\natexlab{a}})}]{AlRa18}%
  \BibitemOpen
  \bibfield  {author} {\bibinfo {author} {\bibfnamefont {T.}\ \bibnamefont
  {Albers}}\ and\ \bibinfo {author} {\bibfnamefont {G.}\ \bibnamefont
  {Radons}},\ }\bibfield  {title} {\enquote {\bibinfo {title} {Exact results
  for the nonergodicity of $d$-dimensional generalized L\'evy walks},}\ }\href
  {\doibase 10.1103/PhysRevLett.120.104501} {\bibfield  {journal} {\bibinfo
  {journal} {Phys. Rev. Lett.}\ }\textbf {\bibinfo {volume} {120}},\ \bibinfo
  {pages} {104501} (\bibinfo {year} {2018}{\natexlab{a}})}\BibitemShut
  {NoStop}%
  \bibitem [{Note2()}]{Note2}%
  \BibitemOpen
  \bibinfo {note} {Without loss of generality, we restrict ourselves
    here and in the following to two-state LWs. We remark, however,
    that the velocity LW model could easily be embedded into our
    general theory that we develop here by appropriately adjusting the
    transition matrix Eq.~(\ref {eq4_14}) in
    Appendix~\ref{subsec:LWrecovery}.}\BibitemShut {Stop}%
\bibitem [{\citenamefont {Hughes}\ \emph {et~al.}(1981)\citenamefont {Hughes},
  \citenamefont {Shlesinger},\ and\ \citenamefont {Montroll}}]{Hughes1981}%
  \BibitemOpen
  \bibfield  {author} {\bibinfo {author} {\bibfnamefont {B.~D.}\ \bibnamefont
  {Hughes}}, \bibinfo {author} {\bibfnamefont {M.~F.}\ \bibnamefont
  {Shlesinger}}, \ and\ \bibinfo {author} {\bibfnamefont {E.~W.}\
  \bibnamefont {Montroll}},\ }\bibfield  {title} {\enquote {\bibinfo {title}
  {Random walks with self-similar clusters},}\ }\href@noop {} {\bibfield
  {journal} {\bibinfo  {journal} {Proc. Natl. Acad. Sci.}\ }\textbf {\bibinfo
  {volume} {78}},\ \bibinfo {pages} {3287--3291} (\bibinfo {year}
  {1981})}\BibitemShut {NoStop}%
\bibitem [{\citenamefont {Shlesinger}\ \emph {et~al.}(1995)\citenamefont
  {Shlesinger}, \citenamefont {Zaslavsky},\ and\ \citenamefont
  {Frisch}}]{Shlesinger1995}%
  \BibitemOpen
  \bibfield  {author} {\bibinfo {author} {\bibfnamefont {M.~F.}\
  \bibnamefont {Shlesinger}}, \bibinfo {author} {\bibfnamefont {G.~M.}\
  \bibnamefont {Zaslavsky}}, \ and\ \bibinfo {author} {\bibfnamefont {U.}\
  \bibnamefont {Frisch}},\ }\href@noop {} {\emph {\bibinfo {title} {L{\'e}vy
  flights and related topics in {Physics}}}},\ \bibinfo {series} {Lecture notes
  in {Physics}}, Vol.\ \bibinfo {volume} {450}\ (\bibinfo  {publisher}
  {Springer-Verlag},\ \bibinfo {year} {1995})\BibitemShut {NoStop}%
\bibitem [{\citenamefont {Sokolov}\ and\ \citenamefont
  {Metzler}(2003)}]{sokolov2003}%
  \BibitemOpen
  \bibfield  {author} {\bibinfo {author} {\bibfnamefont {I.~M.}\ \bibnamefont
  {Sokolov}}\ and\ \bibinfo {author} {\bibfnamefont {R.}~\bibnamefont
  {Metzler}},\ }\bibfield  {title} {\enquote {\bibinfo {title} {Towards
  deterministic equations for {L}{\'e}vy walks: The fractional material
  derivative},}\ }\href@noop {} {\bibfield  {journal} {\bibinfo  {journal}
  {Phys. Rev. E}\ }\textbf {\bibinfo {volume} {67}},\ \bibinfo {pages} {010101}
  (\bibinfo {year} {2003})}\BibitemShut {NoStop}%
\bibitem [{\citenamefont {Friedrich}\ \emph {et~al.}(2006)\citenamefont
  {Friedrich}, \citenamefont {Jenko}, \citenamefont {Baule},\ and\
  \citenamefont {Eule}}]{Friedrich2006}%
  \BibitemOpen
  \bibfield  {author} {\bibinfo {author} {\bibfnamefont {R.}~\bibnamefont
  {Friedrich}}, \bibinfo {author} {\bibfnamefont {F.}~\bibnamefont {Jenko}},
  \bibinfo {author} {\bibfnamefont {A.}~\bibnamefont {Baule}}, \ and\ \bibinfo
  {author} {\bibfnamefont {S.}~\bibnamefont {Eule}},\ }\bibfield  {title}
  {\enquote {\bibinfo {title} {{Anomalous Diffusion of Inertial, Weakly Damped
  Particles}},}\ }\href@noop {} {\bibfield  {journal} {\bibinfo  {journal}
  {Phys. Rev. Lett.}\ }\textbf {\bibinfo {volume} {96}},\ \bibinfo {pages}
  {230601} (\bibinfo {year} {2006})}\BibitemShut {NoStop}%
\bibitem [{\citenamefont {Cairoli}\ and\ \citenamefont
  {Baule}(2015)}]{cairoli2015}%
  \BibitemOpen
  \bibfield  {author} {\bibinfo {author} {\bibfnamefont {A.}~\bibnamefont
  {Cairoli}}\ and\ \bibinfo {author} {\bibfnamefont {A.}~\bibnamefont
  {Baule}},\ }\bibfield  {title} {\enquote {\bibinfo {title} {{Anomalous
  Processes with General Waiting Times: Functionals and Multipoint
  Structure}},}\ }\href@noop {} {\bibfield  {journal} {\bibinfo  {journal}
  {Phys. Rev. Lett.}\ }\textbf {\bibinfo {volume} {115}},\ \bibinfo {pages}
  {110601} (\bibinfo {year} {2015})}\BibitemShut {NoStop}%
\bibitem [{\citenamefont {Campos}\ \emph {et~al.}(2015)\citenamefont {Campos},
  \citenamefont {Abad}, \citenamefont {M\'endez}, \citenamefont {Yuste},\ and\
  \citenamefont {Lindenberg}}]{CAMYL15}%
  \BibitemOpen
  \bibfield  {author} {\bibinfo {author} {\bibfnamefont {D.}~\bibnamefont
  {Campos}}, \bibinfo {author} {\bibfnamefont {E.}~\bibnamefont {Abad}},
  \bibinfo {author} {\bibfnamefont {V.}~\bibnamefont {M\'endez}}, \bibinfo
  {author} {\bibfnamefont {S.~B.}\ \bibnamefont {Yuste}}, \ and\ \bibinfo
  {author} {\bibfnamefont {K.}~\bibnamefont {Lindenberg}},\ }\bibfield  {title}
  {\enquote {\bibinfo {title} {Optimal search strategies of space-time coupled
  random walkers with finite lifetimes},}\ }\href@noop {} {\bibfield  {journal}
  {\bibinfo  {journal} {Phys. Rev. E}\ }\textbf {\bibinfo {volume} {91}},\
  \bibinfo {pages} {052115} (\bibinfo {year} {2015})}\BibitemShut {NoStop}%
\bibitem [{\citenamefont {Cox}\ and\ \citenamefont {Miller}(1965)}]{CoMi65}%
  \BibitemOpen
  \bibfield  {author} {\bibinfo {author} {\bibfnamefont {D.~R.}\ \bibnamefont
  {Cox}}\ and\ \bibinfo {author} {\bibfnamefont {H.~D.}\ \bibnamefont
  {Miller}},\ }\href@noop {} {\emph {\bibinfo {title} {The theory of stochastic
  processes}}}\ (\bibinfo  {publisher} {Methuen},\ \bibinfo {address}
  {London},\ \bibinfo {year} {1965})\BibitemShut {NoStop}%
\bibitem [{\citenamefont {Lee}\ and\ \citenamefont {Wang}(2003)}]{Lee2003}%
  \BibitemOpen
  \bibfield  {author} {\bibinfo {author} {\bibfnamefont {E.~T.}\ \bibnamefont
  {Lee}}\ and\ \bibinfo {author} {\bibfnamefont {J.}\ \bibnamefont {Wang}},\
  }\href@noop {} {\emph {\bibinfo {title} {Statistical methods for survival
  data analysis}}},\ Vol.\ \bibinfo {volume} {476}\ (\bibinfo  {publisher}
             {John Wiley \& Sons},\ \bibinfo {year} {2003})\BibitemShut {NoStop}%
  \bibitem [{Note3()}]{Note3}%
  \BibitemOpen
  \bibinfo {note} { To be complete, we need also to supplement the model with 
  regularity conditions at infinity with respect to both $\tau$ and $x$. 
  Specifically, we assume that $p_\pm(x,t,\tau)$, 
  for any $x \in {\mathbb R}$, $t >0$, 
  decay faster than any polynomials 
  for $\tau \rightarrow \infty$, and analogously for $|x|\rightarrow \infty$, 
  for any $t,\tau \geq 0$,
  \begin{equation*}
  \lim_{\tau \rightarrow \infty} \tau^q \, p_\pm(x,\tau,t)  =
  \lim_{|x| \rightarrow \infty} x^q \, p_\pm(x,\tau,t)=0
  \qquad \forall q=0,1,2,\dots \:.
  %\label{eq:regul}
  \end{equation*}
  These equations are  satisfied {\em a fortiori} if the initial 
  conditions admit compact support both in space and in $\tau$, 
  owing to the finite velocity of propagation and to the physical meaning of $\tau$. 
  In particular, for the initial conditions Eq.~(\ref{eq:ic}), 
  and assuming, e.g., that $p_0(x)=0$ for $|x| > a$, 
  we have $p_\pm(x,\tau,t)=0$ for $|x| > a + b \, t$ and likewise 
  $p_\pm(x,\tau,t)=0$ for $\tau>t$. 
  Consequently, Eqs.~(\ref{eq:ppdw}) and 
  the integrals entering in Eqs.~(\ref{eq:bc}), 
  (\ref{eq2_8}) and (\ref{eq:gpdw}), {\em ipso facto} are limited to the 
  closed interval $[0,t]$.
}\BibitemShut {Stop}%
\bibitem [{\citenamefont {Laskin}(2003)}]{Lask03}%
  \BibitemOpen
  \bibfield  {author} {\bibinfo {author} {\bibfnamefont {N.}\ \bibnamefont
  {Laskin}},\ }\bibfield  {title} {\enquote {\bibinfo {title} {{Fractional
  Poisson process}},}\ }\href@noop {} {\bibfield  {journal} {\bibinfo
  {journal} {Commun. Nonlin. Sci. Num. Simul.}\ }\textbf {\bibinfo {volume}
  {8}},\ \bibinfo {pages} {201--213} (\bibinfo {year} {2003})}\BibitemShut
  {NoStop}%
\bibitem [{\citenamefont {Scalas}\ \emph {et~al.}(2004)\citenamefont {Scalas},
  \citenamefont {Gorenflo},\ and\ \citenamefont {Mainardi}}]{SGM04}%
  \BibitemOpen
  \bibfield  {author} {\bibinfo {author} {\bibfnamefont {E.}\ \bibnamefont
  {Scalas}}, \bibinfo {author} {\bibfnamefont {R.}\ \bibnamefont
  {Gorenflo}}, \ and\ \bibinfo {author} {\bibfnamefont {F.}\
  \bibnamefont {Mainardi}},\ }\bibfield  {title} {\enquote {\bibinfo {title}
  {{Uncoupled continuous-time random walks: Solution and limiting behavior of
  the master equation}},}\ }\href@noop {} {\bibfield  {journal} {\bibinfo
  {journal} {Phys. Rev. E}\ }\textbf {\bibinfo {volume} {69}},\ \bibinfo
  {pages} {011107} (\bibinfo {year} {2004})}\BibitemShut {NoStop}%
  \bibitem [{Note4()}]{Note4}%
  \BibitemOpen
  \bibinfo {note} {A related model, called a fractional Poisson 
  process, was studied in the literature with $T(\tau)$ as a generalised 
  Mittag-Leffler function exhibiting power law tails \cite{Lask03,SGM04}. 
  We remark that in the following we relax the 
  condition of $\tau$ being positive to $\tau\in\mathbb{R}$.
  }\BibitemShut {Stop}%
\bibitem [{\citenamefont {Strogatz}(2018)}]{Str18}%
  \BibitemOpen
  \bibfield  {author} {\bibinfo {author} {\bibfnamefont {S.~H.}\ \bibnamefont
  {Strogatz}},\ }\href@noop {} {\emph {\bibinfo {title} {Nonlinear Dynamics and
  Chaos: With Applications to Physics, Biology, Chemistry, and Engineering}}},\
  Studies in nonlinearity\ (\bibinfo  {publisher} {CRC Press},\ \bibinfo
  {address} {Boca Raton},\ \bibinfo {year} {2018})\BibitemShut {NoStop}%
\bibitem [{\citenamefont {Giona}\ \emph
  {et~al.}(2017{\natexlab{c}})\citenamefont {Giona}, \citenamefont
  {Brasiello},\ and\ \citenamefont {Crescitelli}}]{GBC17a}%
  \BibitemOpen
  \bibfield  {author} {\bibinfo {author} {\bibfnamefont {M.}\
  \bibnamefont {Giona}}, \bibinfo {author} {\bibfnamefont {A.}\
  \bibnamefont {Brasiello}}, \ and\ \bibinfo {author} {\bibfnamefont
  {S.}\ \bibnamefont {Crescitelli}},\ }\bibfield  {title} {\enquote
  {\bibinfo {title} {Stochastic foundations of undulatory transport phenomena:
  generalized Poisson-Kac processes --- part ii irreversibility, norms and
  entropies},}\ }\href@noop {} {\bibfield  {journal} {\bibinfo  {journal}
  {Journal of Physics A: Mathematical and Theoretical}\ }\textbf {\bibinfo
  {volume} {50}},\ \bibinfo {pages} {335003} (\bibinfo {year}
  {2017}{\natexlab{c}})}\BibitemShut {NoStop}%
\bibitem [{\citenamefont {Giona}\ \emph
  {et~al.}(2017{\natexlab{d}})\citenamefont {Giona}, \citenamefont
  {Brasiello},\ and\ \citenamefont {Crescitelli}}]{GBC17c}%
  \BibitemOpen
  \bibfield  {author} {\bibinfo {author} {\bibfnamefont {M.}\
  \bibnamefont {Giona}}, \bibinfo {author} {\bibfnamefont {A.}\
  \bibnamefont {Brasiello}}, \ and\ \bibinfo {author} {\bibfnamefont
  {S.}\ \bibnamefont {Crescitelli}},\ }\bibfield  {title} {\enquote
  {\bibinfo {title} {Stochastic foundations of undulatory transport phenomena:
  generalized Poisson-Kac processes --- part iii extensions and applications to
  kinetic theory and transport},}\ }\href@noop {} {\bibfield  {journal}
  {\bibinfo  {journal} {Journal of Physics A: Mathematical and Theoretical}\
  }\textbf {\bibinfo {volume} {50}},\ \bibinfo {pages} {335004} (\bibinfo
  {year} {2017}{\natexlab{d}})}\BibitemShut {NoStop}%
  \bibitem [{\citenamefont {Cox}(1962)}]{Cox1962}%
  \BibitemOpen
  \bibfield  {author} {\bibinfo {author} {\bibfnamefont {D.~R.}~\bibnamefont {Cox}},\ }
  \href@noop {} {\emph {\bibinfo {title} {Renewal Theory}}}\
  (\bibinfo  {publisher} {John Wiley \& Sons},\ \bibinfo
  {address} {New York},\ \bibinfo {year} {1962})\BibitemShut {NoStop}%
\bibitem [{Note5()}]{Note5}%
  \BibitemOpen
  \bibinfo {note} {We remark that all the stochastic processes $X(t)$
  possessing finite propagation velocity and bounded $\lambda (\tau )$
  described by Eq.~(\ref {eq:generalised PKv}) possess Lipshitz trajectories.
  Therefore, no issues arise with the definition of stochastic
  integrals.}\BibitemShut {Stop}%  
\bibitem [{\citenamefont {Bielecki}\ \emph {et~al.}(2012)\citenamefont
  {Bielecki}, \citenamefont {Jakubowski},\ and\ \citenamefont
  {Nieweglowki}}]{cmc1}%
  \BibitemOpen
  \bibfield  {author} {\bibinfo {author} {\bibfnamefont {R.}~\bibnamefont
  {Bielecki}}, \bibinfo {author} {\bibfnamefont {J.}~\bibnamefont
  {Jakubowski}}, \ and\ \bibinfo {author} {\bibfnamefont {M.}~\bibnamefont
  {Nieweglowki}},\ }\bibfield  {title} {\enquote {\bibinfo {title} {Intricacies
  of dependence between components of multivariate Markov chains: weak Markov
  consistency and weak Markov copulae},}\ }\href@noop {} {\bibfield  {journal}
  {\bibinfo  {journal} {Electron. J. Probab.}\ }\textbf {\bibinfo {volume}
  {18}},\ \bibinfo {pages} {1--21} (\bibinfo {year} {2012})}\BibitemShut
  {NoStop}%
\bibitem [{\citenamefont {Bielecki}\ \emph {et~al.}(2017)\citenamefont
  {Bielecki}, \citenamefont {Jakubowski},\ and\ \citenamefont
  {Nieweglowki}}]{cmc2}%
  \BibitemOpen
  \bibfield  {author} {\bibinfo {author} {\bibfnamefont {R.}~\bibnamefont
  {Bielecki}}, \bibinfo {author} {\bibfnamefont {J.}~\bibnamefont
  {Jakubowski}}, \ and\ \bibinfo {author} {\bibfnamefont {M.}~\bibnamefont
  {Nieweglowki}},\ }\bibfield  {title} {\enquote {\bibinfo {title} {Conditional
  Markov chains: Properties, construction and structured dependence},}\
  }\href@noop {} {\bibfield  {journal} {\bibinfo  {journal} {Stoc. Proces.
  Appl.}\ }\textbf {\bibinfo {volume} {127}},\ \bibinfo {pages} {1125--1170}
  (\bibinfo {year} {2017})}\BibitemShut {NoStop}%
\bibitem [{\citenamefont {Karlis}\ and\ \citenamefont
  {Xekalaki}(2005)}]{multicount}%
  \BibitemOpen
  \bibfield  {author} {\bibinfo {author} {\bibfnamefont {D.}~\bibnamefont
  {Karlis}}\ and\ \bibinfo {author} {\bibfnamefont {E.}~\bibnamefont
  {Xekalaki}},\ }\bibfield  {title} {\enquote {\bibinfo {title} {Mixed Poisson
  distributions},}\ }\href@noop {} {\bibfield  {journal} {\bibinfo  {journal}
  {Int. Stat. Rev.}\ }\textbf {\bibinfo {volume} {73}},\ \bibinfo {pages}
  {35--58} (\bibinfo {year} {2005})}\BibitemShut {NoStop}%
\bibitem [{\citenamefont {Hayflick}\ and\ \citenamefont
  {Moorhead}(1961)}]{Hayflick1961}%
  \BibitemOpen
  \bibfield  {author} {\bibinfo {author} {\bibfnamefont {L.}\ \bibnamefont
  {Hayflick}}\ and\ \bibinfo {author} {\bibfnamefont {P.~S.}\ \bibnamefont
  {Moorhead}},\ }\bibfield  {title} {\enquote {\bibinfo {title} {The serial
  cultivation of human diploid cell strains},}\ }\href@noop {} {\bibfield
  {journal} {\bibinfo  {journal} {Exp. Cell Res.}\ }\textbf {\bibinfo {volume}
  {25}},\ \bibinfo {pages} {585--621} (\bibinfo {year} {1961})}\BibitemShut
  {NoStop}%
\bibitem [{\citenamefont {Hayflick}(1965)}]{Hayflick1965}%
  \BibitemOpen
  \bibfield  {author} {\bibinfo {author} {\bibfnamefont {L.}\ \bibnamefont
  {Hayflick}},\ }\bibfield  {title} {\enquote {\bibinfo {title} {The limited in
  vitro lifetime of human diploid cell strains},}\ }\href@noop {} {\bibfield
  {journal} {\bibinfo  {journal} {Exp. Cell Res.}\ }\textbf {\bibinfo {volume}
  {37}},\ \bibinfo {pages} {614--636} (\bibinfo {year} {1965})}\BibitemShut
  {NoStop}%
\bibitem [{\citenamefont {Harley}\ \emph {et~al.}(1990)\citenamefont {Harley},
  \citenamefont {Futcher},\ and\ \citenamefont {Greider}}]{Harley1990}%
  \BibitemOpen
  \bibfield  {author} {\bibinfo {author} {\bibfnamefont {C.~B.}\
  \bibnamefont {Harley}}, \bibinfo {author} {\bibfnamefont {A.~B.}\
  \bibnamefont {Futcher}}, \ and\ \bibinfo {author} {\bibfnamefont {C.~W.}\
  \bibnamefont {Greider}},\ }\bibfield  {title} {\enquote {\bibinfo {title}
  {Telomeres shorten during ageing of human fibroblasts},}\ }\href@noop {}
  {\bibfield  {journal} {\bibinfo  {journal} {Nature}\ }\textbf {\bibinfo
  {volume} {345}},\ \bibinfo {pages} {458--460} (\bibinfo {year}
  {1990})}\BibitemShut {NoStop}%
\bibitem [{\citenamefont {Krtolica}\ \emph {et~al.}(2001)\citenamefont
  {Krtolica}, \citenamefont {Parrinello}, \citenamefont {Lockett},
  \citenamefont {Desprez},\ and\ \citenamefont {Campisi}}]{Krtolica2001}%
  \BibitemOpen
  \bibfield  {author} {\bibinfo {author} {\bibfnamefont {A.}\ \bibnamefont
  {Krtolica}}, \bibinfo {author} {\bibfnamefont {S.}\ \bibnamefont
  {Parrinello}}, \bibinfo {author} {\bibfnamefont {S.}\ \bibnamefont
  {Lockett}}, \bibinfo {author} {\bibfnamefont {P.~-Y.}\ \bibnamefont
  {Desprez}}, \ and\ \bibinfo {author} {\bibfnamefont {J.}\ \bibnamefont
  {Campisi}},\ }\bibfield  {title} {\enquote {\bibinfo {title} {Senescent
  fibroblasts promote epithelial cell growth and tumorigenesis: a link between
  cancer and aging},}\ }\href@noop {} {\bibfield  {journal} {\bibinfo
  {journal} {Proc. Nat. Acad. Sci.}\ }\textbf {\bibinfo {volume} {98}},\
  \bibinfo {pages} {12072--12077} (\bibinfo {year} {2001})}\BibitemShut
  {NoStop}%
\bibitem [{\citenamefont {Wang}\ \emph {et~al.}(2009)\citenamefont {Wang},
  \citenamefont {Anthony}, \citenamefont {Bae},\ and\ \citenamefont
  {Granick}}]{Wang2009}%
  \BibitemOpen
  \bibfield  {author} {\bibinfo {author} {\bibfnamefont {B.}~\bibnamefont
  {Wang}}, \bibinfo {author} {\bibfnamefont {S.~M.}\ \bibnamefont
  {Anthony}}, \bibinfo {author} {\bibfnamefont {S.~C.}\ \bibnamefont
  {Bae}}, \ and\ \bibinfo {author} {\bibfnamefont {S.}\ \bibnamefont
  {Granick}},\ }\bibfield  {title} {\enquote {\bibinfo {title} {Anomalous yet
  Brownian},}\ }\href@noop {} {\bibfield  {journal} {\bibinfo  {journal}
  {Proceedings of the National Academy of Sciences}\ }\textbf {\bibinfo
  {volume} {106}},\ \bibinfo {pages} {15160--15164} (\bibinfo {year}
  {2009})}\BibitemShut {NoStop}%
\bibitem [{\citenamefont {Wang}\ \emph
  {et~al.}(2012{\natexlab{b}})\citenamefont {Wang}, \citenamefont {Kuo},
  \citenamefont {Bae},\ and\ \citenamefont {Granick}}]{Wang2012}%
  \BibitemOpen
  \bibfield  {author} {\bibinfo {author} {\bibfnamefont {B.}~\bibnamefont
  {Wang}}, \bibinfo {author} {\bibfnamefont {J.}\ \bibnamefont {Kuo}},
  \bibinfo {author} {\bibfnamefont {S.~C.}\ \bibnamefont {Bae}}, \ and\
  \bibinfo {author} {\bibfnamefont {S.}\ \bibnamefont {Granick}},\
  }\bibfield  {title} {\enquote {\bibinfo {title} {When Brownian diffusion is
  not Gaussian},}\ }\href@noop {} {\bibfield  {journal} {\bibinfo  {journal}
  {Nat. Mater.}\ }\textbf {\bibinfo {volume} {11}},\ \bibinfo {pages}
  {481--485} (\bibinfo {year} {2012}{\natexlab{b}})}\BibitemShut {NoStop}%
\bibitem [{\citenamefont {Toyota}\ \emph {et~al.}(2011)\citenamefont {Toyota},
  \citenamefont {Head}, \citenamefont {Schmidt},\ and\ \citenamefont
  {Mizuno}}]{Toyota2011}%
  \BibitemOpen
  \bibfield  {author} {\bibinfo {author} {\bibfnamefont {T.}\
  \bibnamefont {Toyota}}, \bibinfo {author} {\bibfnamefont {D.~A.}\
  \bibnamefont {Head}}, \bibinfo {author} {\bibfnamefont {C.~F.}\
  \bibnamefont {Schmidt}}, \ and\ \bibinfo {author} {\bibfnamefont {D.}\
  \bibnamefont {Mizuno}},\ }\bibfield  {title} {\enquote {\bibinfo {title}
  {Non-Gaussian athermal fluctuations in active gels},}\ }\href@noop {}
  {\bibfield  {journal} {\bibinfo  {journal} {Soft Matter}\ }\textbf {\bibinfo
  {volume} {7}},\ \bibinfo {pages} {3234--3239} (\bibinfo {year}
  {2011})}\BibitemShut {NoStop}%
\bibitem [{\citenamefont {Valentine}\ \emph {et~al.}(2001)\citenamefont
  {Valentine}, \citenamefont {Kaplan}, \citenamefont {Thota}, \citenamefont
  {Crocker}, \citenamefont {Gisler}, \citenamefont {Prud'homme}, \citenamefont
  {Beck},\ and\ \citenamefont {Weitz}}]{Valentine2001}%
  \BibitemOpen
  \bibfield  {author} {\bibinfo {author} {\bibfnamefont {M.~T.}\ \bibnamefont
  {Valentine}}, \bibinfo {author} {\bibfnamefont {P.~D.}\ \bibnamefont
  {Kaplan}}, \bibinfo {author} {\bibfnamefont {D.}~\bibnamefont {Thota}},
  \bibinfo {author} {\bibfnamefont {J.~C.}\ \bibnamefont {Crocker}}, \bibinfo
  {author} {\bibfnamefont {T.}~\bibnamefont {Gisler}}, \bibinfo {author}
  {\bibfnamefont {R.~K.}\ \bibnamefont {Prud'homme}}, \bibinfo {author}
  {\bibfnamefont {M.}~\bibnamefont {Beck}}, \ and\ \bibinfo {author}
  {\bibfnamefont {D.~A.}\ \bibnamefont {Weitz}},\ }\bibfield  {title} {\enquote
  {\bibinfo {title} {Investigating the microenvironments of inhomogeneous soft
  materials with multiple particle tracking},}\ }\href {\doibase
  10.1103/PhysRevE.64.061506} {\bibfield  {journal} {\bibinfo  {journal} {Phys.
  Rev. E}\ }\textbf {\bibinfo {volume} {64}},\ \bibinfo {pages} {061506}
  (\bibinfo {year} {2001})}\BibitemShut {NoStop}%
\bibitem [{\citenamefont {e~Silva}\ \emph {et~al.}(2014)\citenamefont
  {e~Silva}, \citenamefont {Stuhrmann}, \citenamefont {Betz},\ and\
  \citenamefont {Koenderink}}]{e2014time}%
  \BibitemOpen
  \bibfield  {author} {\bibinfo {author} {\bibfnamefont {M.~S.}\
  \bibnamefont {e~Silva}}, \bibinfo {author} {\bibfnamefont {B.}\
  \bibnamefont {Stuhrmann}}, \bibinfo {author} {\bibfnamefont {T.}\
  \bibnamefont {Betz}}, \ and\ \bibinfo {author} {\bibfnamefont {G.~H.}\
  \bibnamefont {Koenderink}},\ }\bibfield  {title} {\enquote {\bibinfo {title}
  {Time-resolved microrheology of actively remodeling actomyosin networks},}\
  }\href@noop {} {\bibfield  {journal} {\bibinfo  {journal} {New J. Phys.}\
  }\textbf {\bibinfo {volume} {16}},\ \bibinfo {pages} {075010} (\bibinfo
  {year} {2014})}\BibitemShut {NoStop}%
\bibitem [{\citenamefont {Samanta}\ and\ \citenamefont
  {Chakrabarti}(2016)}]{Samanta2016}%
  \BibitemOpen
  \bibfield  {author} {\bibinfo {author} {\bibfnamefont {N.}\
  \bibnamefont {Samanta}}\ and\ \bibinfo {author} {\bibfnamefont {R.}\
  \bibnamefont {Chakrabarti}},\ }\bibfield  {title} {\enquote {\bibinfo {title}
  {Tracer diffusion in a sea of polymers with binding zones: mobile vs. frozen
  traps},}\ }\href@noop {} {\bibfield  {journal} {\bibinfo  {journal} {Soft
  Matter}\ }\textbf {\bibinfo {volume} {12}},\ \bibinfo {pages} {8554--8563}
  (\bibinfo {year} {2016})}\BibitemShut {NoStop}%
\bibitem [{\citenamefont {Leptos}\ \emph {et~al.}(2009)\citenamefont {Leptos},
  \citenamefont {Guasto}, \citenamefont {Gollub}, \citenamefont {Pesci},\ and\
  \citenamefont {Goldstein}}]{Leptos2009}%
  \BibitemOpen
  \bibfield  {author} {\bibinfo {author} {\bibfnamefont {K.~C.}\
  \bibnamefont {Leptos}}, \bibinfo {author} {\bibfnamefont {J.~S.}\
  \bibnamefont {Guasto}}, \bibinfo {author} {\bibfnamefont {J.~P.}\
  \bibnamefont {Gollub}}, \bibinfo {author} {\bibfnamefont {A.~I.}\
  \bibnamefont {Pesci}}, \ and\ \bibinfo {author} {\bibfnamefont {R.~E.}\
  \bibnamefont {Goldstein}},\ }\bibfield  {title} {\enquote {\bibinfo {title}
  {Dynamics of enhanced tracer diffusion in suspensions of swimming eukaryotic
  microorganisms},}\ }\href@noop {} {\bibfield  {journal} {\bibinfo  {journal}
  {Phys. Rev. Lett.}\ }\textbf {\bibinfo {volume} {103}},\ \bibinfo {pages}
  {198103} (\bibinfo {year} {2009})}\BibitemShut {NoStop}%
\bibitem [{\citenamefont {Hapca}\ \emph {et~al.}(2009)\citenamefont {Hapca},
  \citenamefont {Crawford},\ and\ \citenamefont {Young}}]{Hapca2009}%
  \BibitemOpen
  \bibfield  {author} {\bibinfo {author} {\bibfnamefont {S.}\ \bibnamefont
  {Hapca}}, \bibinfo {author} {\bibfnamefont {J.~W.}\ \bibnamefont
  {Crawford}}, \ and\ \bibinfo {author} {\bibfnamefont {I.~M.}\ \bibnamefont
  {Young}},\ }\bibfield  {title} {\enquote {\bibinfo {title} {Anomalous
  diffusion of heterogeneous populations characterized by normal diffusion at
  the individual level},}\ }\href@noop {} {\bibfield  {journal} {\bibinfo
  {journal} {J. R. Soc. Interface}\ }\textbf {\bibinfo {volume} {6}},\ \bibinfo
  {pages} {111--122} (\bibinfo {year} {2009})}\BibitemShut {NoStop}%
\bibitem [{\citenamefont {Witzel}\ \emph {et~al.}(2009)\citenamefont {Witzel},
  \citenamefont {G\"otz},\ and\ \citenamefont {Lanoisel{\'e}e},\ and\ \citenamefont {Franosch},\   and\ \citenamefont {Grebenkov} ,\ and\ \citenamefont {Heinrich}}]{Witzel2019}%
  \BibitemOpen
  \bibfield  {author} {\bibinfo {author} {\bibfnamefont {P.}\ \bibnamefont
  {Witzel}}, \bibinfo {author} {\bibfnamefont {M.}\ \bibnamefont
  {G{\"o}tz}},\ \bibinfo {author} {\bibfnamefont {Y.}\ \bibnamefont
  {Lanoisel{\'e}e}},\ \bibinfo {author} {\bibfnamefont {T.}\ \bibnamefont
      {Franosch}},\ \bibinfo {author} {\bibfnamefont {D.~S.}\ \bibnamefont
      {Grebenkov}},\ and\ \bibinfo {author} {\bibfnamefont {D.}\ \bibnamefont
  {Heinrich}},\ }\bibfield  {title} {\enquote {\bibinfo {title} {Heterogeneities shape passive intracellular transport},
  }\ }\href@noop {} {\bibfield  {journal} {\bibinfo
  {journal} {Biophys. J.}\ }\textbf {\bibinfo {volume} {117}},\ \bibinfo
  {pages} {203--213} (\bibinfo {year} {2019})}\BibitemShut {NoStop}%
\bibitem [{\citenamefont {Beck}\ and\ \citenamefont {Cohen}(2003)}]{Beck2003}%
  \BibitemOpen
  \bibfield  {author} {\bibinfo {author} {\bibfnamefont {C.}\
  \bibnamefont {Beck}}\ and\ \bibinfo {author} {\bibfnamefont {E.~G.~D.}\
  \bibnamefont {Cohen}},\ }\bibfield  {title} {\enquote {\bibinfo {title}
  {Superstatistics},}\ }\href@noop {} {\bibfield  {journal} {\bibinfo
  {journal} {Phys. A}\ }\textbf {\bibinfo {volume} {322}},\ \bibinfo {pages}
  {267--275} (\bibinfo {year} {2003})}\BibitemShut {NoStop}%
\bibitem [{\citenamefont {Chubynsky}\ and\ \citenamefont
  {Slater}(2014)}]{Chubynsky2014}%
  \BibitemOpen
  \bibfield  {author} {\bibinfo {author} {\bibfnamefont {M.~V.}\
  \bibnamefont {Chubynsky}}\ and\ \bibinfo {author} {\bibfnamefont {G.~W.}\
  \bibnamefont {Slater}},\ }\bibfield  {title} {\enquote {\bibinfo {title}
  {Diffusing diffusivity: a model for anomalous, yet Brownian, diffusion},}\
  }\href@noop {} {\bibfield  {journal} {\bibinfo  {journal} {Phys. Rev. Lett.}\
  }\textbf {\bibinfo {volume} {113}},\ \bibinfo {pages} {098302} (\bibinfo
  {year} {2014})}\BibitemShut {NoStop}%
\bibitem [{\citenamefont {Kanazawa}\ \emph {et~al.}(2020)\citenamefont
  {Kanazawa}, \citenamefont {Sano}, \citenamefont {Cairoli},\ and\
  \citenamefont {Baule}}]{Kanazawa2020}%
  \BibitemOpen
  \bibfield  {author} {\bibinfo {author} {\bibfnamefont {K.}\ \bibnamefont
  {Kanazawa}}, \bibinfo {author} {\bibfnamefont {T.~G.}\ \bibnamefont
  {Sano}}, \bibinfo {author} {\bibfnamefont {A.}\ \bibnamefont {Cairoli}},
  \ and\ \bibinfo {author} {\bibfnamefont {A.}\ \bibnamefont {Baule}},\
  }\bibfield  {title} {\enquote {\bibinfo {title} {Loopy L{\'e}vy flights
  enhance tracer diffusion in active suspensions},}\ }\href@noop {} {\bibfield
  {journal} {\bibinfo  {journal} {Nature}\ }\textbf {\bibinfo {volume} {579}},\
  \bibinfo {pages} {364--367} (\bibinfo {year} {2020})}\BibitemShut {NoStop}%
\bibitem [{\citenamefont {Giona}\ and\ \citenamefont
  {Pucci}(2019)}]{ming_giona}%
  \BibitemOpen
  \bibfield  {author} {\bibinfo {author} {\bibfnamefont {M.}~\bibnamefont
  {Giona}}\ and\ \bibinfo {author} {\bibfnamefont {L.}~\bibnamefont {Pucci}},\
  }\bibfield  {title} {\enquote {\bibinfo {title} {Hyperbolic heat/mass
  transport and stochastic modelling-three simple problems},}\ }\href@noop {}
  {\bibfield  {journal} {\bibinfo  {journal} {Mathematics in Engineering}\
  }\textbf {\bibinfo {volume} {1}},\ \bibinfo {pages} {224} (\bibinfo {year}
  {2019})}\BibitemShut {NoStop}%
\bibitem [{\citenamefont {Ben-Avraham}\ and\ \citenamefont
  {Havlin}(2000)}]{Ben2000}%
  \BibitemOpen
  \bibfield  {author} {\bibinfo {author} {\bibfnamefont {D.}\ \bibnamefont
  {Ben-Avraham}}\ and\ \bibinfo {author} {\bibfnamefont {S.}\ \bibnamefont
  {Havlin}},\ }\href@noop {} {\emph {\bibinfo {title} {Diffusion and reactions
  in fractals and disordered systems}}}\ (\bibinfo  {publisher} {Cambridge
  university press},\ \bibinfo {year} {2000})\BibitemShut {NoStop}%
\bibitem [{\citenamefont {Giona}\ \emph {et~al.}(2016)\citenamefont {Giona},
  \citenamefont {Brasiello},\ and\ \citenamefont {Crescitelli}}]{GBC16b}%
  \BibitemOpen
  \bibfield  {author} {\bibinfo {author} {\bibfnamefont {M.}~\bibnamefont
  {Giona}}, \bibinfo {author} {\bibfnamefont {A.}~\bibnamefont {Brasiello}}, \
  and\ \bibinfo {author} {\bibfnamefont {S.}~\bibnamefont {Crescitelli}},\
  }\bibfield  {title} {\enquote {\bibinfo {title} {On the influence of
  reflective boundary conditions on the statistics of Poisson-Kac diffusion
  processes},}\ }\href@noop {} {\bibfield  {journal} {\bibinfo  {journal}
  {Physica A}\ }\textbf {\bibinfo {volume} {450}},\ \bibinfo {pages} {148--164}
  (\bibinfo {year} {2016})}\BibitemShut {NoStop}%
\bibitem [{\citenamefont {Eule}\ \emph {et~al.}(2012)\citenamefont {Eule},
  \citenamefont {Zaburdaev}, \citenamefont {Friedrich},\ and\ \citenamefont
  {Geisel}}]{Eule2012}%
  \BibitemOpen
  \bibfield  {author} {\bibinfo {author} {\bibfnamefont {S.}~\bibnamefont
  {Eule}}, \bibinfo {author} {\bibfnamefont {V.}~\bibnamefont {Zaburdaev}},
  \bibinfo {author} {\bibfnamefont {R.}~\bibnamefont {Friedrich}}, \ and\
  \bibinfo {author} {\bibfnamefont {T.}~\bibnamefont {Geisel}},\ }\bibfield
  {title} {\enquote {\bibinfo {title} {Langevin description of superdiffusive
  L{\'e}vy processes},}\ }\href@noop {} {\bibfield  {journal} {\bibinfo
  {journal} {Physical Review E}\ }\textbf {\bibinfo {volume} {86}},\ \bibinfo
  {pages} {041134} (\bibinfo {year} {2012})}\BibitemShut {NoStop}%
\bibitem [{\citenamefont {Wang}\ \emph {et~al.}(2019)\citenamefont {Wang},
  \citenamefont {Chen},\ and\ \citenamefont {Deng}}]{Wang2019}%
  \BibitemOpen
  \bibfield  {author} {\bibinfo {author} {\bibfnamefont {X.}\ \bibnamefont
  {Wang}}, \bibinfo {author} {\bibfnamefont {Y.}\ \bibnamefont {Chen}}, \ and\
  \bibinfo {author} {\bibfnamefont {W.}\ \bibnamefont {Deng}},\ }\bibfield
  {title} {\enquote {\bibinfo {title} {L{\'e}vy-walk-like Langevin dynamics},}\
  }\href@noop {} {\bibfield  {journal} {\bibinfo  {journal} {New Journal of
  Physics}\ }\textbf {\bibinfo {volume} {21}},\ \bibinfo {pages} {013024}
  (\bibinfo {year} {2019})}\BibitemShut {NoStop}%
\bibitem [{\citenamefont {Magdziarz}\ \emph {et~al.}(2012)\citenamefont
  {Magdziarz}, \citenamefont {Szczotka},\ and\ \citenamefont
  {{\.Z}ebrowski}}]{Magdziarz2012}%
  \BibitemOpen
  \bibfield  {author} {\bibinfo {author} {\bibfnamefont {M.}\ \bibnamefont
  {Magdziarz}}, \bibinfo {author} {\bibfnamefont {W.}\
  \bibnamefont {Szczotka}}, \ and\ \bibinfo {author} {\bibfnamefont {P.}\
  \bibnamefont {{\.Z}ebrowski}},\ }\bibfield  {title} {\enquote {\bibinfo
  {title} {Langevin picture of L\'evy walks and their extensions},}\ }\href@noop
  {} {\bibfield  {journal} {\bibinfo  {journal} {Journal of Statistical
  Physics}\ }\textbf {\bibinfo {volume} {147}},\ \bibinfo {pages} {74--96}
  (\bibinfo {year} {2012})}\BibitemShut {NoStop}%
\bibitem [{\citenamefont {Cercignani}(1988)}]{cercignani}%
  \BibitemOpen
  \bibfield  {author} {\bibinfo {author} {\bibfnamefont {C.}~\bibnamefont {Cercignani}},\ 
  }\href@noop {} {\emph {\bibinfo {title} {The Boltzmann Equation and Its Applications}}}\ 
  (\bibinfo  {publisher} {Springer Verlag},\ \bibinfo
  {address} {New York},\ \bibinfo {year} {1988})\BibitemShut {NoStop}%
\bibitem [{\citenamefont {Chandrasekhar}(1960)}]{chandrasekhar}%
  \BibitemOpen
  \bibfield  {author} {\bibinfo {author} {\bibfnamefont {S.}~\bibnamefont {Chandrasekhar}},\ 
  }\href@noop {} {\emph {\bibinfo {title} {Radiative Transfer}}}\ 
  (\bibinfo  {publisher} {Dover Publ.},\ \bibinfo
  {address} {New York},\ \bibinfo {year} {1960})\BibitemShut {NoStop}%  
\bibitem [{\citenamefont {Giona}\ \emph {et~al.}(2016)\citenamefont {Giona},
  \citenamefont {Brasiello},\ and\ \citenamefont {Crescitelli}}]{brasiello}%
  \BibitemOpen
  \bibfield  {author} {\bibinfo {author} {\bibfnamefont {M.}\ \bibnamefont
  {Giona}}, \bibinfo {author} {\bibfnamefont {A.}\ \bibnamefont {Brasiello}}, \
  and\ \bibinfo {author} {\bibfnamefont {S.}\ \bibnamefont {Crescitelli}},\
  }\bibfield  {title} {\enquote {\bibinfo {title} {On the influence of reflective boundary 
  conditions on the statistics of Poisson–Kac diffusion processes},}\ 
  }\href@noop {} {\bibfield  {journal} {\bibinfo  {journal} {Phys. A}\ 
  }\textbf {\bibinfo {volume} {450}},\ \bibinfo {pages} {148--164}
  (\bibinfo {year} {2016})}\BibitemShut {NoStop}%
\bibitem [{\citenamefont {Adrover}\ \emph {et~al.}(2021)\citenamefont {Adrover},
  \citenamefont {Venditti},\ and\ \citenamefont {Giona}}]{gels}%
  \BibitemOpen
  \bibfield  {author} {\bibinfo {author} {\bibfnamefont {A.}\ \bibnamefont
  {Adrover}}, \bibinfo {author} {\bibfnamefont {C.}\ \bibnamefont {Venditti}}, \
  and\ \bibinfo {author} {\bibfnamefont {M.}\ \bibnamefont {Giona}},\
  }\bibfield  {title} {\enquote {\bibinfo {title} {Swelling and Drug Release in Polymers 
  through the Theory of Poisson–Kac Stochastic Processes},}\ 
  }\href@noop {} {\bibfield  {journal} {\bibinfo  {journal} {Gels}\ 
  }\textbf {\bibinfo {volume} {7}},\ \bibinfo {pages} {32--51}
  (\bibinfo {year} {2021})}\BibitemShut {NoStop}%   
\bibitem [{\citenamefont {Bardou}(2002)}]{lasercooling}%
  \BibitemOpen
  \bibfield  {author} {\bibinfo {author} {\bibfnamefont {F.}~\bibnamefont {Bardou}},\ 
  , \bibinfo {author} {\bibfnamefont {J.~-P.}\ \bibnamefont {Bouchaud}}, \
  , \bibinfo {author} {\bibfnamefont {A.}\ \bibnamefont {Aspect}}, \ and \
  , \bibinfo {author} {\bibfnamefont {C.}\ \bibnamefont {Cohen-Tannoudi}} 
  }\href@noop {} {\emph {\bibinfo {title} {L\'evy Statistics and Laser Cooling}}}\ 
  (\bibinfo  {publisher} {Cambridge Univ. Press},\ \bibinfo
  {address} {Cambridge},\ \bibinfo {year} {2002})\BibitemShut {NoStop}%    
\bibitem [{\citenamefont {Stokes}\ (1851)\citenamefont {Stokes}}]{stokes1851}%
  \BibitemOpen
  \bibfield  {author} {\bibinfo {author} {\bibfnamefont {G.~G.}\ \bibnamefont
  {Stokes}}, \ }\bibfield  {title} {\enquote {\bibinfo {title} {On the effect of internal friction
of fluids on the motion of a pendulum},}\ }\href@noop {} {\bibfield  {journal} {\bibinfo  {journal} {Trans. Cambr. Philos. Soc.
}\ }\textbf {\bibinfo {volume} {9}},\ \bibinfo {pages} {8--106}
  (\bibinfo {year} {1851})}\BibitemShut {NoStop}%
\bibitem [{\citenamefont {Landau}(1959)}]{landau}%
  \BibitemOpen
  \bibfield  {author} {\bibinfo {author} {\bibfnamefont {L.~D.}~\bibnamefont {Landau}}\ and 
  , \bibinfo {author} {\bibfnamefont {E.~M.}\ \bibnamefont {Lifshitz}}, \ 
  }\href@noop {} {\emph {\bibinfo {title} {Fluid Mechanics}}}\ 
  (\bibinfo  {publisher} {Pergamon Press},\ \bibinfo
  {address} {London},\ \bibinfo {year} {1959})\BibitemShut {NoStop}%    
\bibitem [{\citenamefont {Jannasch}\ \emph {et~al.}(2011)\citenamefont {Jannasch},
  \citenamefont {Mahamdeh},\ and\ \citenamefont {Sch\"affer}}]{bm1}%
  \BibitemOpen
  \bibfield  {author} {\bibinfo {author} {\bibfnamefont {A.}\ \bibnamefont
  {Jannasch}}, \bibinfo {author} {\bibfnamefont {M.}\ \bibnamefont {Mahamdeh}}, \
  and\ \bibinfo {author} {\bibfnamefont {E.}\ \bibnamefont {Sch\"affer}},\
  }\bibfield  {title} {\enquote {\bibinfo {title} {Inertial Effects of a Small Brownian 
  Particle Cause a Colored Power Spectral Density of Thermal Noise},}\ 
  }\href@noop {} {\bibfield  {journal} {\bibinfo  {journal} {Ann.
  Phys. Rev. Lett.}\ }\textbf {\bibinfo {volume} {107}},\ \bibinfo {pages} {228301}
  (\bibinfo {year} {2011})}\BibitemShut {NoStop}%
\bibitem [{\citenamefont {Franosch}\ \emph {et~al.}(2011)}]{bm2}%
  \BibitemOpen
  \bibfield  {author} {\bibinfo {author} {\bibfnamefont {T.}\ \bibnamefont
  {Franosch}}, \bibinfo {author} {\bibfnamefont {M.}\ \bibnamefont {Grimm}}, \
  \bibinfo {author} {\bibfnamefont {M.}\ \bibnamefont {Belushkin}}, \
  \bibinfo {author} {\bibfnamefont {F.~M.}\ \bibnamefont {Mor}}, \
  \bibinfo {author} {\bibfnamefont {G.}\ \bibnamefont {Foffi}}, \
  \bibinfo {author} {\bibfnamefont {L.}\ \bibnamefont {Forr\'o}}, \
  and\ \bibinfo {author} {\bibfnamefont {S.}\ \bibnamefont {Jeney}},\
  }\bibfield  {title} {\enquote {\bibinfo {title} {Resonances arising from hydrodynamic memory in Brownian motion},
  }\ }\href@noop {} {\bibfield  {journal} {\bibinfo  {journal} {Nature}\ 
  }\textbf {\bibinfo {volume} {478}},\ \bibinfo {pages} {85--88}
  (\bibinfo {year} {2011})}\BibitemShut {NoStop}%
\bibitem [{\citenamefont {Huang}\ \emph {et~al.}(2011)}]{bm3}%
  \BibitemOpen
  \bibfield  {author} {\bibinfo {author} {\bibfnamefont {R.}\ \bibnamefont
  {Huang}}, \bibinfo {author} {\bibfnamefont {I.}\ \bibnamefont {Chavez}}, \
  \bibinfo {author} {\bibfnamefont {K.~M.}\ \bibnamefont {Taute}}, \
  \bibinfo {author} {\bibfnamefont {B.}\ \bibnamefont {Lukic}}, \
  \bibinfo {author} {\bibfnamefont {S.}\ \bibnamefont {Jeney}}, \
  \bibinfo {author} {\bibfnamefont {M.~G.}\ \bibnamefont {Raizen}}, \
  and\ \bibinfo {author} {\bibfnamefont {E.~-L.}\ \bibnamefont {Florin}},\
  }\bibfield  {title} {\enquote {\bibinfo {title} {Direct observation of the full transition 
  from ballistic to diffusive Brownian motion in a liquid},}\ }\href@noop {} {\bibfield  {journal} {\bibinfo  {journal} {Nature}\ 
  }\textbf {\bibinfo {volume} {7}},\ \bibinfo {pages} {576--580}
  (\bibinfo {year} {2011})}\BibitemShut {NoStop}%
\bibitem [{\citenamefont {Toner}\ \emph {et~al.}(2005)\citenamefont {Toner},
  \citenamefont {Tu},\ and\ \citenamefont {Ramaswamy}}]{Toner2005}%
  \BibitemOpen
  \bibfield  {author} {\bibinfo {author} {\bibfnamefont {J.}\ \bibnamefont
  {Toner}}, \bibinfo {author} {\bibfnamefont {Y.}\ \bibnamefont {Tu}}, \
  and\ \bibinfo {author} {\bibfnamefont {Sriram}\ \bibnamefont {Ramaswamy}},\
  }\bibfield  {title} {\enquote {\bibinfo {title} {Hydrodynamics and phases of
  flocks},}\ }\href@noop {} {\bibfield  {journal} {\bibinfo  {journal} {Ann.
  Phys.}\ }\textbf {\bibinfo {volume} {318}},\ \bibinfo {pages} {170--244}
  (\bibinfo {year} {2005})}\BibitemShut {NoStop}%
\bibitem [{\citenamefont {Schweitzer}(2007)}]{Schweitzer2007}%
  \BibitemOpen
  \bibfield  {author} {\bibinfo {author} {\bibfnamefont {F.}\ \bibnamefont
  {Schweitzer}},\ }\href@noop {} {\emph {\bibinfo {title} {Brownian agents and
  active particles: collective dynamics in the natural and social sciences}}}\
  (\bibinfo  {publisher} {Springer},\ \bibinfo {year} {2007})\BibitemShut
  {NoStop}%
\bibitem [{\citenamefont {Ramaswamy}(2010)}]{Ramaswamy2010}%
  \BibitemOpen
  \bibfield  {author} {\bibinfo {author} {\bibfnamefont {S.}\ \bibnamefont
  {Ramaswamy}},\ }\bibfield  {title} {\enquote {\bibinfo {title} {The mechanics
  and statistics of active matter},}\ }\href@noop {} {\bibfield  {journal}
  {\bibinfo  {journal} {Annu. Rev. Condens. Matter Phys.}\ }\textbf {\bibinfo
  {volume} {1}},\ \bibinfo {pages} {323--345} (\bibinfo {year}
  {2010})}\BibitemShut {NoStop}%
\bibitem [{\citenamefont {Vicsek}\ and\ \citenamefont
  {Zafeiris}(2012)}]{Vicsek2012}%
  \BibitemOpen
  \bibfield  {author} {\bibinfo {author} {\bibfnamefont {T.}\
  \bibnamefont {Vicsek}}\ and\ \bibinfo {author} {\bibfnamefont {A.}\
  \bibnamefont {Zafeiris}},\ }\bibfield  {title} {\enquote {\bibinfo {title}
  {Collective motion},}\ }\href@noop {} {\bibfield  {journal} {\bibinfo
  {journal} {Phys. Rep.}\ }\textbf {\bibinfo {volume} {517}},\ \bibinfo {pages}
  {71--140} (\bibinfo {year} {2012})}\BibitemShut {NoStop}%
\bibitem [{\citenamefont {Marchetti}\ \emph {et~al.}(2013)\citenamefont
  {Marchetti}, \citenamefont {Joanny}, \citenamefont {Ramaswamy}, \citenamefont
  {Liverpool}, \citenamefont {Prost}, \citenamefont {Rao},\ and\ \citenamefont
  {Simha}}]{Marchetti2013}%
  \BibitemOpen
  \bibfield  {author} {\bibinfo {author} {\bibfnamefont {M.~~C.}\
  \bibnamefont {Marchetti}}, \bibinfo {author} {\bibfnamefont {J.~~F.}\
  \bibnamefont {Joanny}}, \bibinfo {author} {\bibfnamefont {S.}~\bibnamefont
  {Ramaswamy}}, \bibinfo {author} {\bibfnamefont {T.~~B.}\ \bibnamefont
  {Liverpool}}, \bibinfo {author} {\bibfnamefont {J.}~\bibnamefont {Prost}},
  \bibinfo {author} {\bibfnamefont {M.}\ \bibnamefont {Rao}}, \ and\
  \bibinfo {author} {\bibfnamefont {R.~A.}\ \bibnamefont {Simha}},\
  }\bibfield  {title} {\enquote {\bibinfo {title} {Hydrodynamics of soft active
  matter},}\ }\href@noop {} {\bibfield  {journal} {\bibinfo  {journal} {Rev.
  Mod. Phys.}\ }\textbf {\bibinfo {volume} {85}},\ \bibinfo {pages} {1143}
  (\bibinfo {year} {2013})}\BibitemShut {NoStop}%
\bibitem [{\citenamefont {K\"orner}\ and\ \citenamefont
  {Bergmann}(1998)}]{KoeBe98}%
  \BibitemOpen
  \bibfield  {author} {\bibinfo {author} {\bibfnamefont {C.}~\bibnamefont
  {K\"orner}}\ and\ \bibinfo {author} {\bibfnamefont {H.~W.}\ \bibnamefont
  {Bergmann}},\ }\bibfield  {title} {\enquote {\bibinfo {title} {The physical
  defects of the hyperbolic heat conduction equation},}\ }\href@noop {}
  {\bibfield  {journal} {\bibinfo  {journal} {Appl. Phys.}\ }\textbf {\bibinfo
  {volume} {67}},\ \bibinfo {pages} {397--401} (\bibinfo {year}
  {1998})}\BibitemShut {NoStop}%
\end{thebibliography}
%merlin.mbs apsrev4-1.bst 2010-07-25 4.21a (PWD, AO, DPC) hacked
%Control: key (0)
%Control: author (0) dotless jnrlst
%Control: editor formatted (1) identically to author
%Control: production of article title (0) allowed
%Control: page (1) range
%Control: year (0) verbatim
%Control: production of eprint (0) enabled
%

\end{document}